%% file: dycirc.tex
\documentclass[12pt,titlepage]{article}
\usepackage{graphicx}
\usepackage{graphics}
\usepackage{amsmath}
\usepackage{amsfonts}
\usepackage{amssymb}
\usepackage{latexsym}
\usepackage{setspace}

\usepackage{comment}

\usepackage{natbib}

\newcommand{\ctn}{\cite}
\newtheorem{theorem}{Theorem}

\newtheorem{definition}[theorem]{Definition}

\newcommand{\bi}[1]{\mbox{\boldmath{$ #1 $}}}
\begin{document}

\title{Bayesian Nonparametric Dynamic State Space Modeling with Circular Latent States}
\author{Satyaki Mazumder\footnote{Indian Institute of Science Education and Research, Kolkata} and 
Sourabh Bhattacharya\footnote{Indian Statistical Institute, Kolkata}}
\date{\vspace{-0.5in}}
\maketitle

\begin{abstract}
State space models are well-known for their versatility in modeling dynamic systems 
that arise in various scientific disciplines. Although parametric state space models
are well-studied, nonparametric approaches are much less explored in comparison. In this article
we propose a novel Bayesian nonparametric approach to state space modeling assuming that both
the observational and evolutionary functions are unknown and are varying with time; crucially, 
we assume that the unknown evolutionary equation describes dynamic evolution of some
latent circular random variable.

Based on appropriate kernel convolution of the standard Weiner process 
we model the time-varying observational and evolutionary functions as suitable Gaussian processes
that take both linear and circular variables as arguments. Additionally, for
the time-varying evolutionary function, we wrap the Gaussian process thus constructed around the unit circle
to form an appropriate circular Gaussian process.
We show that our process thus created satisfies desirable
properties.

For the purpose of inference we develop an MCMC based methodology combining Gibbs sampling and Metropolis-Hastings
algorithms. Applications to a simulated dataset, a real wind speed dataset and a real ozone dataset 
demonstrated quite encouraging performances of our model and methodologies. 
\\[2mm]
\end{abstract}

\tableofcontents

\section{Introduction}
\label{intro}

\subsection{Flexibility of state space models}
\label{subsec:flexibility}

The versatility of state space models is clearly reflected from their utility in multifarious disciplines
such as engineering, finance, medicine, ecology, statistics, etc. One reason for such widespread use of
state space models is their inherent flexibility which allows modeling complex dynamic systems through
the underlying latent states associated with an ``evolutionary equation" and an ``observational equation"
that corresponds to the observed dynamic data. That most of the established time series models admit appropriate
state space representations (see, for example, \ctn{Durbin01}, \ctn{Shumway11}) is vindication of the 
enormous flexibility of state space models.

\subsection{A brief discussion on state space models with circular states}
\label{subsection:circular_nonpara}

In reality, there may be strong evidences that the observed time series data depends upon some circular time series. 
For instance, the ozone level time series data depends upon wind direction (see \ctn{Jamma06}, for example).
However, data on wind direction are often not recorded along with ozone level. 
A concrete example of such a real data, on which we 
illustrate our model and methodologies, is provided in Section \ref{real data analysis}.
Other examples (see \ctn{Holzmann06}) include time series data on wind speed (linear) and ocean current (linear) 
which depend upon wind direction (circular);
daily peak load of pollutants (linear) and the time of day when the peak is attained (circular);
speed (linear) and direction change (circular) of movements of objects, organisms and animals, to name a few. 
In Section \ref{real data analysis}, for the purpose of illustration, we apply our model and methodologies on 
another real example on wind speed data.
When both the linear and circular time series data are available, 
\ctn{Holzmann06} consider hidden Markov models in a discrete mixture context to statistically analyse such data sets.
Our aim in this article is to propose a novel nonparametric state space approach when the circular
time series data are unobserved, even though they are known to affect the available linear time series data.

\subsection{Need for nonparametric approaches to state space models}
\label{subsec:need_nonpara}

To date, most of the research on state space models have adhered to the parametric set-up,
assuming known forms (either linear or non-linear) of the observational and evolutionary functions.
Recently \ctn{Ghosh14} considered a Bayesian nonparametric approach to state space modeling, assuming
that these time-varying functional forms are unknown, which they modeled by Gaussian processes. However, 
in their approach, observational as well as evolutionary functions consist of only linear arguments and the functions 
were assumed to take values on the real line $\mathbb R$. In our case both the functions have linear 
as well as circular arguments and moreover, 
the evolutionary function itself is circular. Hence to model our unknown observational and evolutionary functions, 
it is necessary to construct a new Gaussian process which can take time and angle as arguments. Therefore, 
a significantly different approach is taken here to deal with the problem. Moreover, as a by-product of the
nonparametric approach based on Gaussian processes, it turned out that the latent states have a non-Markov, non-Gaussian, 
nonparametric distribution with a complex dependence structure, which is suitable for modeling complex,
realistic, dynamic systems. 
Importantly, using our novel methodology, we are able to retain these advantages for a even more challenging set up.
These are briefly discussed in Section \ref{conclusion}; details will be provided in our future work.


\subsection{A brief overview of the contributions and organisation of this paper}
\label{subsec:contributions}

In this paper we use Gaussian processes for modeling the unknown
observational and evolutionary functions. It is important to note here that both the 
functions have arguments which are linear
as well as circular. Moreover, the evolutionary function itself is circular. Thus, it is clear that the 
approaches of any other paper (for example \ctn{Ghosh14} for dynamic modeling in linear components 
and the references therein) previous to ours
are no longer appropriate in such a framework. Hence, quite substantial
methodological advancement is necessary in our case.


We introduce our Bayesian nonparametric state space model with circular latent states in Section \ref{Univar}.
The first challenge is to define a Gaussian process taking time and angle as arguments. We construct an
appropriate Gaussian process by convolving a suitable kernel with the standard Brownian
motion (Weiner process).
The Gaussian process so defined enjoys desirable smoothness properties; moreover, as the absolute
difference between two time points tends to infinity and/or the absolute difference between two angles
tend to $\pi/2$ indicating orthogonality, the Gaussian process based covariance tends to zero as it should be.
We provide these technical details in the Appendix. 
We provide further details of our Gaussian process with respect to continuity and smoothness properties 
in the supplement \ctn{Mazumder14b}, 
whose sections, figures and tables have the prefix ``S-" when referred to in this paper.
The Gaussian process that we create is an appropriate model for the time-varying observational function, but to model
the evolutionary function which is circular in nature, we convert this Gaussian process into a wrapped
Gaussian process so that it becomes a well-defined circular process. 


To obtain the joint distribution of the latent states, 
in Section \ref{sec:lookup} we employ the ``look-up table" approach
of \ctn{Bhattacharya07}, 
but substantially modified for our circular set-up, which
will play an important role in our MCMC based Bayesian inference. 
A detailed discussion on look-up table is also provided in this section. 
%

In Section \ref{simulation study} we illustrate our model and methodologies with a simulation study, where we simulate
the data set from a highly non-linear dynamic model, but fit our nonparametric model, pretending that the data-generating
mechanism is unknown.
Our experiment shows that even in this highly challenging situation our method successfully 
captures future observations in terms of coverage associated with 95\% highest posterior density credible regions. 
It is observed that 
the posterior densities of the latent variable at different time points are multimodal;
in these cases coloured graphical representation of the posterior densities of the latent variables with
higher intensity on the color standing for higher density regions provide useful visual information,
which we adopt. As we find,
most of the true values of the latent variables fall within the high posterior probability regions.
%
In Section \ref{real data analysis} 
we demonstrate the performance of our dynamic nonparametric 
model in the case of two real time series datasets comprising wind speed and  
the level of ozone present in the atmosphere. 
In the first example, wind direction (the relevant data on circular process) are recorded, 
but we analyse the wind speed data
using our circular latent process model assuming unavailability of this dataset, and assess the fit of the
posterior latent process to the actually available wind direction data. Indeed, the purpose of this exercise
is to demonstrate the effectiveness of our method in capturing the true latent process in a real data set-up.
In the second example, although the ozone level is recorded, the relevant wind direction data are not available,
even though ozone level depends upon wind direction; see the discussion in the following paragraph. 
Hence, we analyse the observed ozone level data considering wind direction as a latent circular process.
As such, in both the experiments on real data, we obtain quite encouraging results, particularly in terms
capturing the wind directions associated with the data sets, and the set aside observed data meant for forecasting,
quite precisely.

The simulation study and the wind speed data analysis, however important and interesting, 
are meant for validation of our model and
methodologies, while our actual interest is in analysis of the ozone data using our ideas.  
Since ozone analysis has been the interest of many researchers so far,
it is worth providing a glimpse of the history of 
such data analysis. It is crucial to note that the way we analyse the data is completely different 
from the previous approaches existing in the literature, for instance, 
\ctn{RT87}, \ctn{S89}, \ctn{Jamma06}, \ctn{HHO08}. Most of these papers except \ctn{S89} 
fit parametric regression models taking ozone data as a dependent variable. 
\ctn{S89} uses extreme value analysis to detect trend in ground level ozone. \ctn{Jamma06} point out that 
ozone level depends on wind direction. None of the other papers take wind direction into consideration. 
One reason for not taking advantage of the information on wind direction is that the 
data on wind direction is circular in nature and therefore, 
the usual statistical techniques are rendered invalid, as rightly pointed out by \ctn{Jamma06}. 
But perhaps the more important reason for not accounting for wind direction is the fact that
such data are often not recorded along with the ozone level.  
None of the previous work available in the literature focuses on such an important issue. 
In this paper we analyze such an ozone data considering wind direction as circular latent (unobserved) variable, 
and ozone level as observed linear variable, using our novel nonparametric model. 
Our work differs from the existing ones in two aspects. We are the first to 
analyze such a data using dynamic modeling in a nonparametric framework. Also, 
we are the first to treat wind direction as a circular latent variable and include it in such an analysis.

\section{Gaussian process based dynamic state space model with circular latent states}
\label{Univar}
We introduce our proposed state space model as follows:
For $t=1,2,\ldots T$, 
\begin{align}
 y_t &= f(t,x_t) + \epsilon_t, ~~ \epsilon_t\sim N(0,\sigma^2_{\mbox{\scriptsize{$\epsilon$}}}),
 \label{eq:eq1}
\\[1ex]
x_t &=  \left\{g(t,x_{t-1}) + \eta_t\right\}~[2\pi], ~~ \eta_t\sim N(0,\sigma^2_{\mbox{\scriptsize{$\eta$}}}),
\label{eq:eq2}
\end{align}
where $\{y_t;~t=1,\ldots,T\}$ is the time series observed on the real line; $\{x_t;~t=0,1,\ldots,T\}$ are the latent
circular states; $f(\cdot,\cdot)$ is the unknown observational function taking values on the real line, 
and $g(\cdot,\cdot)$ is the unknown evolutionary function with values on the circular manifold. 
In (\ref{eq:eq2}), $[2\pi]$ stands for the $\mbox{mod}~2\pi$ operation.
Note that 
\begin{align}
\left\{g(t,x_{t-1}) + \eta_t\right\}~[2\pi]
&=\left\{g(t,x_{t-1})~[2\pi] + \eta_t~[2\pi]\right\}~[2\pi]\notag\\
&=\left\{g^*(t,x_{t-1}) + \eta_t\right\}~[2\pi],
\label{eq:modulo}
\end{align}
where $g^*$ is the linear counterpart of $g$, that is, 
$g^*(t,x_{t-1})$ is the linear random variable
such that $g^*(t,x_{t-1})~[2\pi]=g(t,x_{t-1})$. 
For convenience, we shall often use
representation (\ref{eq:modulo}).
Indeed, for obtaining the distribution of $x_t$, we shall first obtain the distribution
of the linear random variable $g^*(t,x_{t-1}) + \eta_t$ and then apply the $\mbox{mod}~2\pi$ operation
to $g^*(t,x_{t-1}) + \eta_t$ to compute the distribution of the circular variable $x_t$.

Both the observational and the evolutionary functions have arguments $t$, which is linear in nature, and
$x$, which is angular. The linear argument has been brought in to ensure that the functions are time-varying, that is, 
the functions are allowed to freely evolve with time. 

\subsection{Gaussian and wrapped Gaussian process representations of the observational and evolutionary functions}
\label{subsec:gp1}

We consider Gaussian and wrapped Gaussian processes to model $f$ and $g$ independently;
for this purpose we first construct appropriate Gaussian processes for $f$ and $g^*$ by convolving
a suitable kernel with the standard Wiener process. 
The details are provided in Appendix \ref{Gaussian-univar}.
Once we build such Gaussian processes, we can convert that for modeling $g$ into a wrapped 
Gaussian process with the $\mbox{mod}~2\pi$ operation applied to $g^*$. However, since our evolutionary equation 
given by (\ref{eq:eq2}) involves the error term $\eta_t$, we will need to compute the distribution of
$g^*(\cdot,\cdot)+\eta_t$ before applying the $\mbox{mod}~2\pi$ operation.

In the Gaussian process construction detailed in Appendix \ref{Gaussian-univar}  
we assume the mean functions of $f$ and $g^*$ to be of forms
$\mu_f(\cdot,\cdot)$ = $\bi{h}(\cdot,\cdot)'\bi{\beta}_f$ and $\mu_g (\cdot,\cdot)$ = $\bi{h}(\cdot,\cdot)'\bi{\beta}_g$, where
$\bi{h}(t,z)$ = $(1,t,\cos(z),\sin(z))'$; here $z$ is an angular quantity and $\bi{\beta}_f$ and $\bi{\beta}_g$ 
are parameters in $\mathbb{R}^4$. As shown in Appendix \ref{subsec:covariance}, 
for any fixed $(t_1,z_1)$ and 
$(t_2,z_{2})$, where $t_1,t_2$ are linear quantities and $z_1,z_2$ are angular quantities, the forms of the covariances are given by 
$c_f((t_1,z_{1}),(t_2,z_{2}))$ = $\exp \{-\sigma_f^4(t_1-t_2)^2\}\cos(|z_{1}-z_{2}|)$ and
$c_g((t_1,z_{1}),(t_2,z_{2}))$ = $\exp \{-\sigma_g^4(t_1-t_2)^2\}\cos(|z_{1}-z_{2}|)$, 
where $\sigma_f$ and $\sigma_g$ are positive, real valued parameters. 

A very attractive property of our Gaussian process is that whenever 
$|\theta_1-\theta_2|$ = $\pi/2$, implying orthogonality of two directions,
the covariance becomes $0$, the difference in time notwithstanding. 
To see that this is a desirable condition, first note that the sample correlation coefficient 
between two vectors is cosine of the angle between them. So, if
the vectors are orthogonal, then the sample correlation coefficient is zero. This simple intuition
seems to encourage development of correlation functions that have this property. The angular correlation function
of \ctn{Dufour76} satisfies this property, albeit it also involves an infinite sum.
The test statistic proposed in \ctn{Epp71}, given by $\sum_{i=1}^n\cos(\theta_i)$, where $\theta_i$
is the angle between unit vectors $\bi{X}_i$ and $\bi{Y}_i$, also satisfies this property.

Obviously, as the time difference tends to infinity, then also the covariance tends to zero.
That desired continuity and smoothness properties hold for our Gaussian process are proved 
in Section S-1 of the supplement.
Thus, the Gaussian process we constructed seems to have quite reasonable features that are desirable in our
linear-circular context.

A pertinent question that arises in the context of modeling the circular latent variables directly using
wrapped Gaussian process is what if some known transformation of $x_t$, say, $z_t=\psi(x_t)$, 
projecting $x_t$ on the Euclidean space, is considered as the relevant (linear) latent process, which
is then modeled using the linear Gaussian process based idea of \ctn{Ghosh14}? The issue here is that
it is usually feasible to postulate a single Gaussian process, but since there is no unique choice
of the transformation $\psi$, under various such transformations the distribution of the original
latent states $x_t$ would be different. To avoid this undesirable feature we modeled $x_t$ directly 
using wrapped Gaussian process.

\subsection{Bayesian hierarchical structure of our nonparametric model based on circular latent states}
\label{subsec:hierarchy}
Our model admits the following hierarchical representation:
\allowdisplaybreaks
{
\begin{align}
[y_t|f,\bi{\theta}_f,x_t]&\sim N\left(f(t,x_t),\sigma^2_{\epsilon}\right);~t=1,\ldots,T,\label{eq:y_t_dist}\\
[x_t|g,\bi{\theta}_g,x_{t-1}]&\sim N\left(g^*(t,x_{t-1}),\sigma^2_{\eta}\right)[2\pi];~t=1,\ldots,T,\label{eq:x_t_dist}\\
[x_0]&\sim N\left(\mu_{x_0},\sigma^2_{x_0}\right)[2\pi],\label{eq:x_0_dist}\\
[f(\cdot,\cdot)|\bi{\theta}_f]&\sim GP\left(\bi{h}(\cdot,\cdot)'\bi{\beta}_f,\sigma^2_fc_f(\cdot,\cdot)\right),\label{eq:gp_f}\\
[g(\cdot,\cdot)|\bi{\theta}_g]&\sim GP\left(\bi{h}(\cdot,\cdot)'\bi{\beta}_g,\sigma^2_gc_g(\cdot,\cdot)\right)[2\pi],\label{eq:gp_g}\\
[\bi{\beta}_f,\sigma^2_f,\bi{\beta}_g,\sigma^2_g,\sigma^2_{\epsilon},\sigma^2_{\eta}]
&=[\bi{\beta}_f,\sigma^2_f][\bi{\beta}_g,\sigma^2_g][\sigma^2_{\epsilon},\sigma^2_{\eta}],\label{eq:theta_prior}
\end{align}
}where $\bi{\theta}_f = (\bi{\beta}_f,\sigma_f,\sigma_{\epsilon})'$ and 
$\bi{\theta}_g = (\bi{\beta}_g,\sigma_g,\sigma_{\eta})'.$
In the above, GP stands for ``Gaussian Process". 
%
%
%
%
Integrating out $f(\cdot,\cdot)$ from the above hierarchical structure we obtain that 
given $x_1,\ldots,x_T$,  $\bi{D}_T=(y_1,\ldots,y_T)'$  
has the multivariate normal distribution of dimension $T$ with mean 
\begin{equation}
\label{eq3: mean y|all}
\bi{\mu}_{y_{t}} = \bi{H}_{D_{T}}\bi{\beta}_f
\end{equation}
and covariance matrix 
\begin{equation}
\label{eq4: cov y|all}
\bi{\Sigma}_{y_{t}} = \sigma^2_f \bi{A}_{f}+\sigma^2_{\mbox{\scriptsize $\epsilon$}}\bi{I}_{T},
\end{equation}
with $\bi{H}_{D_{T}}'$ = $(\bi{h}(1,x_1),\ldots,\bi{h}(T,x_T))$ and the 
$(i,j)$-th element of $\bi{A}_{f}$ being $c_f((i,x_{i}),(j,x_{j}))$. 

For obtaining the joint distribution of the latent circular state variables, we consider the 
``look-up" table approach, but before introducing this, which we discuss in details in Section \ref{sec:lookup},
in the next section we provide details regarding the prior distributions of the parameters associated
with the above hierarchical structure.

\subsection{Prior specifications}
\label{Priors}
We assume the following prior distributions.
\begin{align}
 [x_0]&\sim \mbox{von Mises} (\mu_0,\sigma_0^2)
\\[1ex]
 [\sigma^2_{\mbox{\scriptsize $\epsilon$}}] & \propto (\sigma^2_{\mbox{\scriptsize $\epsilon$}})^{\left(-\frac{\alpha_{\mbox{\tiny $\epsilon$}}+2}{2}\right)}\mbox{ exp }\left\{-\frac{\gamma_{\mbox{\tiny $\epsilon$}}}{2\sigma^2_{\mbox{\scriptsize $\epsilon$}}}\right\}; ~~ \alpha_{\mbox{\tiny $\epsilon$}},\,\gamma_{\mbox{\tiny $\epsilon$}} >0
\\[1ex]
 [\sigma^2_{\mbox{\scriptsize $\eta$}}] & \propto (\sigma^2_{\mbox{\scriptsize $\eta$}})^{\left(-\frac{\alpha_{\mbox{\tiny $\eta$}}+2}{2}\right)}\mbox{ exp }\left\{-\frac{\gamma_{\mbox{\tiny $\eta$}}}{2\sigma^2_{\mbox{\scriptsize $\eta$}}}\right\}; ~~ \alpha_{\mbox{\tiny $\eta$}},\,\gamma_{\mbox{\tiny $\eta$}} > 0
 \\[1ex]
  [\sigma^2_{g}] & \propto (\sigma^2_{g})^{\left(-\frac{\alpha_{\mbox{\tiny $g$}}+2}{2}\right)}\mbox{ exp }\left\{-\frac{\gamma_{\mbox{\tiny $g$}}}{2\sigma^2_{g}}\right\}; ~~ \alpha_{\mbox{\tiny $g$}},\,\gamma_{\mbox{\tiny $g$}} >0
 \\[1ex]
 [\sigma^2_{f}] & \propto (\sigma^2_{f})^{\left(-\frac{\alpha_{\mbox{\tiny $f$}}+2}{2}\right)}\mbox{ exp }\left\{-\frac{\gamma_{\mbox{\tiny $f$}}}{2\sigma^2_{f}}\right\}; ~~ \alpha_{\mbox{\tiny $f$}},\,\gamma_{\mbox{\tiny $f$}} >0
\\[1ex]
 [\bi{\beta}_{f}] &\sim N(\bi{\beta}_{f,0},\Sigma_{\beta_{f,0}})
\\[1ex]
[\bi{\beta}_{g}] &\sim N(\bi{\beta}_{g,0},\Sigma_{\beta_{g,0}}).
\end{align}
Choice of the prior parameters are discussed in 
Sections \ref{simulation study} and \ref{real data analysis}.

\section{Look-up table approach to representing the distribution of the latent circular time series}
\label{sec:lookup}

For obtaining the joint distribution of the latent circular variables, we employ the look-up table approach of 
\ctn{Bhattacharya07}, but because of the circular nature of the latent states, appropriate modifications
are necessary (for treatment of look-up table in linear dynamic system one may see \ctn{Ghosh14}). 
In the next section we briefly discuss the intuition behind look-up table idea for our circular set-up.

\subsection{Intuition behind the look-up table approach} 
\label{subsec:lookup}

For illustrative purposes let $\eta_t=0$ for all $t$, yielding the model
$x_t=g^*(t,x_{t-1})~[2\pi]$. 
Let us first assume that the entire linear process $g^*(\cdot)$ is available. This means that for every input 
$u=(t,z)$, where $t>0$ and $z$ lies on the unit circle, 
the corresponding $g^*(u)$ is available, thus constituting a look-up table, with the first column representing
$u$ and the second column representing the corresponding $g^{*}(u)$.
Conditional on $(t,x_{t-1})$, $x_t=g^*(t,x_{t-1})$ can be obtained 
by simply picking the input $(t,x_{t-1})$ from the first column of the look-up table, locating  
the corresponding output value $g^*(t,x_{t-1})$ in the second column of the look-up table, and then finally
reporting $x_t=g^*(t,x_{t-1})~[2\pi]$. 
In practice, we will simulate the Gaussian process $g^*$ on a fine enough grid of inputs, and conditional
on this simulated process, will simulate from the conditional distribution of 
$g^*(t,x_{t-1})$, given $x_{t-1}$, before applying the modulo $2\pi$ operation.
By making the grid as fine as required, this strategy can be made to approximate $x_t$ as accurately as desired;
this is formalized in \ctn{Ghosh14} and easily goes through in our circular set-up.
By repeating the aforementioned procedure for each $t$, the joint distribution of the circular state variables 
can be approximated as closely as desired. Details of our strategy are provided in the next section.

\subsection{Details of the lookup table approach in our circular context}
\label{subsec:lookup_details}

We consider a set of grid points in the interval $[0,2\pi]$; let this set 
be denoted by $\bi{G}_z=\{z_1,\ldots,z_n\}$. Let $\bi{D}_z$ = $(g^*(1,z_1),\ldots,g^*(n,z_n))$.
Note that
$\bi{D}_z$ has a joint 
multivariate normal distribution of dimension $n$ with mean vector 
\begin{equation}
\label{eq5:mean D_z}
E[\bi{D}_z|\bi{\beta}_g,\sigma^2_g] = \bi{H}_{D_{z}}\bi{\beta}_{g},
\end{equation}
and covariance matrix 
\begin{equation}
\label{eq6:cov D_z}
V[\bi{D}_z|\bi{\beta}_g,\sigma^2_g] = \sigma^2_g \bi{A}_{g,D_{z}},
\end{equation}
where $\bi{H}_{D_{z}}'$ = $(\bi{h}(1,z_1),\ldots,\bi{h}(n,z_n))$ and the $(i,j)$-th element of $\bi{A}_{g,D_{z}}$ is 
$c_g((t_i,z_i),(t_j,z_j))$. 
%
The conditional distribution of $\bi{D}_z$ given $(x_0,g^*(1,x_0))$, $\bi{\beta}_g$ and $\sigma^2_g$ 
is an $n$-variate normal with mean vector
 
\begin{equation}
\label{eq8:mean D_z|g}
 E[\bi{D}_z|\bi{\beta}_g,\sigma^2_g, x_0,g^*(1,x_0)]= \bi{H}_{D_{z}}\bi{\beta}_{g} 
 + \bi{s}_{g,D_z}(1,x_0) (g^*(1,x_0)-\bi{h}(1,x_0)'\bi{\beta}_g)
\end{equation}
and conditional variance
\begin{equation}
\label{eq9:cov D_z|g}
 \mbox{Var}\left[\bi{D}_z|\bi{\beta}_g,\sigma^2_g, x_0,g^*(1,x_0)\right] = \sigma^2_g\left(\bi{A}_{g,D_{z}} 
 - \bi{s}_{g,D_z}(1,x_0) (\bi{s}_{g,D_z}(1,x_0))'\right),
\end{equation}
where $\bi{s}_{g,D_z}(\cdot,\cdot) = (c_{g}((\cdot,\cdot),(t_1,z_1)),\ldots,c_g((\cdot,\cdot),(t_n,z_n)))'$.
\par

The conditional distribution of $g^*(t,x_{t-1})$ given $\bi{D}_z$ and $x_{t-1}$
is a normal distribution with mean 
\begin{equation}
\label{eq10:mean g|D_z}
E[g^*(t,x_{t-1})|\bi{D}_z,x_{t-1},\bi{\beta}_g,\sigma^2_g] = \bi{h}(t,x_{t-1})'\bi{\beta}_g + (\bi{s}_{g,D_z}(t,x_{t-1}))' 
\bi{A}_{g,D_{z}}^{-1} (\bi{D}_z-\bi{H}_{D_{z}}\bi{\beta}_{g})
\end{equation}
and variance 
\begin{equation}
\label{eq11:var g|D_z}
\mbox{Var}\left[g^*(t,x_{t-1})|\bi{D}_z,x_{t-1},\bi{\beta}_g,\sigma^2_g \right] = 
\sigma_g^2\left(1-(\bi{s}_{g,D_z}(t,x_{t-1}))' \bi{A}_{g,D_{z}}^{-1}\bi{s}_{g,D_z}(t,x_{t-1})\right). 
\end{equation}

With the above distributional details, our procedure of representing the circular latent states in terms of the 
auxiliary random vector $\bi{D}_z$, conditional on $\bi{\beta}_g$ and $\sigma^2_g$, can be described as follows.  
\begin{enumerate}
 \item $x_0\sim \pi^*$, where $\pi^*$ is some appropriate prior distribution on the unit circle.
 \item Given $x_0$, $\bi{\beta}_g$ and $\sigma^2_g$, 
 $x_1$ = $g^*(1,x_0)~[2\pi]$ = $g(1,x_0)$, where $g^*(1,x_0)$ has a normal distribution 
 with mean $\bi{h}(1,x_0)'\bi{\beta}_g$ and variance ${\sigma_g^2}$.
 \item Given $x_0$, $x_1$, $\bi{\beta}_g$ and $\sigma^2_g$,
 $[\bi{D}_z|x_0,g^*(1,x_0),\bi{\beta}_g,\sigma^2_g]$ is a multivariate 
 normal distribution with mean (\ref{eq8:mean D_z|g}) and covariance matrix (\ref{eq9:cov D_z|g}).
 \item For $t$ = $2,3,\ldots$, $x^*_t\sim [g^*(t,x_{t-1})|\bi{D}_z,x_{t-1},\bi{\beta}_g,\sigma^2_g]$ which is a normal distribution
 with mean and variance given by (\ref{eq10:mean g|D_z}) and (\ref{eq11:var g|D_z}) respectively; 
 $x_t$ is related to $x^*_t$ via $x_t$ = $x^*_t~[2\pi]$.
\end{enumerate}


\subsection{Joint distribution of 
of the latent circular variables induced by the look-up table}
\label{Look up table for x(T+1)}
Using the look-up table approach the joint distribution of $(\bi{D}_z,x_0, x_1,x_2,\ldots,x_T,x_{T+1})$ given $\bi{\beta}_g$, 
$\sigma^2_{\mbox{\scriptsize{$\eta$}}}$ and $\sigma^2_{g}$ is as follows:
\begin{align}
[x_0,x_1,\ldots,x_{T+1},\bi{D}_z|\bi{\beta}_g,\sigma^2_{\mbox{\scriptsize{$\eta$}}},\sigma^2_{g}] 
&= [x_0][x_1=\left\{g^*(1,x_0)+\eta_1\right\}~[2\pi]|x_0,\sigma^2_{\mbox{\scriptsize{$\eta$}}},\sigma^2_g]
\left[\bi{D}_z|x_0,g^*(1,x_0),\right. 
\notag\\
& \qquad \left. \bi{\beta}_g,\sigma^2_g\right] \times\prod_{t=1}^T \left[x_{t+1}=\left\{g^*((t+1),x_t)+\eta_{t+1}\right\}~[2\pi]|\bi{\beta}_g,\sigma^2_g, \right. 
\notag\\
&\qquad \left. \bi{D}_z,x_t,
\sigma^2_{\mbox{\scriptsize{$\eta$}}}\right]
.
\label{eq:joint1}
\end{align}
In the above, $[x_0]\sim \pi^*$ is a prior distribution on the unit circle, 
$[x_1=\left\{g^*(1,x_0)+\eta_1\right\}~[2\pi]|x_0,$ $\bi{\beta}_g, \sigma^2_{\mbox{\scriptsize{$\eta$}}},\sigma^2_g]$ 
follows a wrapped normal distribution, derived from 
$[x^*_1=g^*(1,x_0)+\eta_1|x_0,\bi{\beta}_g, \sigma^2_{\mbox{\scriptsize{$\eta$}}},$ $\sigma^2_g]$, 
which is a normal distribution with mean $\mu_g(1,x_0)$ = $\bi{h}(1,x_0)'\bi{\beta}_g$ and variance 
${\sigma_g^2}+\sigma_{\mbox{\scriptsize $\eta$}}^2$. 
As already noted in Section \ref{sec:lookup}, $[\bi{D}_z|x_0,g^*(1,x_0),\bi{\beta}_g,\sigma^2_g]$ 
is multivariate normal with mean and covariance matrix given by (\ref{eq8:mean D_z|g}) and (\ref{eq9:cov D_z|g}) respectively, 
and finally the conditional distribution $[x_{t+1}=\left\{g^*((t+1),x_t)+\eta_{t+1}\right\}~[2\pi]|\bi{\beta}_g,
\sigma^2_g,\bi{D}_z,x_t,\sigma^2_{\mbox{\scriptsize{$\eta$}}}]$ is again a wrapped normal distribution 
derived from $[x^*_{t+1}=g^*((t+1),x_t)+\eta_{t+1}|\bi{\beta}_g,\sigma^2_g, \bi{D}_z,x_t,
\sigma^2_{\mbox{\scriptsize{$\eta$}}}]$, which is a normal distribution with mean $\mu_{x_{t}}$ given 
by (\ref{eq10:mean g|D_z}) and variance
\begin{equation}
 \label{eq12:var g(x_t)|D_z}
\sigma^2_{x_{t}} = \sigma^2_{\mbox{\scriptsize{$\eta$}}} + 
{\sigma_g^2}\left(1-(\bi{s}_{g,D_z}(t,x_{t-1}))' \bi{A}_{g,D_{z}}^{-1}\bi{s}_{g,D_z}(t,x_{t-1})\right).
\end{equation}


For explicit derivations of the conditional distributions associated with (\ref{eq:joint1}) it is necessary to
bring in some more auxiliary variables. To be specific, note that $x^*_t=x_t+2\pi K_t$, where
$K_t=\langle x^*_t/2\pi\rangle$, where, for any $u$, $\langle u\rangle$ denotes the greatest integer not exceeding $u$. Note that
for each $t$, $K_t$ can take values in the set $\left\{\cdots,-2,-1,0,1,2,\cdots\right\}$.
Here we view the wrapped number $K_t$ as a random variable; see also \ctn{Ravindran11}.

Note that $x_1\mbox{ given }(g^*(1,x_0),\bi{\beta}_g,\sigma^2_{\mbox{\scriptsize $\eta$}},\sigma^2_g, K_1)$ 
has the following distribution:
\begin{equation}
\label{eq17:x_1 | g,beta_g,sigma_eta,sigma_g, K_1}
[x_1|g^*(1,x_0),\bi{\beta}_g,\sigma^2_{\mbox{\scriptsize $\eta$}},\sigma^2_g, K_1]
=\frac{\frac{1}{\sqrt{2\pi}\sigma_{\mbox{\scriptsize $\eta$}}}
\exp\left(-\frac{1}{2\sigma^2_{\mbox{\scriptsize $\eta$}}}(x_1+2\pi K_1-g^*(1,x_0))^2\right)I_{[0,2\pi]}(x_1)}
{\Phi\left(\frac{2\pi (K_1+1)-g^*(1,x_0)}{\sigma_{\mbox{\scriptsize $\eta$}}}\right)
-\Phi\left(\frac{2\pi K_1-g^*(1,x_0)}{\sigma_{\mbox{\scriptsize $\eta$}}}\right)}
\end{equation}
and the distribution of $K_1\mbox{ given } (g^*(1,x_0),\sigma^2_{\mbox{\scriptsize $\eta$}})$ is
\begin{equation}
\label{eq18:K_1 | g*,beta_g,sigma_eta,sigma_g}
[K_1|g^*(1,x_0),\sigma^2_{\mbox{\scriptsize $\eta$}},\sigma^2_g] 
= \Phi\left(\frac{2\pi (K_1+1)-g^*(1,x_0)}{\sigma_{\mbox{\scriptsize $\eta$}}}\right)
-\Phi\left(\frac{2\pi K_1-g^*(1,x_0)}{\sigma_{\mbox{\scriptsize $\eta$}}}\right),
\end{equation}
where $\Phi(\cdot)$ is the cumulative distribution function of standard normal distribution. 

Similarly, for $t=2,\ldots, T+1$,
the distributions of 
$x_t\mbox{ given }(\bi{\beta}_g,\sigma^2_{\mbox{\scriptsize $\eta$}},\sigma^2_g, \bi{D}_z,x_{t-1},K_t)$ and 
$K_t$ given $(\bi{\beta}_g,\sigma^2_{\mbox{\scriptsize $\eta$}}, \sigma^2_g, \bi{D}_z,x_{t-1})$, respectively, are
\begin{equation}
\label{eq19:x_t| beta_g,sigma_eta,sigma_g, D_z,x_t-1,K_t}
[x_t|\bi{\beta}_g,\sigma^2_{\mbox{\scriptsize $\eta$}},\sigma^2_g, \bi{D}_z,x_{t-1},K_t] 
= \frac{\frac{1}{\sqrt{2\pi}\sigma_{x_{t}}}\exp\left(-\frac{1}{2\sigma^2_{x_{t}}}(x_t+2\pi K_t-\mu_{x_{t}})^2 \right)
I_{[0,2\pi]}(x_t)}{\Phi\left(\frac{2\pi (K_t+1)-\mu_{x_{t}}}{\sigma_{x_{t}}}\right)
-\Phi\left(\frac{2\pi K_t-\mu_{x_{t}}}{\sigma_{x_{t}}}\right)}
\end{equation}

\begin{equation}
\label{eq20: K_t|beta_g,sigma_eta,sigma_g, D_z,x_t-1}
[K_t|\bi{\beta}_g,\sigma^2_{\mbox{\scriptsize $\eta$}},\sigma^2_g, \bi{D}_z,x_{t-1}] 
= \Phi\left(\frac{2\pi (K_t+1)-\mu_{x_{t}}}{\sigma_{x_{t}}}\right)-\Phi\left(\frac{2\pi K_t-\mu_{x_{t}}}{\sigma_{x_{t}}}\right),
\end{equation}
where $\mu_{x_{t}}$ and $\sigma_{x_{t}}$ are given by (\ref{eq10:mean g|D_z}) and 
(\ref{eq12:var g(x_t)|D_z}), respectively. 

Thus, using the conditionals (\ref{eq17:x_1 | g,beta_g,sigma_eta,sigma_g, K_1}), (\ref{eq18:K_1 | g*,beta_g,sigma_eta,sigma_g}),
(\ref{eq19:x_t| beta_g,sigma_eta,sigma_g, D_z,x_t-1,K_t}) and (\ref{eq20: K_t|beta_g,sigma_eta,sigma_g, D_z,x_t-1}), the
joint distribution of the latent circular variables, conditional on $\bi{\beta}_g$, $\sigma^2_{\eta}$ and $\sigma^2_g$
can be represented as
\allowdisplaybreaks
{
\begin{align}
&[x_0,x_1,\ldots,x_{T+1}|\bi{\beta}_g,\sigma^2_{\mbox{\scriptsize{$\eta$}}},\sigma^2_{g}]\notag\\
&=\sum_{K_1,\ldots,K_{t+1}}
\int [x_0,x_1,\ldots,x_{T+1},\bi{D}_z,K_1,\ldots,K_{T+1}|\bi{\beta}_g,\sigma^2_{\mbox{\scriptsize{$\eta$}}},\sigma^2_{g}]
d \bi{D}_z\notag\\
&=\sum_{K_1,\ldots,K_{t+1}}\int [x_0][\bi{D}_z|x_0,g^*(1,x_0),\bi{\beta}_g,\sigma^2_g][g^*(1,x_0)|x_0,\bi{\beta}_g,\sigma^2_g]\notag\\
&\qquad\times[x_1|x_0,g^*(1,x_0),K_1,\sigma^2_{\mbox{\scriptsize{$\eta$}}},\sigma^2_g]
[K_1|x_0,g^*(1,x_0),\sigma^2_{\mbox{\scriptsize{$\eta$}}},\sigma^2_g]
\notag\\
& \qquad \times\prod_{t=1}^T [x_{t+1}|\bi{\beta}_g,\sigma^2_g, \bi{D}_z,x_t,
\sigma^2_{\mbox{\scriptsize{$\eta$}}},K_{t+1}]
[K_{t+1}|\bi{\beta}_g,\sigma^2_{\mbox{\scriptsize $\eta$}},\sigma^2_g, \bi{D}_z,x_{t}]d g^*(1,x_0)d \bi{D}_z.
\label{eq:joint2}
\end{align}
}
\subsection{Advantages of the look-up table approach}
\label{subsec:advantages_look_up_table}

\ctn{Ghosh14} provide ample details on the accuracy of the look-up table approach. In particular, they prove
a theorem on the accuracy of the approximation of the distribution of the latent states using the look-up table,
show that the joint distribution of the latent states is non-Markovian, even though conditionally on $\bi{D}_z$,
the latent states have a Markov structure. Quite importantly, \ctn{Ghosh14} point out that this approach leads
to great computational savings and remarkable numerical stability of the associated MCMC algorithm thanks to the fact
that $\bi{A}_{g,D_z}$ needs to be inverted only once, even before beginning the MCMC simulations, and that the set of 
grid points $\bi{G}_z$ can be chosen so that $\bi{A}_{g,D_z}$ is invertible.
These advantages clearly remain valid even in our circular set-up.

\section{Simulation study}
\label{simulation study}

\subsection{True model}
\label{subsec:true_model}

We now illustrate the performance of our model and methodologies using a simulation study. For this purpose
we simulate a set of observations of size $101$ from the following nonlinear dynamic model:
\begin{align*}
\label{nonlinear model}
y_{t} &= \tan^2 (\theta_t)/20 + v_{t};
\\[1ex]
\tan{\left(\frac{\theta_t-\pi}{2}\right)} 
&= \alpha \tan{\left(\frac{\theta_{t-1}-\pi}{2}\right)} 
+ \frac{\beta \tan \left(\frac{\theta_{t-1}-\pi}{2}\right)}{1+\tan^2\left(\frac{\theta_{t-1}-\pi}{2}\right)} 
+ \gamma \cos(1.2 (t-1)) + u_{t},
\end{align*}
for $t=1,\ldots,101$, where $u_{t}$ and $v_{t}$ are normally distributed with means zero and variances 
$\sigma^2_{\eta}$ and $\sigma^2_{\epsilon}$. We set the values of $\alpha$, $\beta$ and $\gamma$ to be 
$0.05$, $0.1$ and $0.2$, respectively; we fix the values of both $\sigma_{\eta}$ and $\sigma_{\epsilon}$ at  $0.1$.
We consider the first $100$ observations of $y_t$ as known, and set aside the last observation for the 
purpose of forecasting. 

\subsection{Choices of prior parameters and the grid $\bi{G}_{z}$}
\label{subsec:prior_chocies}
In this experiment we consider a four-variate normal prior distribution for $\bi{\beta}_f$ with mean 
$(0,0,0,0)'$ and the identity matrix as the covariance matrix. For $\bi{\beta}_g$ we choose a four-variate normal 
prior with mean vector
$(2.5,0.04,1.0,1.0)'$. The choice of the covariance matrix for $\bi{\beta}_g$ is discussed in the next paragraph. 
Choices of these prior parameters ensured adequate mixing of our MCMC algorithm.

Observe that in (\ref{eq:eq2}) of our proposed model, in the mean function of the underlying Gaussian process $g$, 
the third and the fourth components of $\bi{\beta}_{g}$ are multiplied by $\cos(x_{t-1})$ and $\sin(x_{t-1})$, respectively,
so that an identifiability problem crops up.  
To counter this problem, we set the third and the fourth components of $\bi{\beta}_{g}$ to be 1, throughout the experiment. 
Therefore, the covariance matrix for $\bi{\beta}_g$ is chosen to be a diagonal matrix with the entries $(1.0,1.0,0.0,0.0)'$.

For $\sigma_{\epsilon}$ and $\sigma_{f}$ we consider inverse gamma priors with (shape, scale) 
parameters $(4.01,0.005\times 5.01)$ 
and $(4.01,0.1\times 5.01)$, respectively, so that the mode of $\sigma_{\epsilon}$ is $0.005$ and that of $\sigma_{f}$ 
is $0.1$. We choose the first parameter of the inverse gamma distribution to be equal to $4.01$ so that 
the variance is $200$ times the square of the mean of the inverse gamma distribution, which are in this case 
$0.012$ and $0.25$, respectively. The choices of second prior parameters for $\sigma_{\epsilon}$ and $\sigma_{f}$ 
yielded adequate mixing of our MCMC algorithm. 

Finally we divide the interval $[0,2\pi]$ into $100$ sub-intervals and choose one point from each of the sub-intervals;
these values constitute the second component of the two dimensional grid $\bi{G}_{z}$. For the first component 
of $\bi{G}_{z}$, we select a random number uniformly from each of the 100 subintervals 
$\left[\frac{2\pi i}{100},\frac{2\pi(i+1)}{100}\right]$, $i=0,\ldots,99$. 
%

\subsection{Brief discussion related to impropriety of the posteriors of some unknowns and the remedy}
\label{subsec:impropriety}
An interesting feature associated with our model is the impropriety of the posteriors of $\sigma_g$, $\sigma_{\eta}$
and $K_1,\ldots,K_{T+1}$, when they are all allowed to be random.
In a nutshell, for any value of $K_t$, exactly the same value of the circular variable $x_t$ is obtained by the
$\mbox{mod}~2\pi$ operation applied to $x^*_t=x_t+2\pi K_t$. Thus, given $x_t$, it is not possible to constrain
$K_t$ unless both $\sigma_g$ and $\sigma_{\eta}$ are bounded. Boundedness of $\sigma_g$ and $\sigma_{\eta}$ would ensure
that $x^*_t$ has finite variance, which would imply finite variability of $K_t$.

Since it is unclear how to select a bounded prior for $\sigma_g$ and $\sigma_{\eta}$, we obtain the maximum likelihood
estimates (MLEs) of these variances and plug in these estimates in our model. To obtain the MLEs, we implemented
the simulated annealing methodology (see, for example, \ctn{Robert04}, \ctn{Liu01})
where at each iteration we proposed new values of these variances, then integrated out all the other parameters using
averages of Monte Carlo simulations, given the proposed values of $\sigma_g$ and $\sigma_{\eta}$, so that
we obtain the integrated likelihood given the proposed variances; then we calculated the
acceptance ratio, and finally decreased the temperature parameter of our simulated annealing algorithm 
before proceeding to the next iteration. The MLEs turned out to be $\hat\sigma_g=0.1258$ and $\hat\sigma_{\eta}=0.1348$.

\subsection{MCMC details}
\label{subsec:mcmc_details}
As detailed in Section S-2 of the supplement, our MCMC algorithm updates some parameters using Gibbs steps, 
and the remaining using random walk Metropolis-Hastings steps.
To update $\sigma_{\epsilon}$ and $\sigma_f$ we implemented normal random walk with variance $0.05$; 
$x_0$ is updated using von-Mises distribution with $\kappa$ = $3.0$, and for updating $x_t$ a mixture of 
two von-Mises distributions with $\kappa=0.5$ and $\kappa=3.0$ is used for $t=1,\ldots, T$. 
The wrapping variables $K_t;~t=1,\ldots,T$, are updated 
using the discrete normal random walk with variance $1.0$. 
All these choices are made very painstakingly after carefully adjudging the mixing properties of many pilot
MCMC runs. The rest of the parameters are updated using 
Gibbs steps, as detailed in Section S-2 of the supplement. 

With the above choices of the prior parameters and $\bi{G}_{z}$, and with the above MCMC updating procedure of the parameters, 
we performed $2,10,000$ MCMC simulations with a burn-in period consisting of the first $1,50,000$ iterations. 
The time taken to run $2,10,000$ MCMC simulations in a desktop computer with $i7$ processors is 20 hours and 34 minutes.

\subsection{Results of our simulation study}
\label{subsec:simulation_results}

\begin{figure}[htp]
\centering
\includegraphics[height=1.5in,width=1.5in]{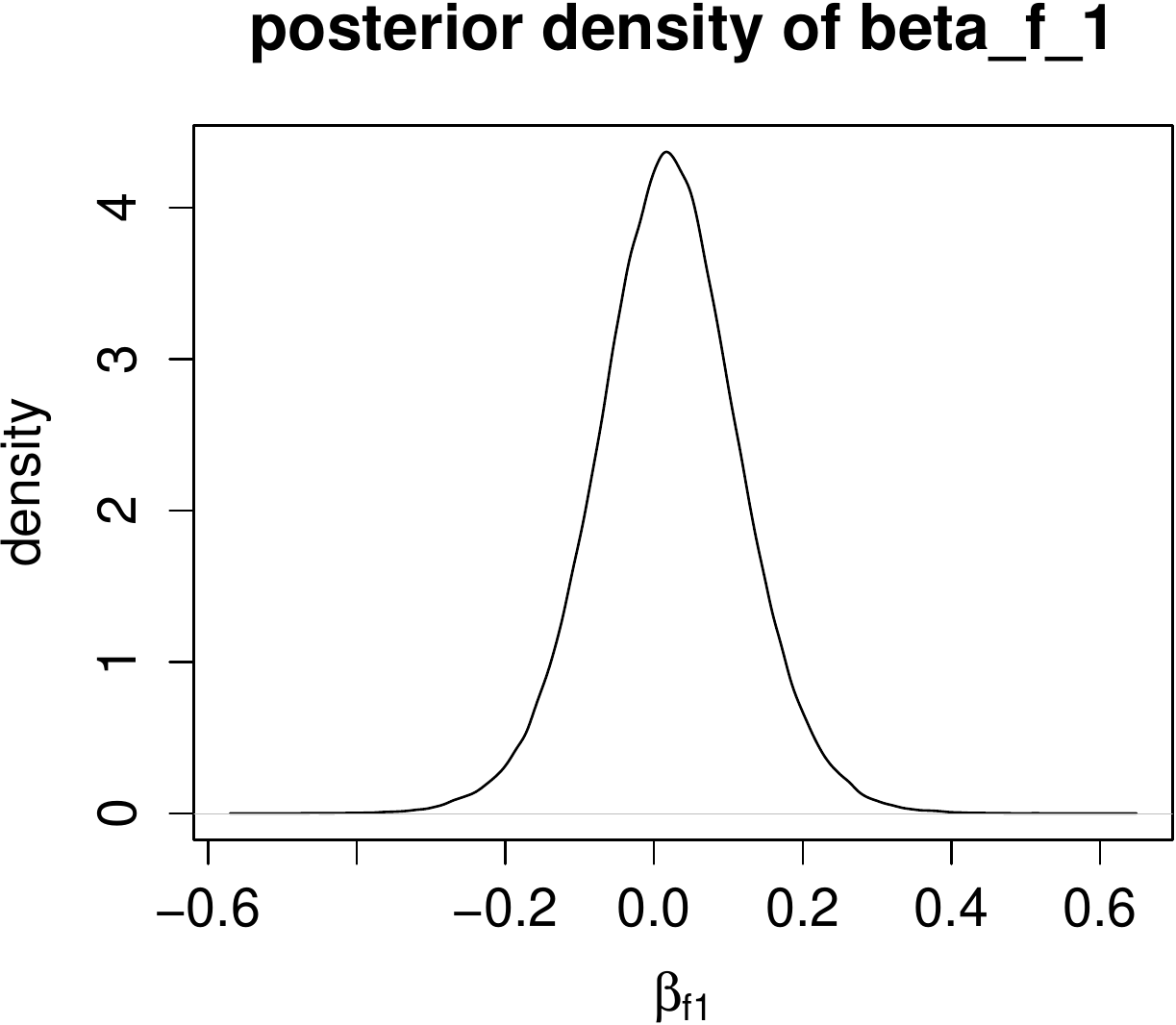}
\includegraphics[height=1.5in,width=1.5in]{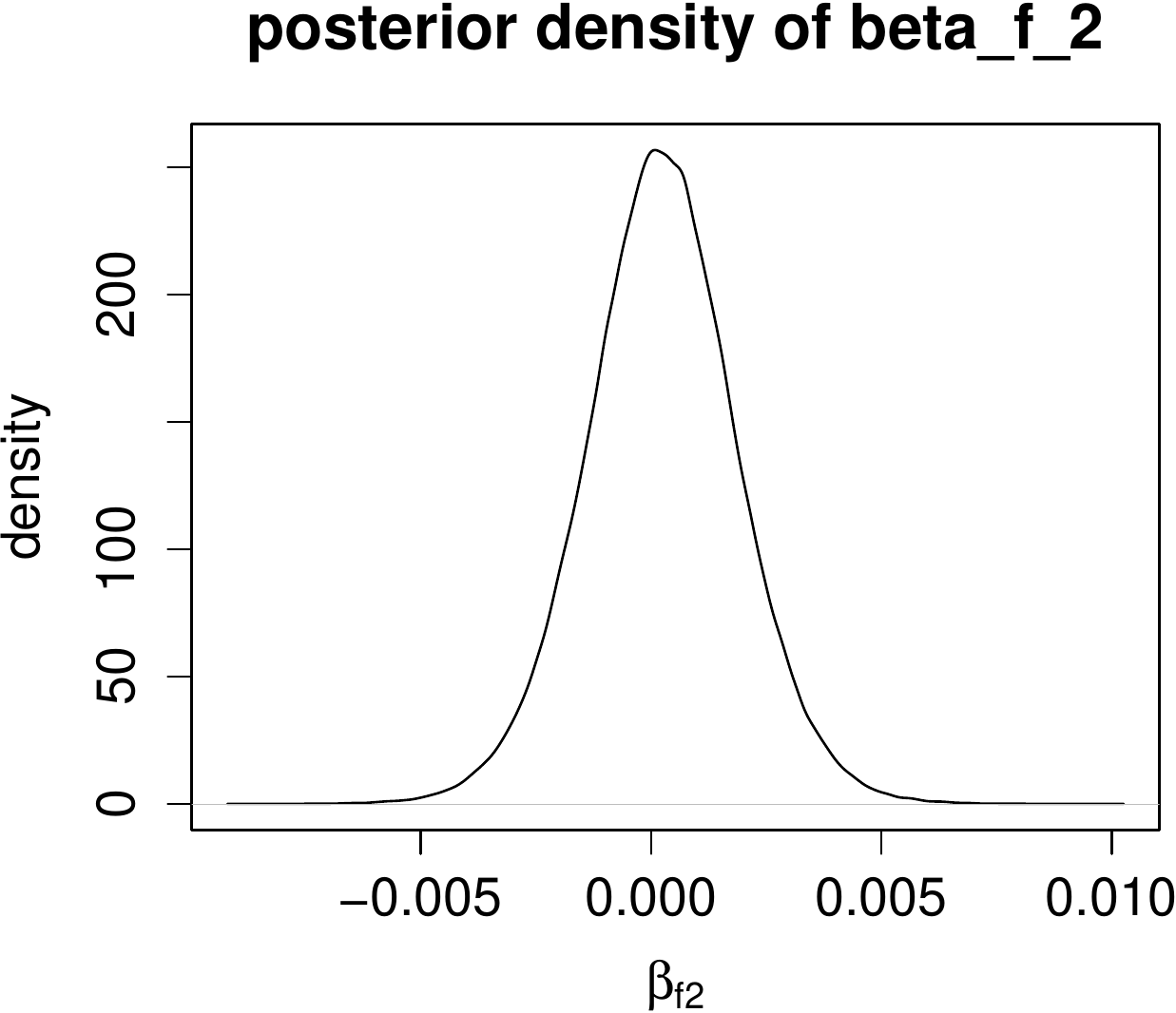}
\includegraphics[height=1.5in,width=1.5in]{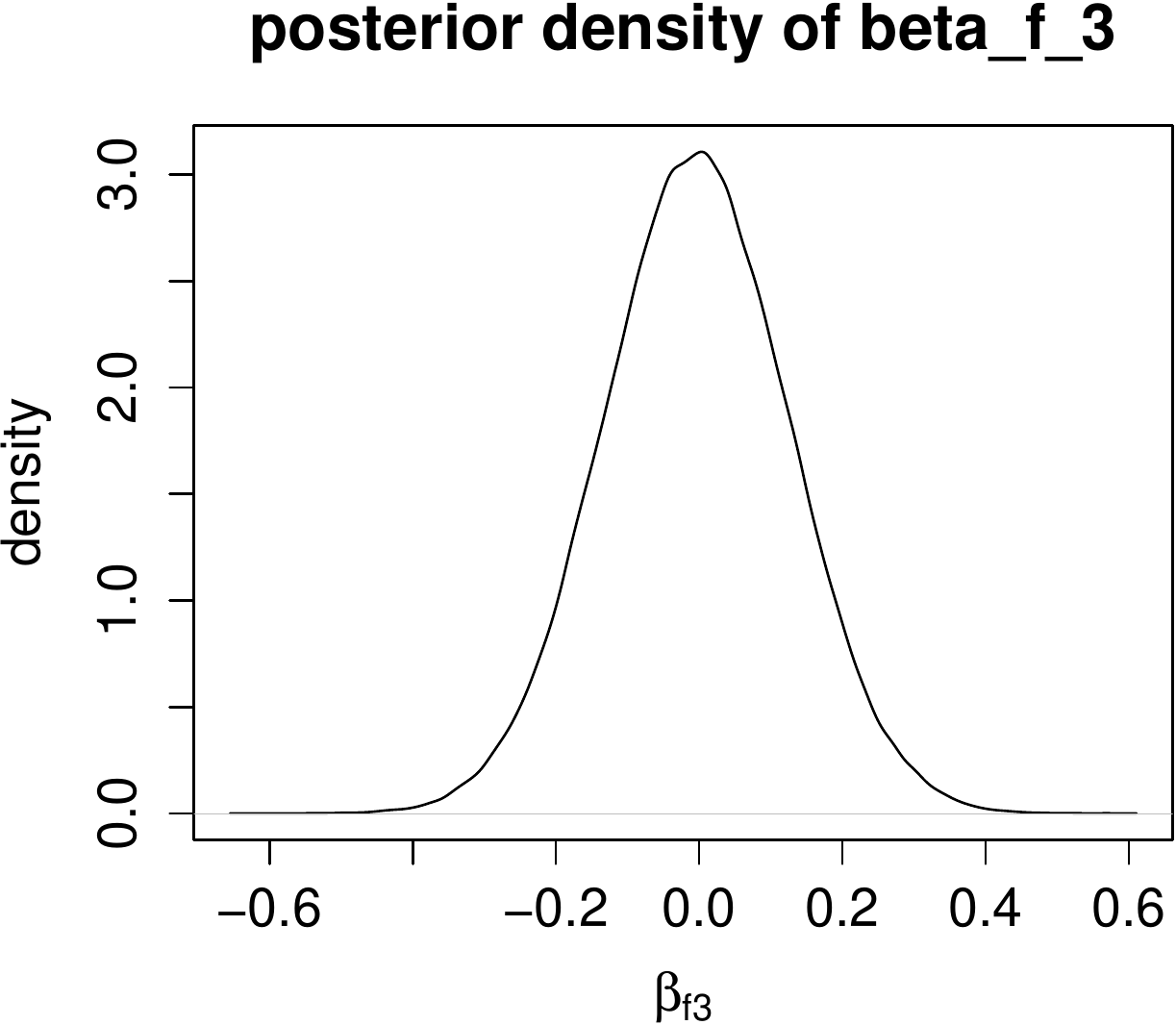}
\includegraphics[height=1.5in,width=1.5in]{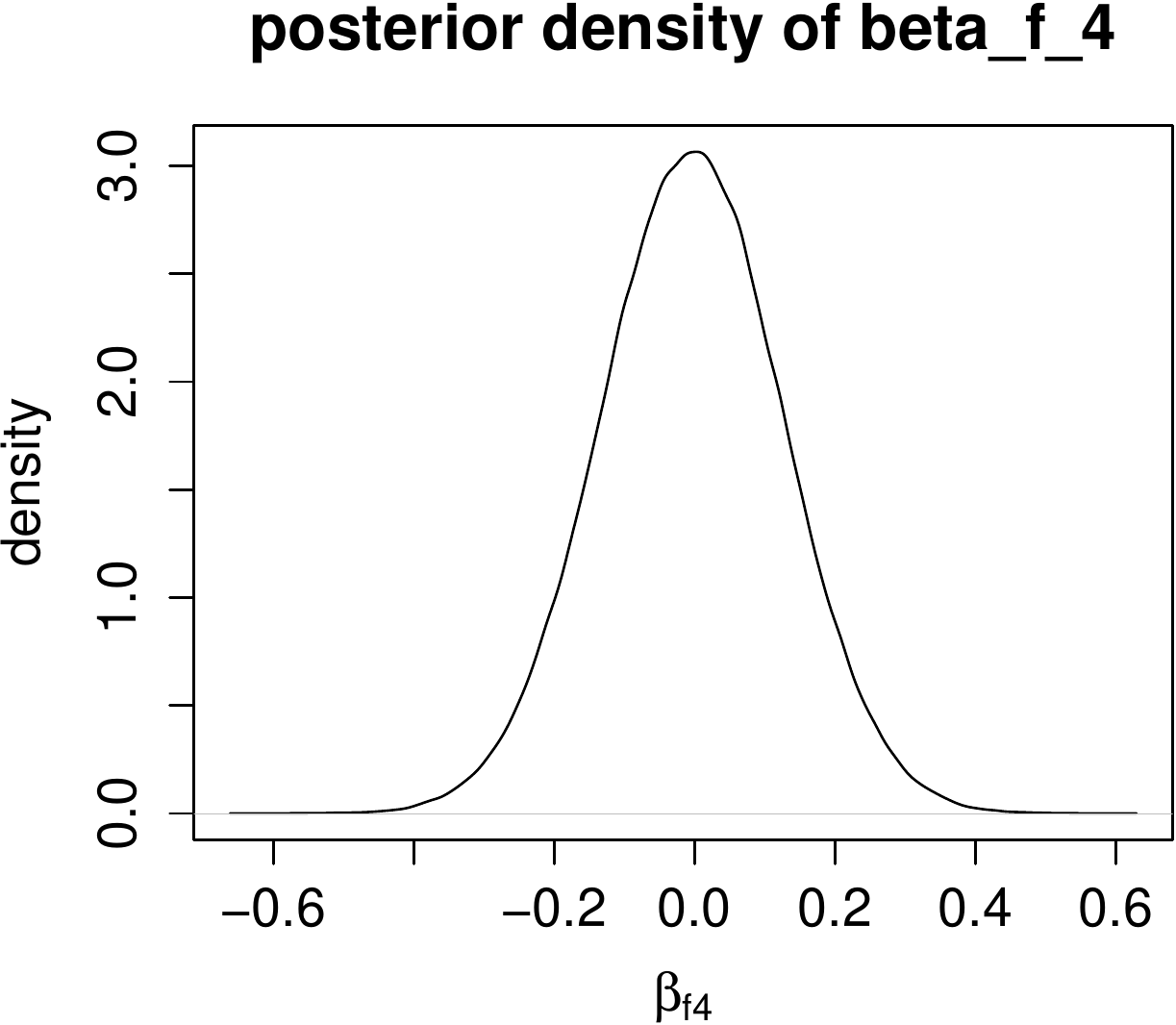}
\caption{Posterior densities of the four components of $\bi{\beta}_f$.}
\label{Fig00:Post of beta_f components in sim data}
\end{figure}
\begin{figure}[htp]
\centering
\includegraphics[height=2in,width=2in]{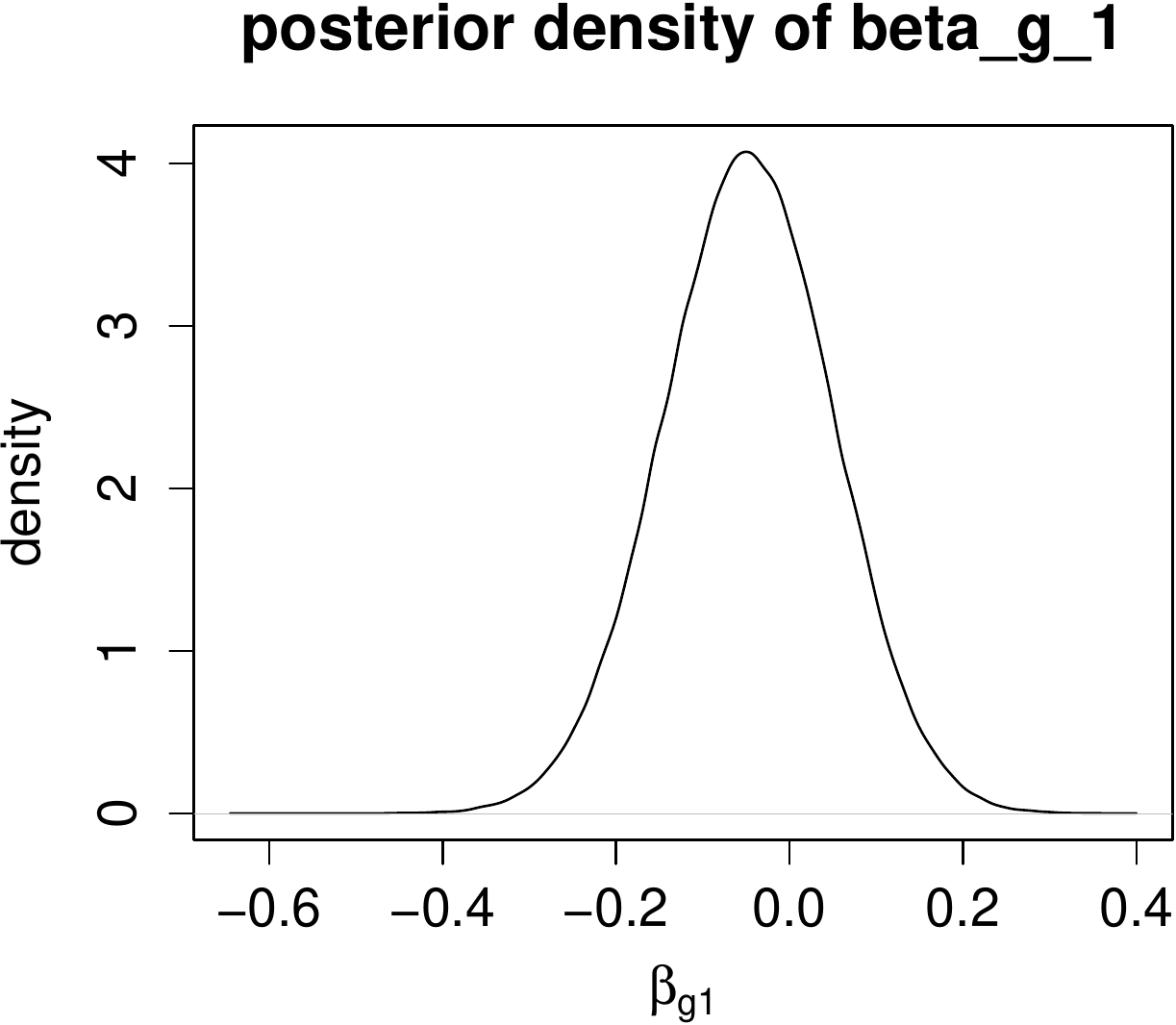}
\includegraphics[height=2in,width=2in]{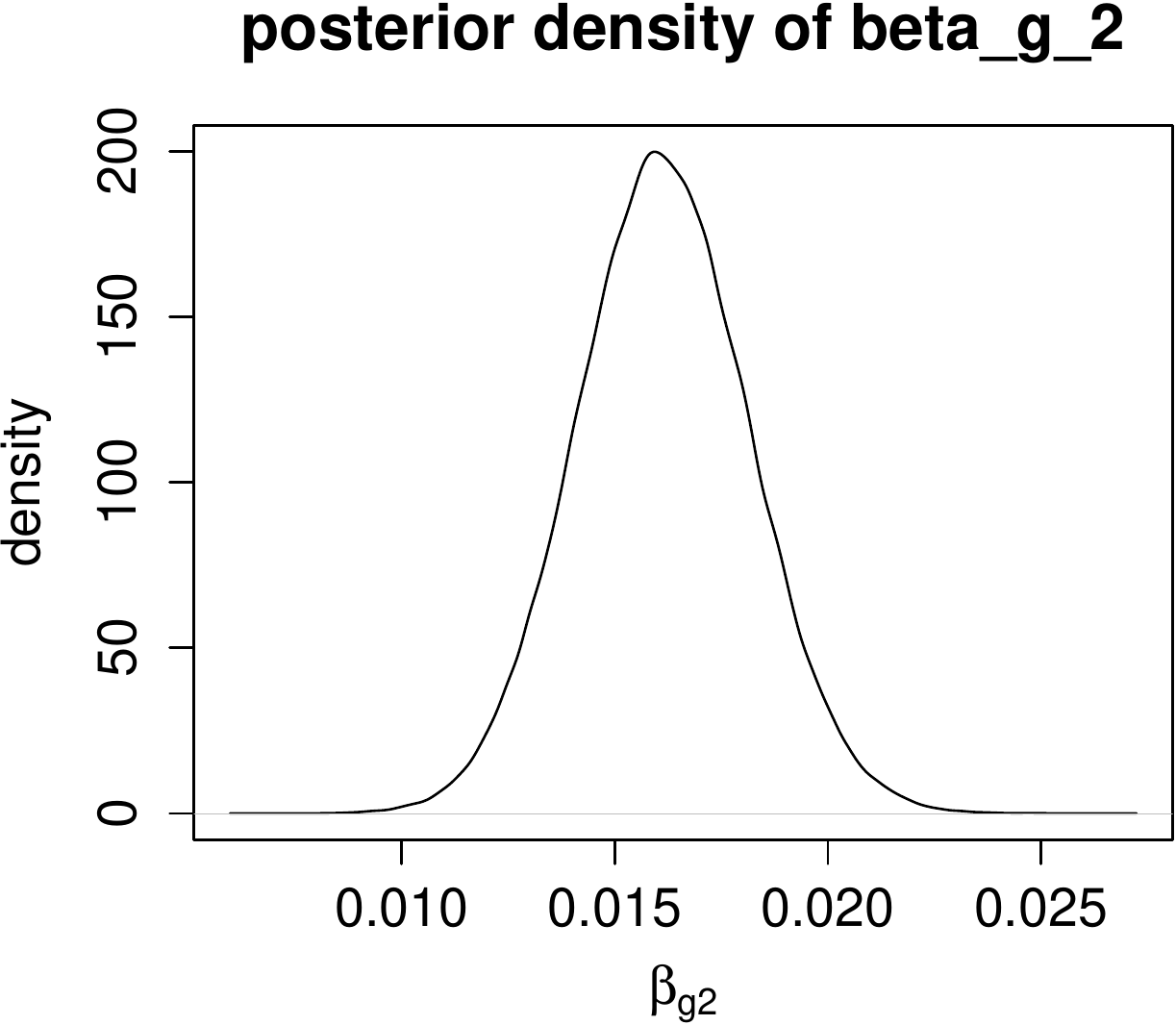}
\includegraphics[height=2in,width=2in]{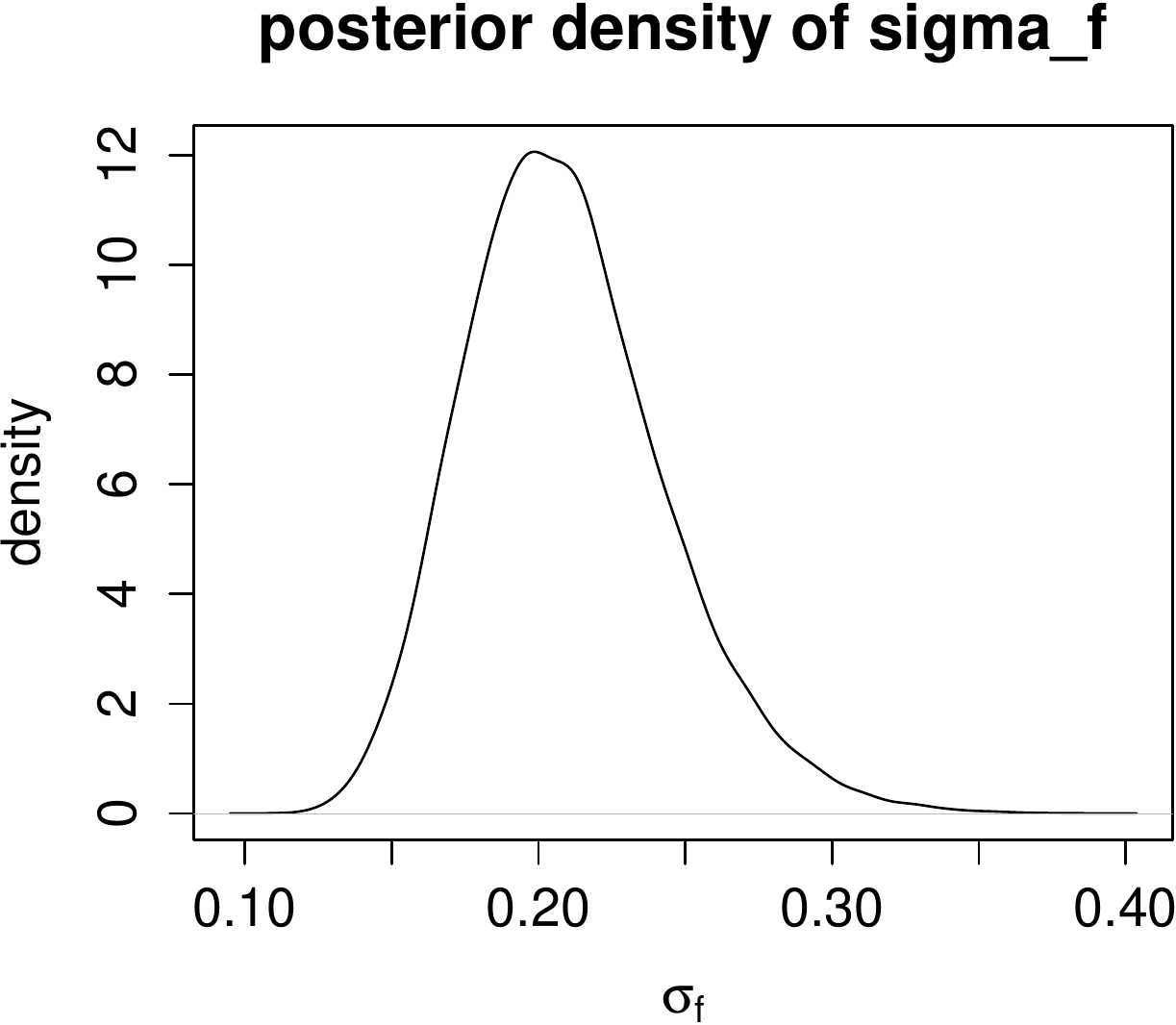}
\caption{Posterior densities of the first and second components of $\bi{\beta}_g$ and the posterior density of $\sigma_f$.}
\label{Fig01:Post of beta_g and sigma_f in sim data}
\end{figure}

The posterior densities of the components of $\bi{\beta}_f$ are provided in 
Figure \ref{Fig00:Post of beta_f components in sim data}. Figure \ref{Fig01:Post of beta_g and sigma_f in sim data} 
displays the posterior densities of the first two components of $\bi{\beta}_g$, and the posterior density of $\sigma_f$. 
Figure \ref{Fig1:Post of sigma_e, x_last in sim data} depicts the posterior density of the $\sigma_{\epsilon}$ and 
$x_{101}$. The horizontal bold black lines denote the 95\% highest posterior density credible intervals 
and the vertical lines denote the true values. 
Observe that the true values in each of the cases fall well within the intervals. 

As already mentioned, it is seen that the densities of most of $x_t$, $t=1,\ldots,T,$ has multiple modes, so that
a plot of the posterior probability distribution of the latent process for each time point, 
rather than ordinary credible regions, is appropriate. Such a plot for the latent time series $x_1,\dots,x_T$ is displayed in 
Figure \ref{Fig2: fitted latent process}, where regions with progressively higher densities are 
represented by more progressively intense colors. 
Quite encouragingly, most of the true values are seen to lie in the high probability regions.

Figure \ref{Fig3: Post of y_last in sim data} depicts the posterior predictive density corresponding to $y_{101}$; 
the true value is well within the 95\% highest posterior density credible interval of the predictive density. 
A trace plot of $y_{101}$ for last 60,000 thousand iteration is also provided in Figure 
\ref{Fig3: Post of y_last in sim data} as a sample trace plot to show the convergence of our MCMC iterations. 

Thus, our model performs quite encouragingly, in spite of the true model being highly non-linear and assumed
to be unknown. As a result, we expect our model to perform adequately in general situations.

\begin{figure}[htp]
\centering
\includegraphics[height=2in,width=2in]{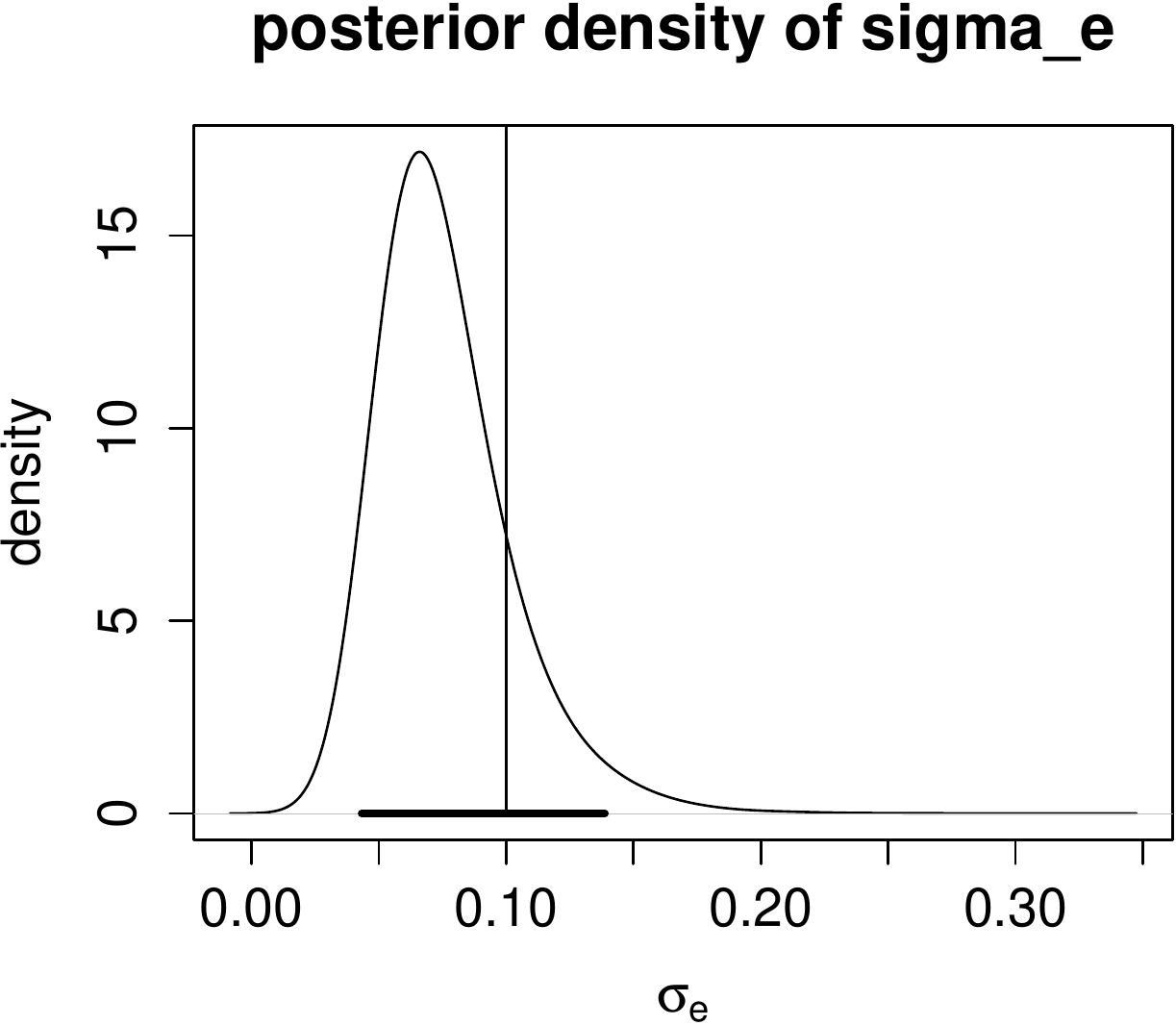}
\includegraphics[height=2in,width=2in]{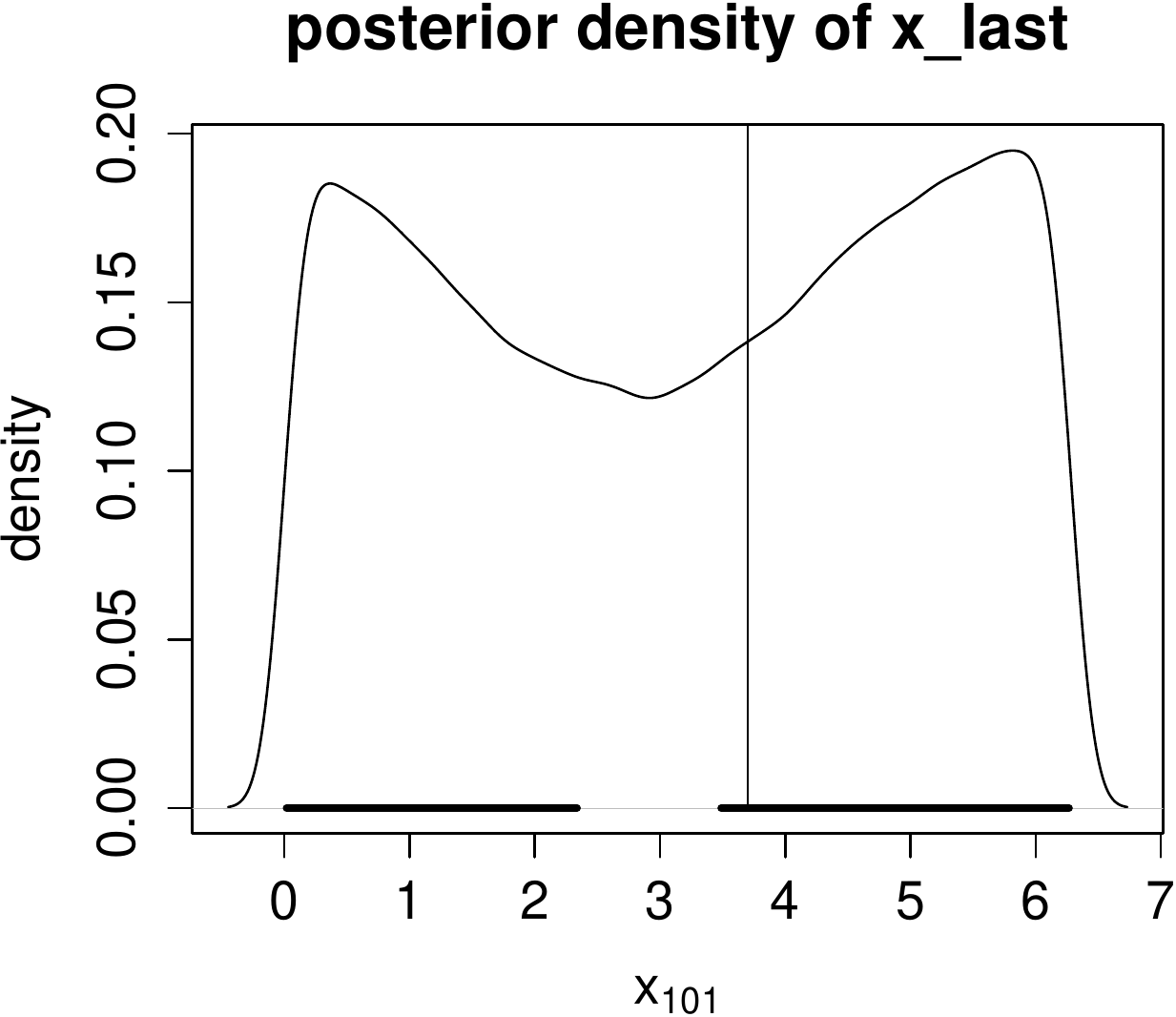}
\caption{Posterior densities of $\sigma_{\epsilon}$ and the $x_{101}$.}
\label{Fig1:Post of sigma_e, x_last in sim data}
\end{figure}


\begin{figure}[htp]
\centering
\includegraphics[height = 3in,width=5in]{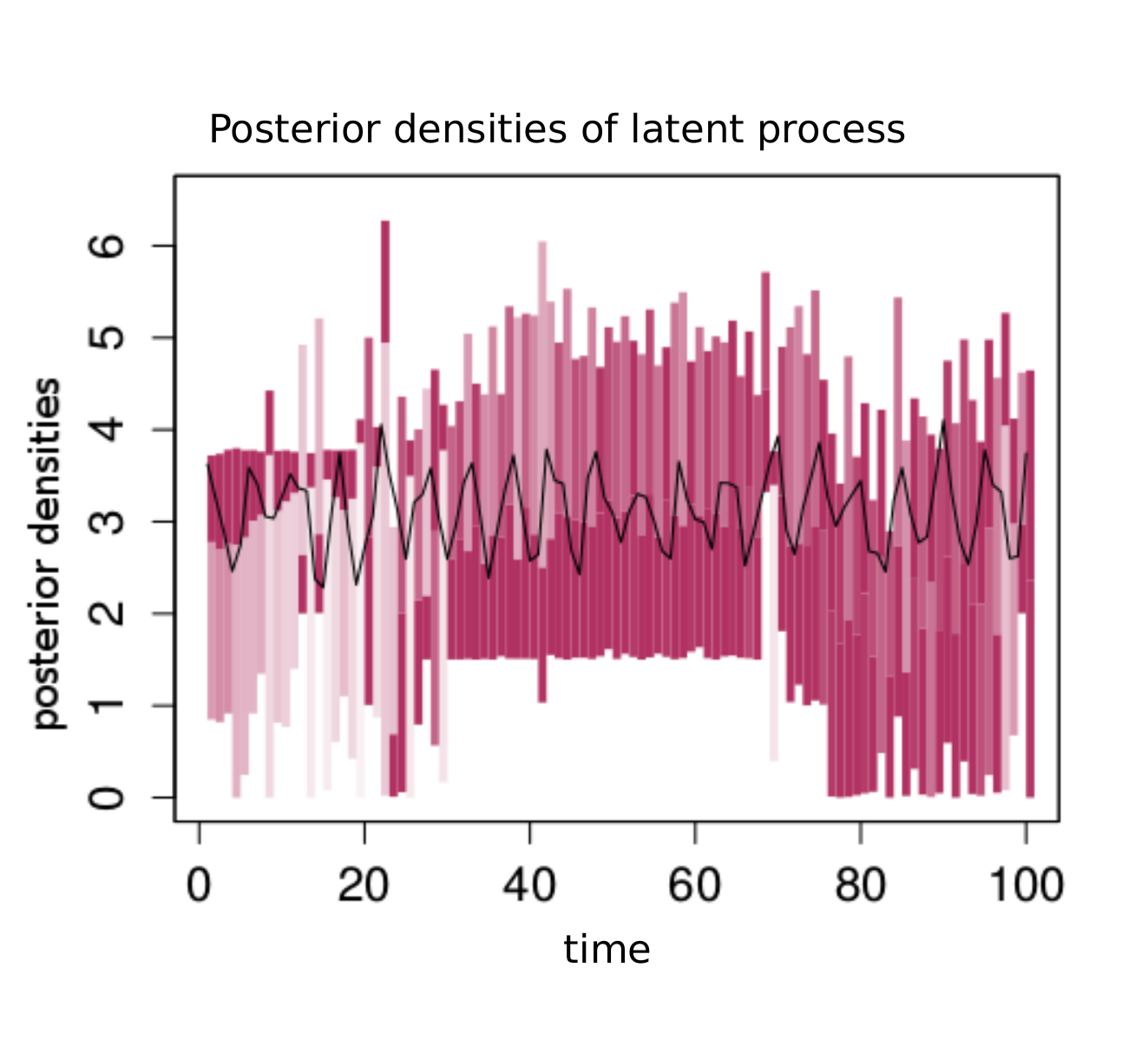}
\caption{Depiction of the posterior densities of the latent circular process
$\left\{x_t;t=1,\ldots,T\right\}$; higher the intensity of the color, higher is the posterior
density. The black line denotes the true time series.}
\label{Fig2: fitted latent process}
\end{figure}

\begin{figure}[htp]
\centering
\includegraphics[height=3in,width=3in]{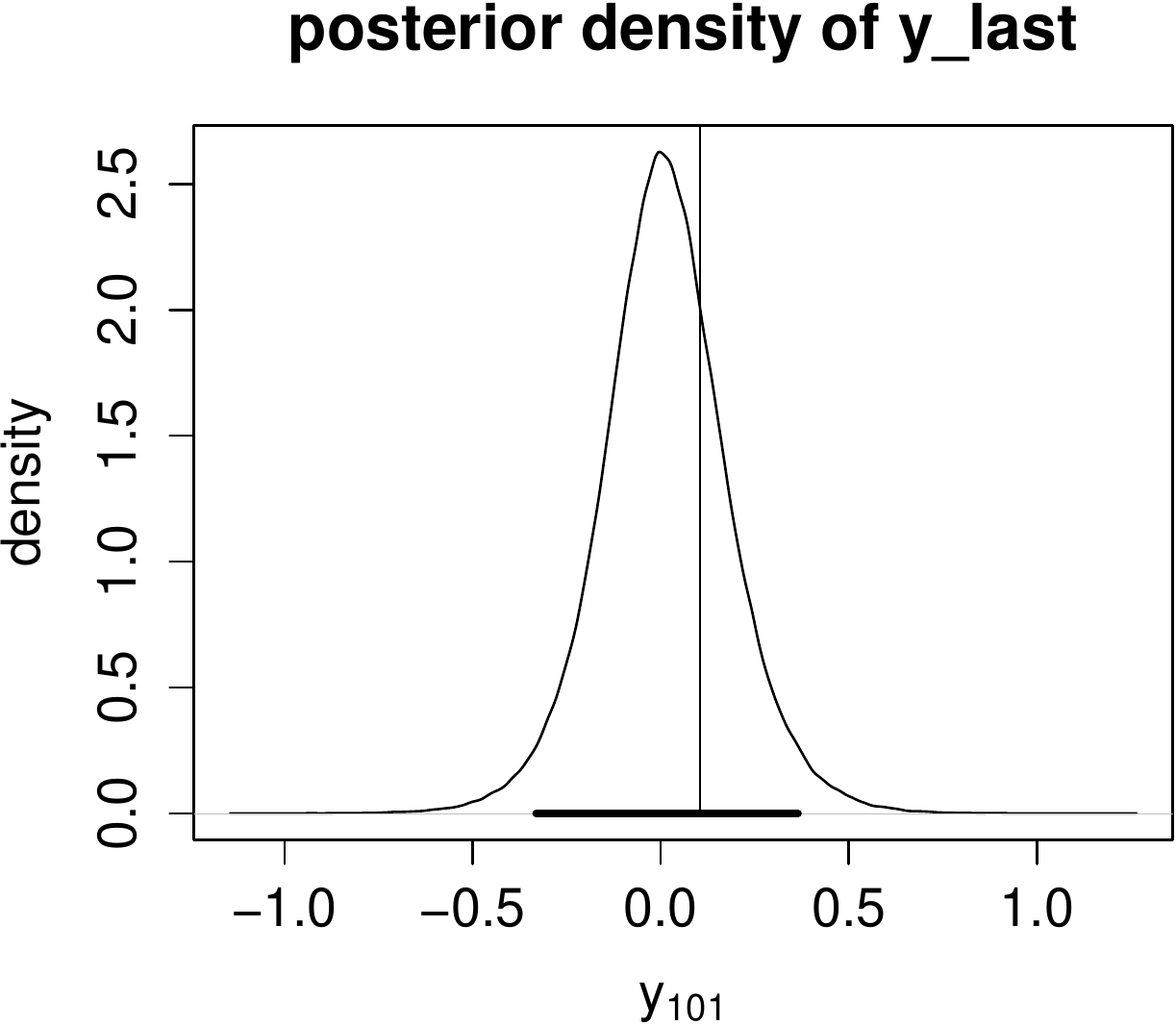} 
\includegraphics[height=3in,width=3in]{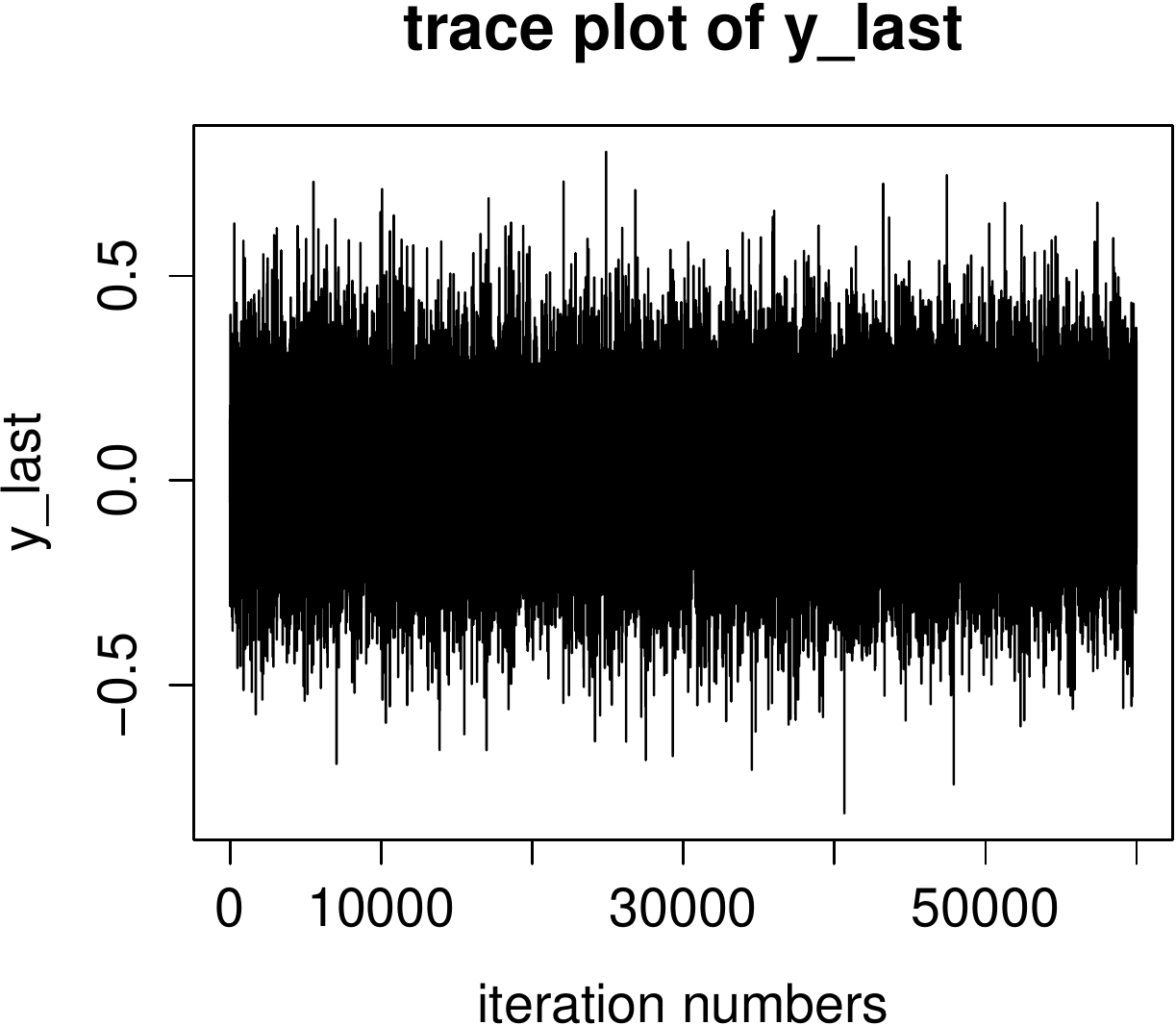}
\caption{From left\/: The first plot is that of the posterior predictive density corresponding to $y_{101}$, 
where the vertical line denotes the true value and 
the bold black horizontal line denotes the 95\% highest posterior density credible interval. 
The second plot is trace plot of $y_{101}$ for the last 60,000 observations.}
\label{Fig3: Post of y_last in sim data}
\end{figure}

 \section{Real data analysis}
\label{real data analysis}

\subsection{Wind speed data}
\label{subsec:wind speed}
\subsubsection{Brief description of the data}
\label{description of wind speed}

\begin{figure}[htp]
\centering
\includegraphics[height=3in,width=3in]{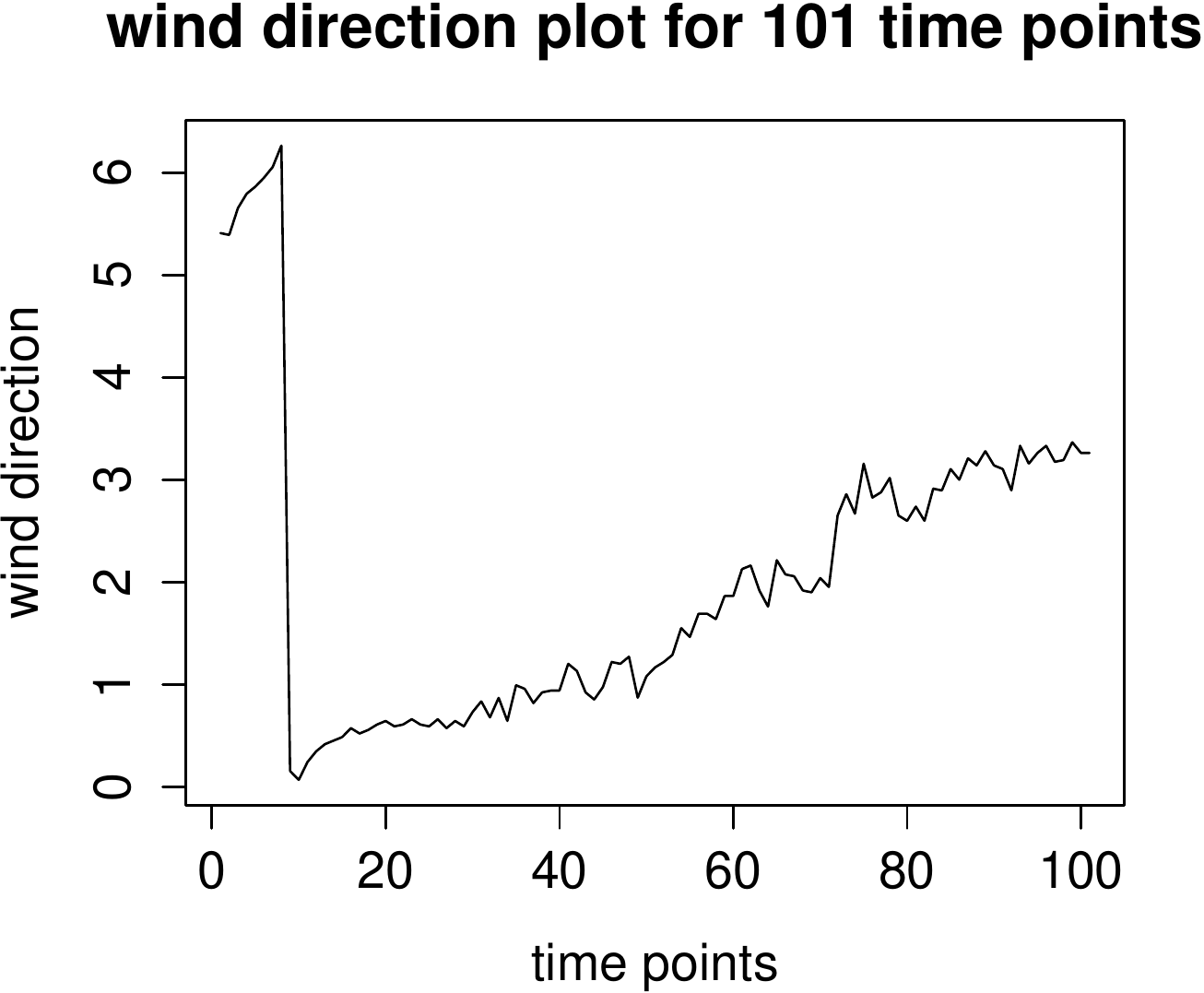}
\includegraphics[height=3in,width=3in]{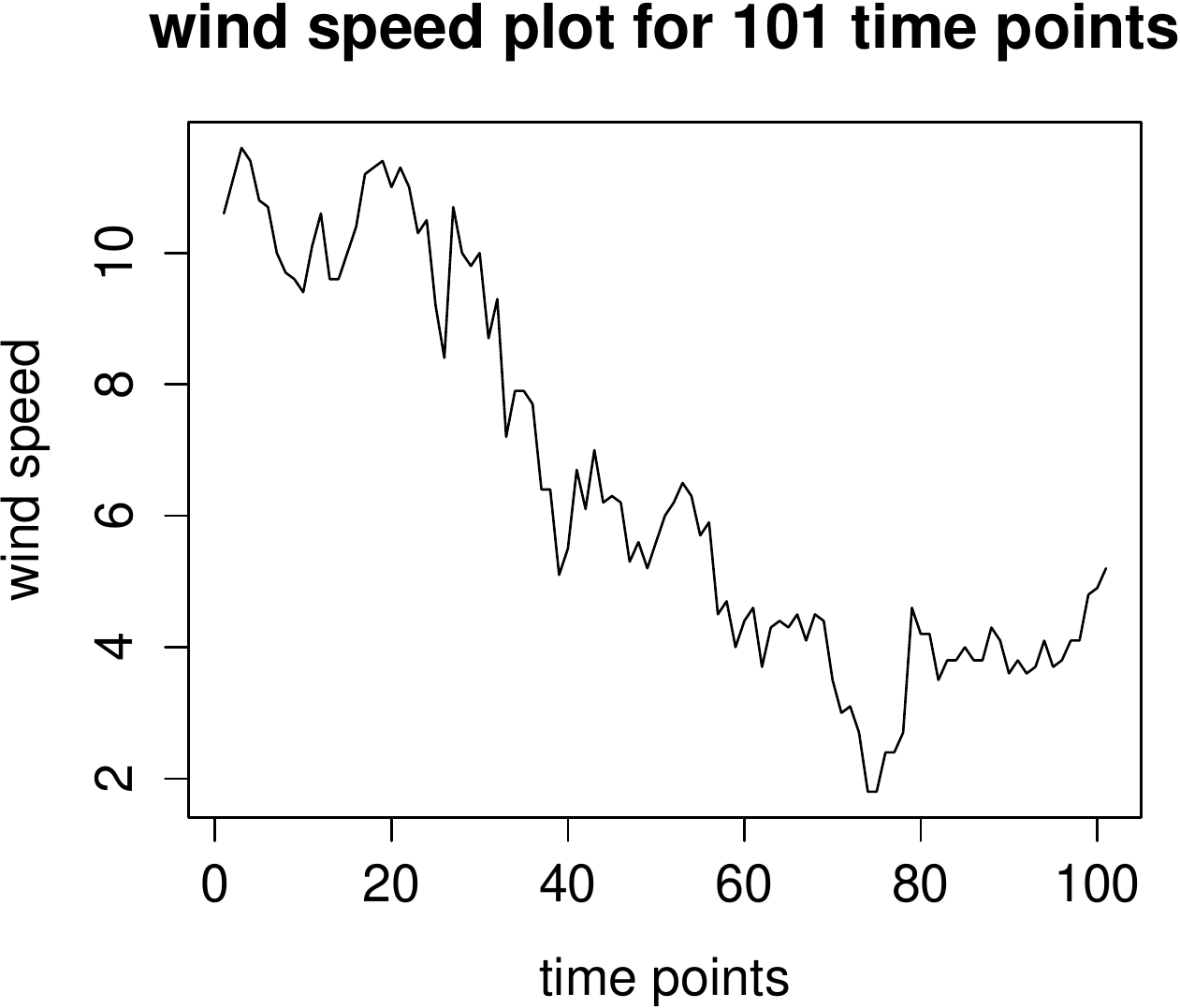}
\caption{Plot of the wind direction and wind speed data}
\label{Fig:Wind direction plot}
\end{figure}

In Section \ref{simulation study} we have demonstrated the effectiveness of our ideas with a simulation study.
Now we validate our model and methodologies on a real data using historic wind speed and wind direction data 
recorded by the National Oceanic and Atmospheric Administration's National Data Buoy Center
(http://www.ndbc.noaa.gov/historical data.shtml); the website has been very kindly brought to our notice by
the Associate Editor. Similar to \ctn{Marzio12b}, standard meteorological data obtained 
at the year 2009, monitored 
at station 41010 (120NM East of Cape Canaveral), which is
automatically recorded every 30 minutes (at 20 and 50 past each hour), are considered for our 
analysis.
Here we collect the wind speed and wind direction data for 101 times points starting from 1st January, 2009. Wind
directions originally recorded in degrees are converted to radians, ranging from 0 to $2\pi$. Wind speed
data are observed in meter per second. It is important to mention that for our analysis we use 
100 wind speed data points assuming that the wind direction data have not been observed. The main
purpose is here to demonstrate that our method is well-equipped to capture the recorded, real, wind direction data,
considered to be latent with respect to our model. A plot of true wind direction data is given in 
Figure \ref{Fig:Wind direction plot} along with the plot of observed wind speed data.  
\subsubsection{Prior choices and MCMC implementations}
\label{prior for wind direction}

We chose the prior parameters so as to obtain reasonable prediction of the future observation (set aside as $y_{101}$), 
and to obtain adequate mixing of our MCMC algorithm.
As such we specify the prior means of $\bi{\beta}_f$ and $\bi{\beta}_g$ to be $(0,0,0,0)'$ and 
$(1,1,1,1)'$, respectively. 
The prior covariance matrix for $\bi{\beta}_f$ has been chosen to be an identity matrix of order $4\times 4$ 
and for $\bi{\beta}_g$ it has been taken to be a diagonal matrix of order $4\times 4$, 
with diagonal elements $0.01, 0.01, 0.0 \mbox{ and }0.0$, respectively. 
Following the discussion in Section \ref{simulation study} we fixed the third and fourth components of $\bi{\beta}_g$  
at $1$ throughout the experiment to avoid identifiability issues. The shape parameters for 
$\sigma_e$ and $\sigma_f$ in the respective inverse gamma prior distributions are chosen to be 4.01 and the scale parameters 
are chosen to be $0.01\times 5.01$ and $0.001\times 5.01$, respectively, so that the prior modes 
for $\sigma_e$ and $\sigma_f$ are $0.01$ and $0.001$, respectively. The choice of the first 
parameter of inverse gamma is justified in Section \ref{simulation study}.

The MLEs of $\sigma_{\eta}$ and $\sigma_{g}$,
obtained using simulated annealing method, are 0.1455 and 0.1258. With all these prior choices our 
MCMC algorithm as detailed in section S-2
of the supplement has been used. As mentioned in Section \ref{subsec:mcmc_details} here also we use the same 
mixture of von-Mises to update $x_t$, $t=1,\ldots,100$. We implemented 2,50,000 MCMC iterations, where the last 1,00,000
iterations have been taken for the analysis after discarding a burn in of period 1,50,000. The time taken to implement
2,50,000 iterations on our $i7$ machine is 19 hours 58 minutes.
\subsubsection{Results}
\label{results of wind speed}
\begin{figure}[htp]
\centering
\includegraphics[height=1.5in,width=1.5in]{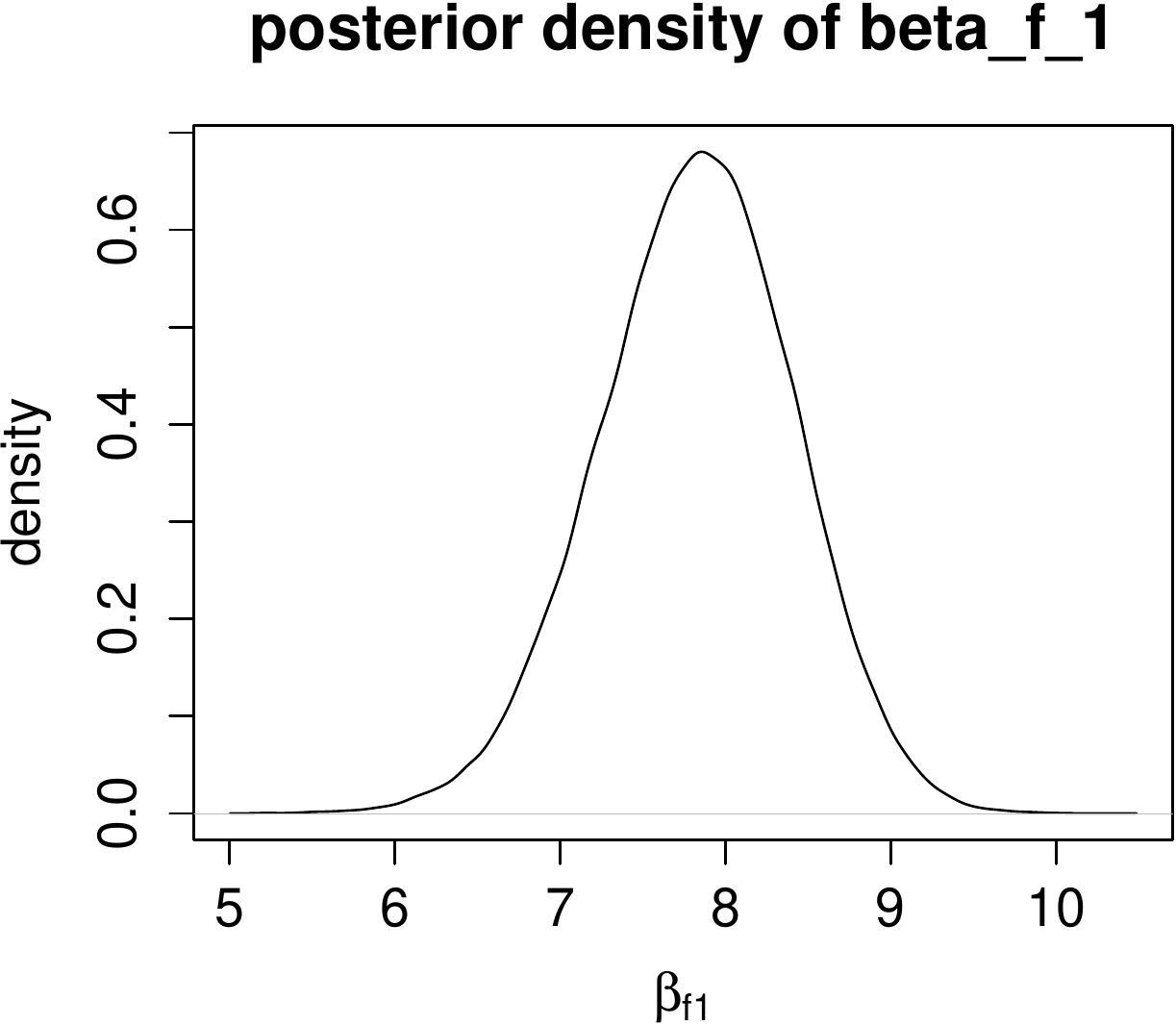}
\includegraphics[height=1.5in,width=1.5in]{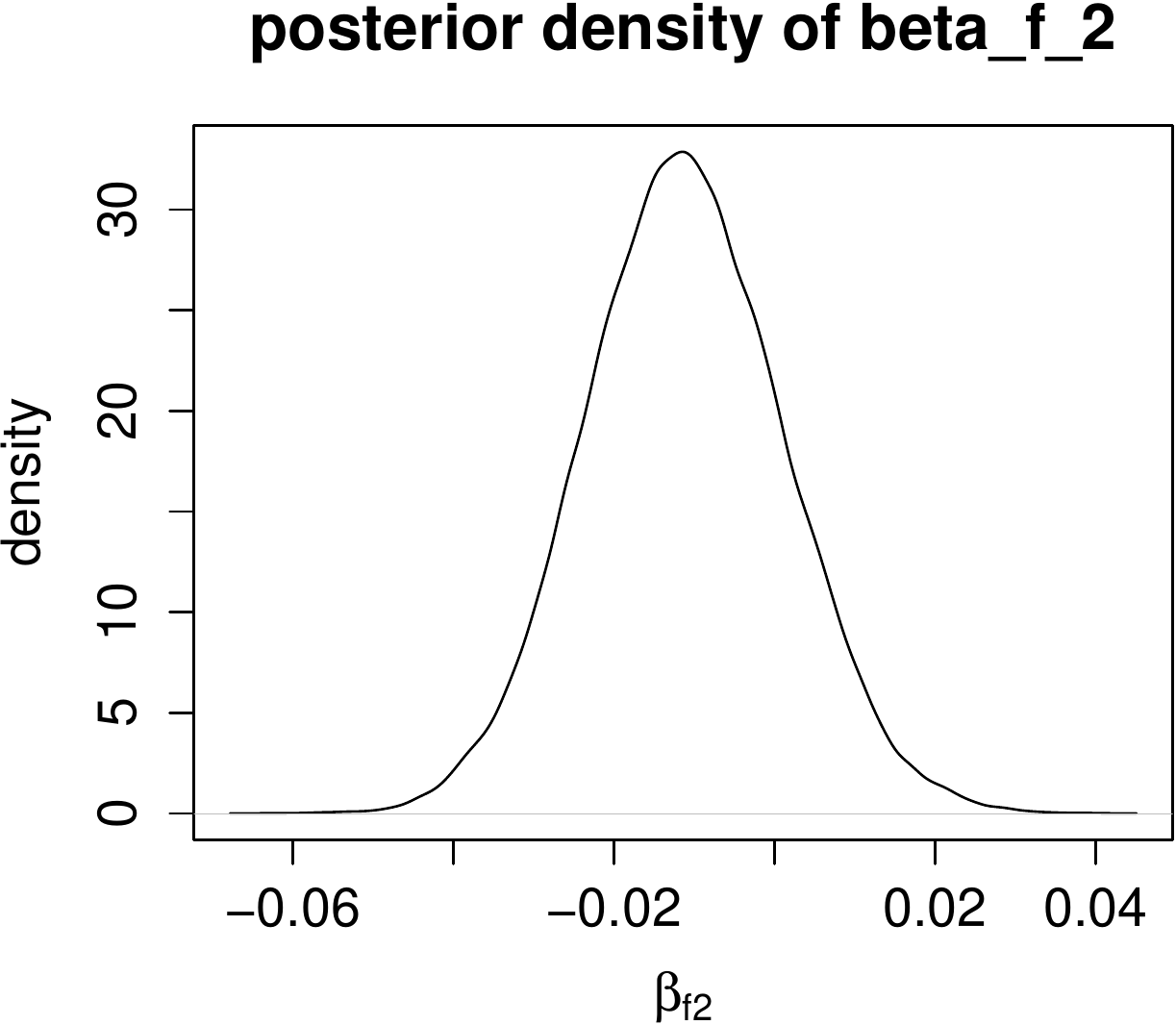}
\includegraphics[height=1.5in,width=1.5in]{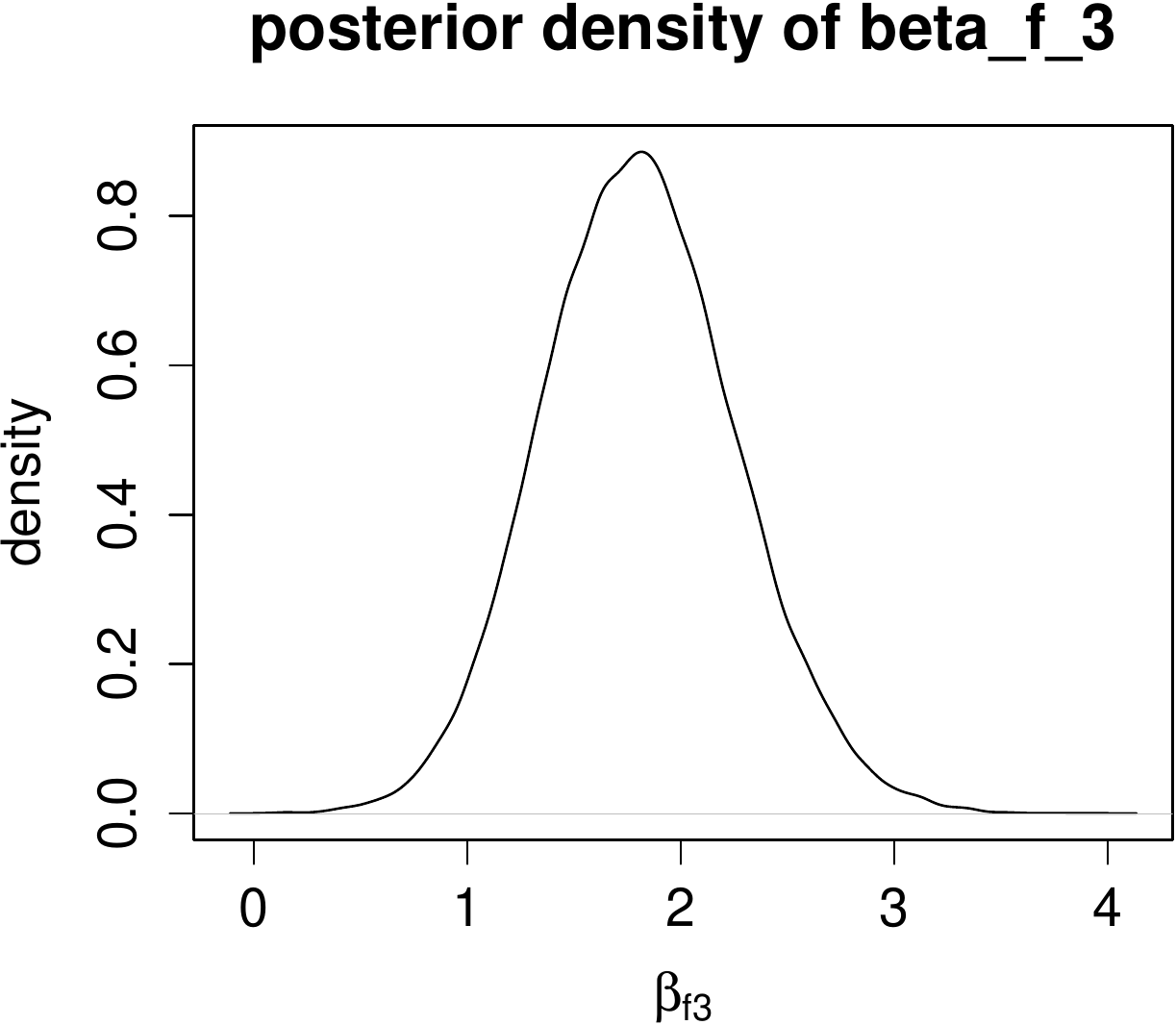}
\includegraphics[height=1.5in,width=1.5in]{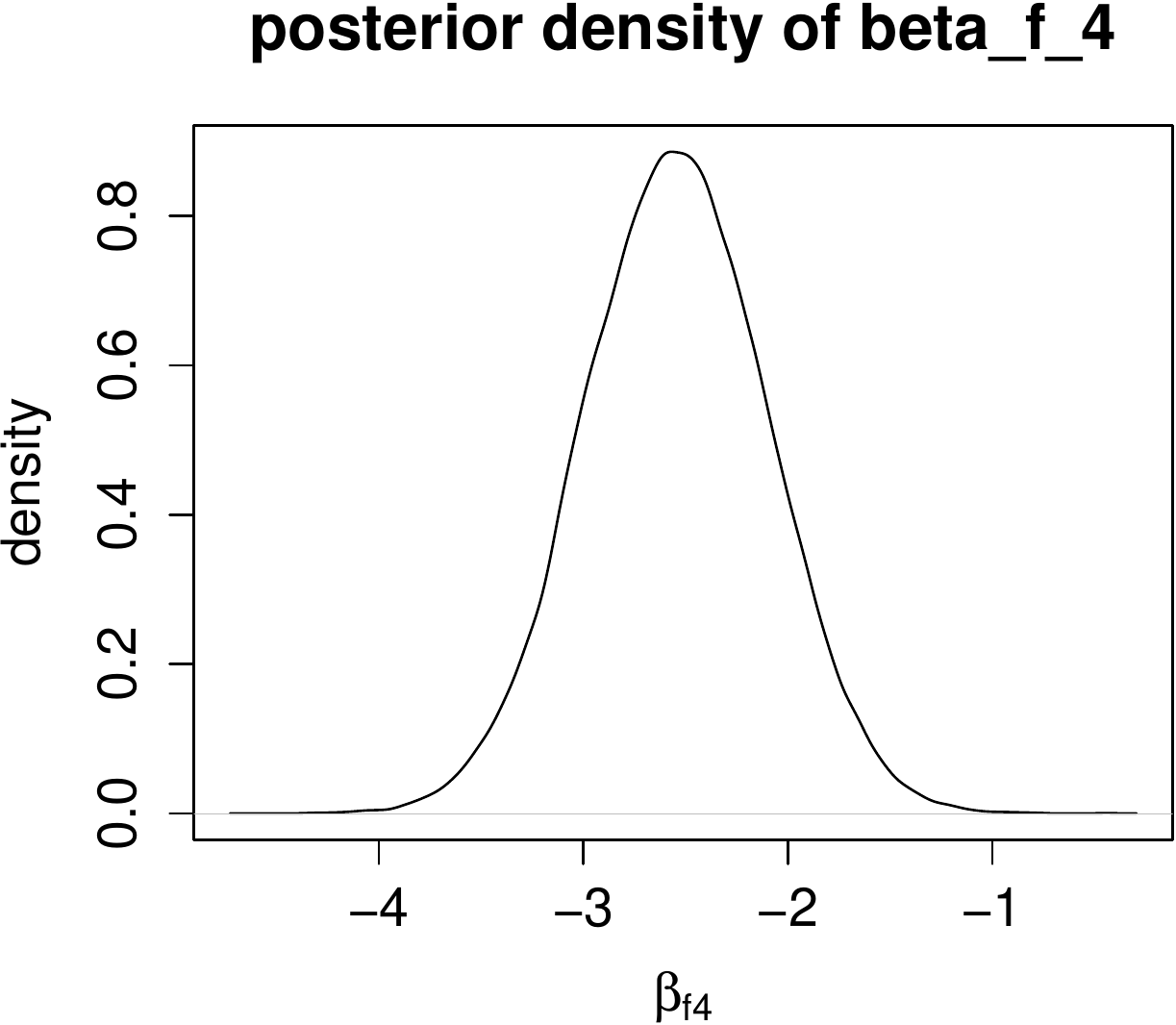}
\caption{Posterior densities of the four components of $\bi{\beta}_{f}$ for the wind speed data.}
\label{Fig:Post of beta_f for wind speed data}
\end{figure}

\begin{figure}[htp]
\centering
\includegraphics[height=2in,width=2in]{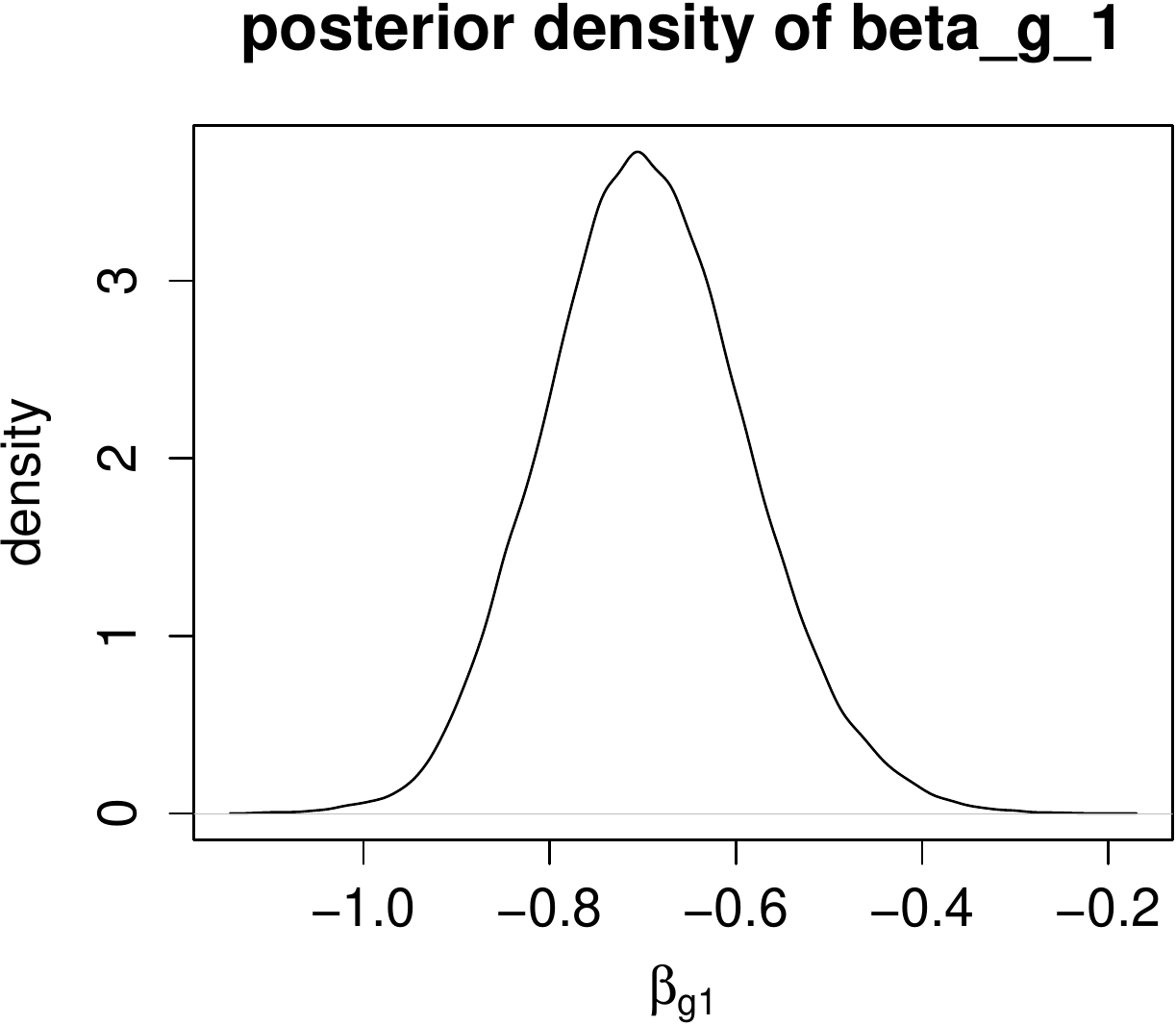}
\includegraphics[height=2in,width=2in]{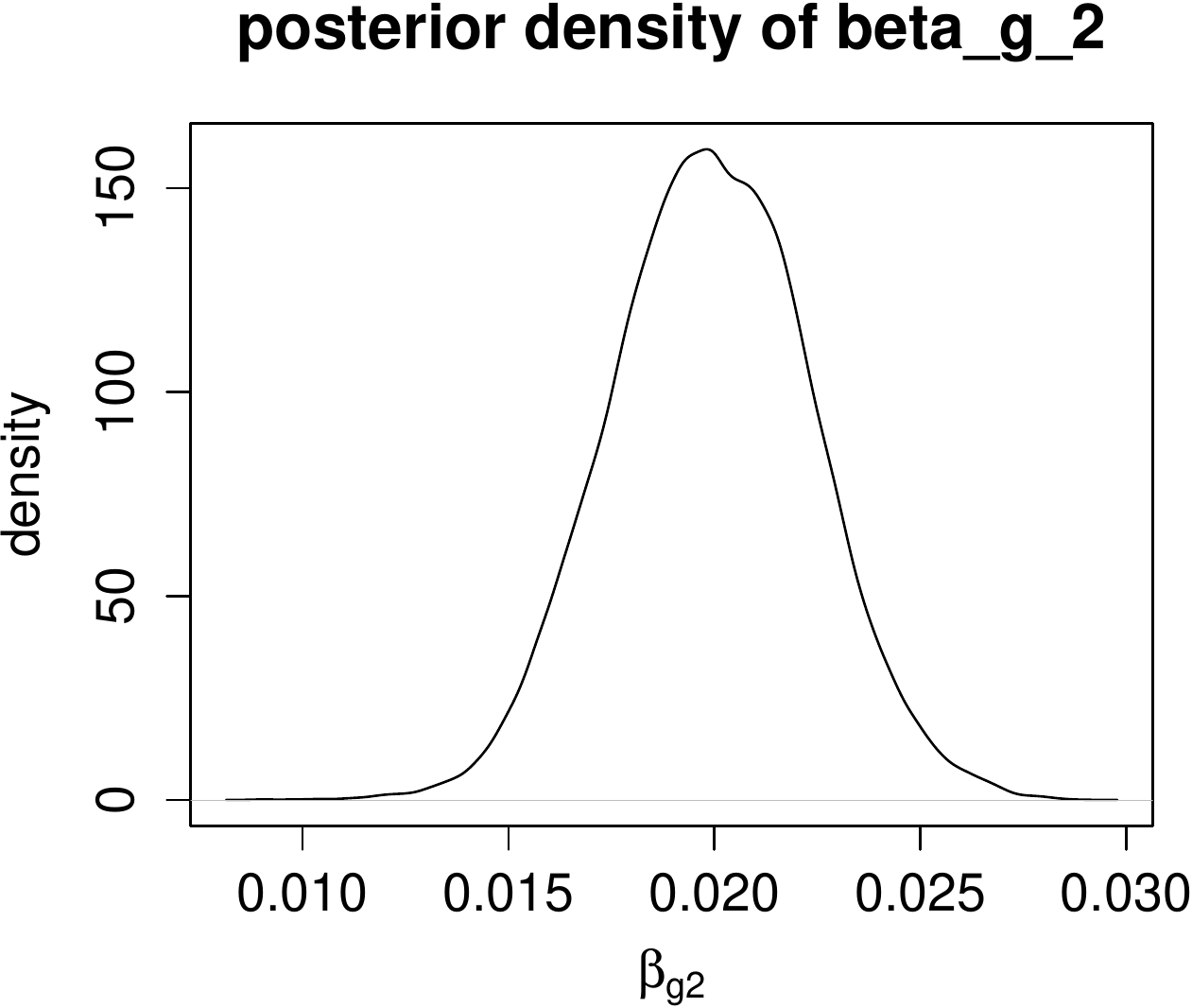}
\caption{Posterior densities of the first two components of $\bi{\beta}_{g}$ for the wind speed data.}
\label{Fig:Post of beta_g for wind speed data}
\end{figure}

\begin{figure}[htp]
\centering
\includegraphics[height=2in,width=2in]{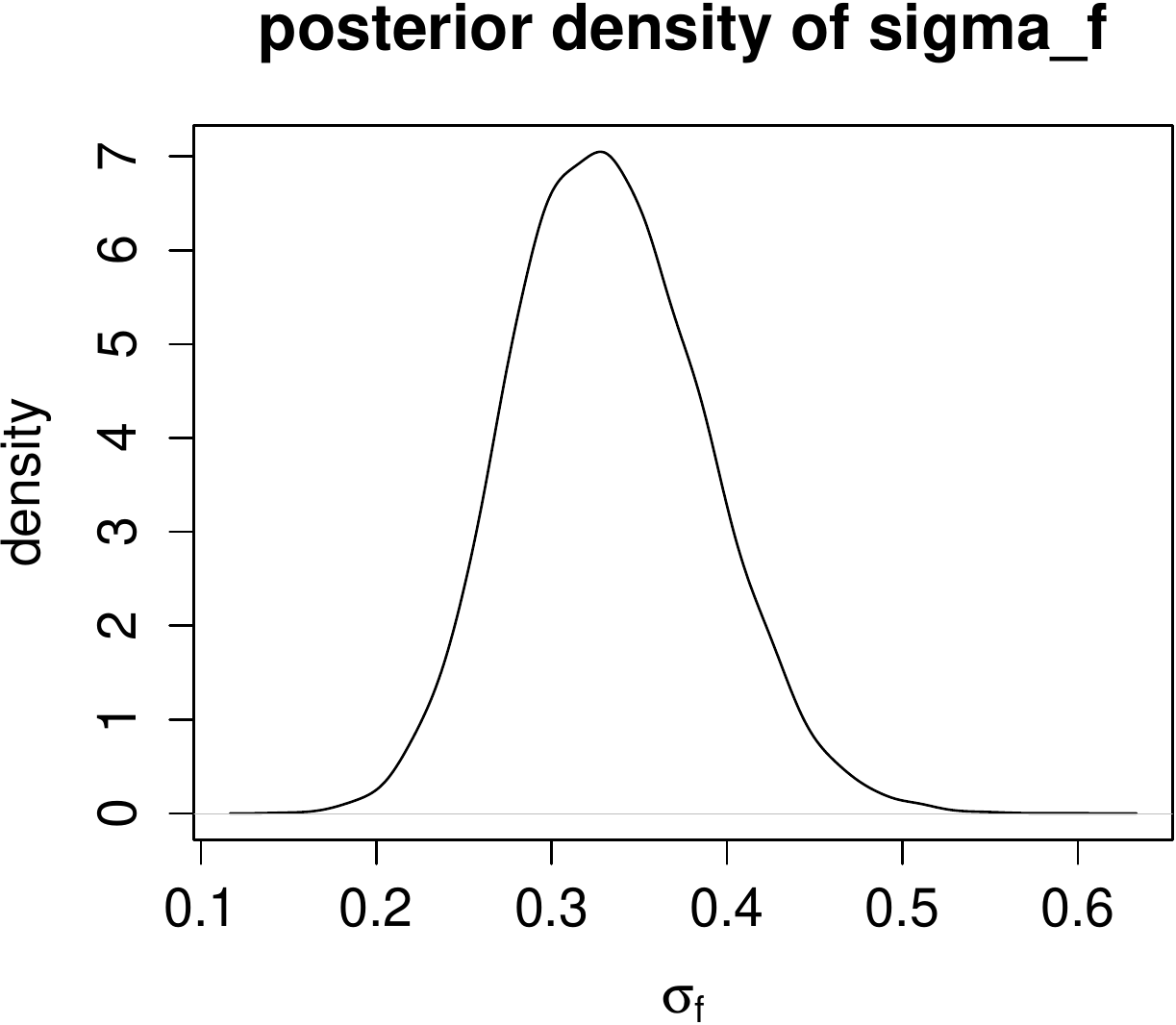}
\includegraphics[height=2in,width=2in]{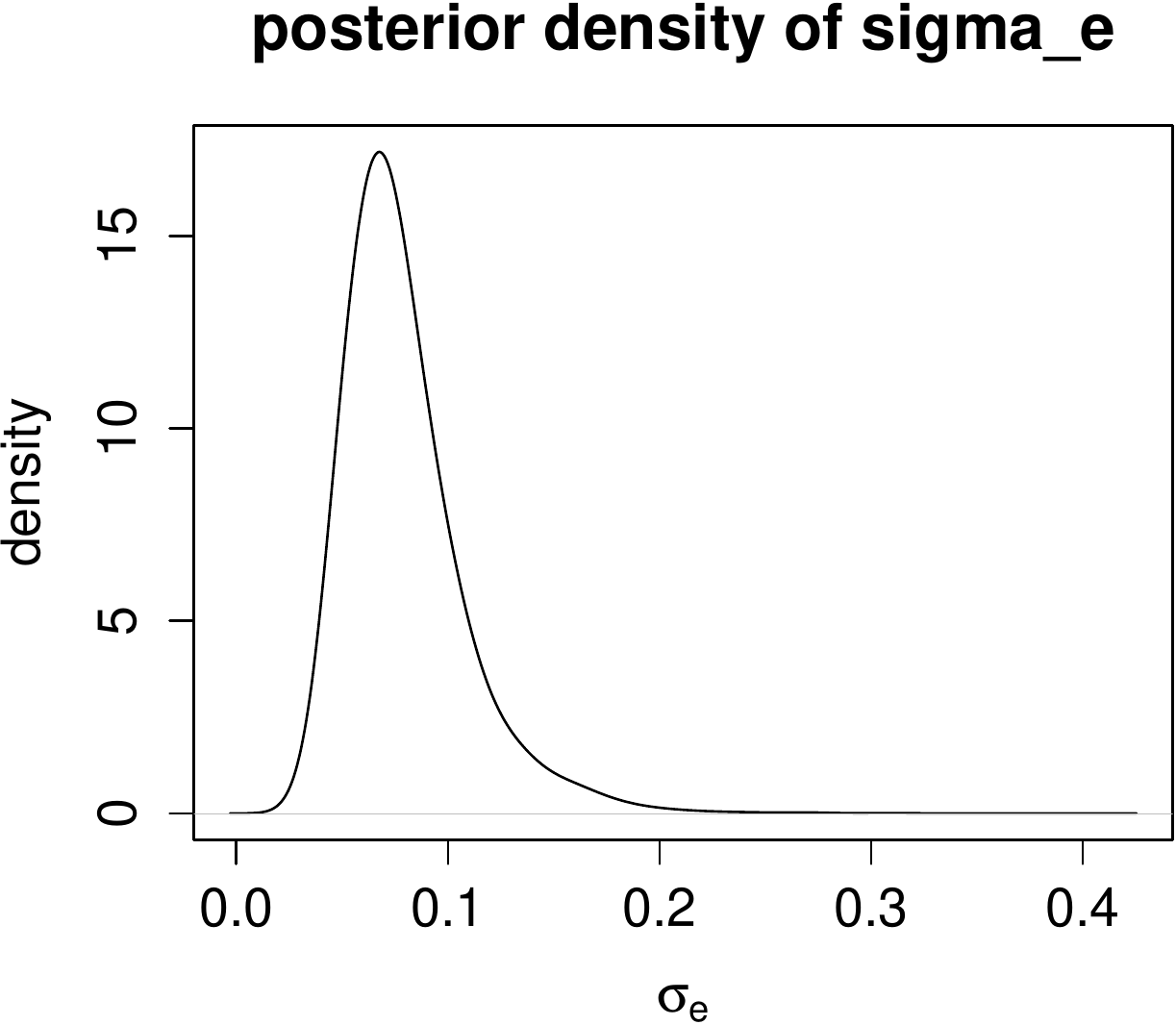}
\caption{Posterior densities of $\sigma_{f}$ and $\sigma_e$ for the wind speed data.}
\label{Fig:Post of sigma_f and sigma_e for wind speed data}
\end{figure}

\begin{figure}[htp]
\centering
\includegraphics[height=3in,width=5in]{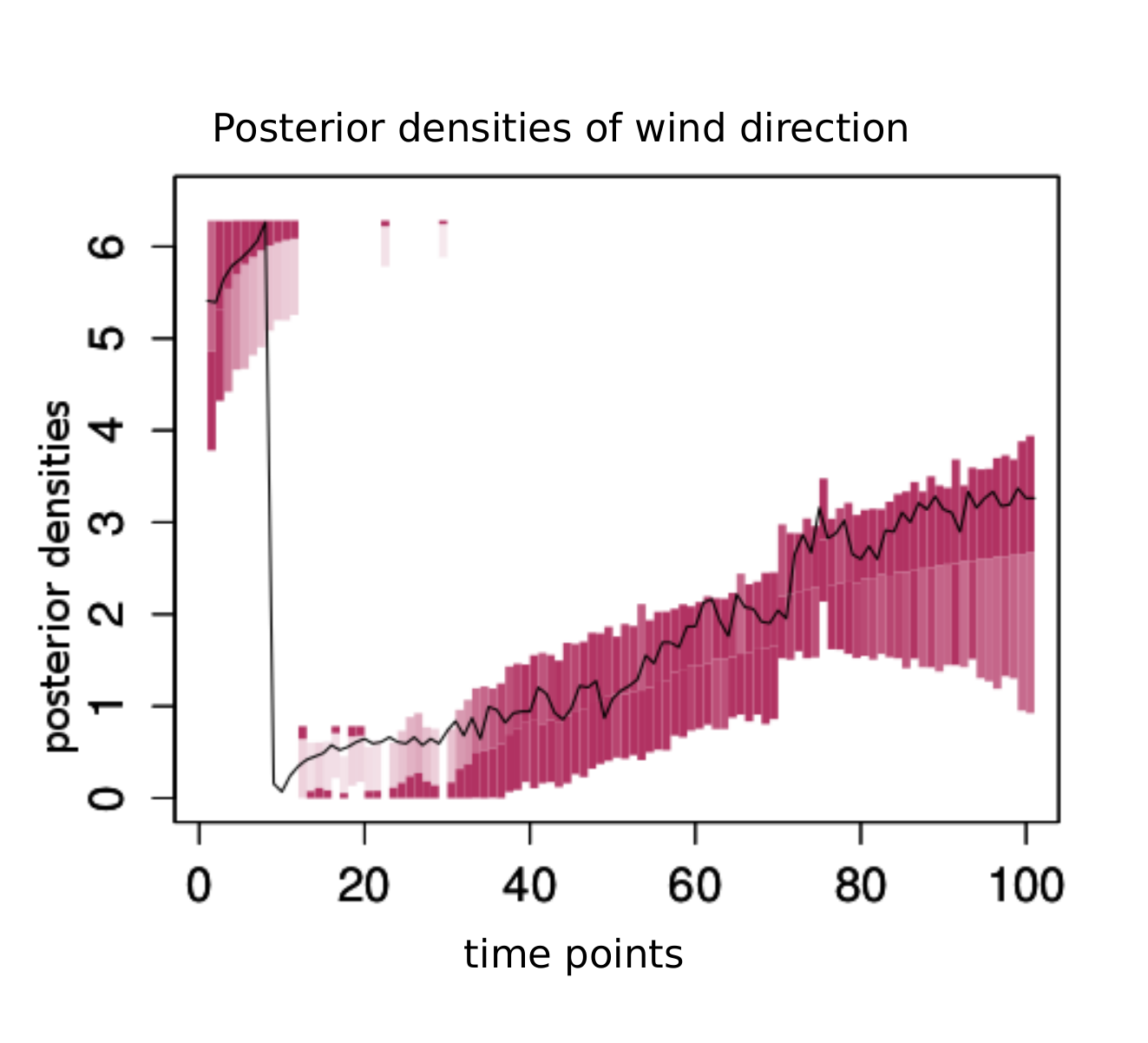}
\caption{Representation of the marginal posterior densities of the latent variables corresponding to
wind directions as a color plot; progressively higher densities are represented by progressively intense colors. 
The black line represents the true wind direction data.}
\label{Fig:latent_x for wind speed}
\end{figure}

\begin{figure}[htp]
\centering
\includegraphics[height=3in,width=3in]{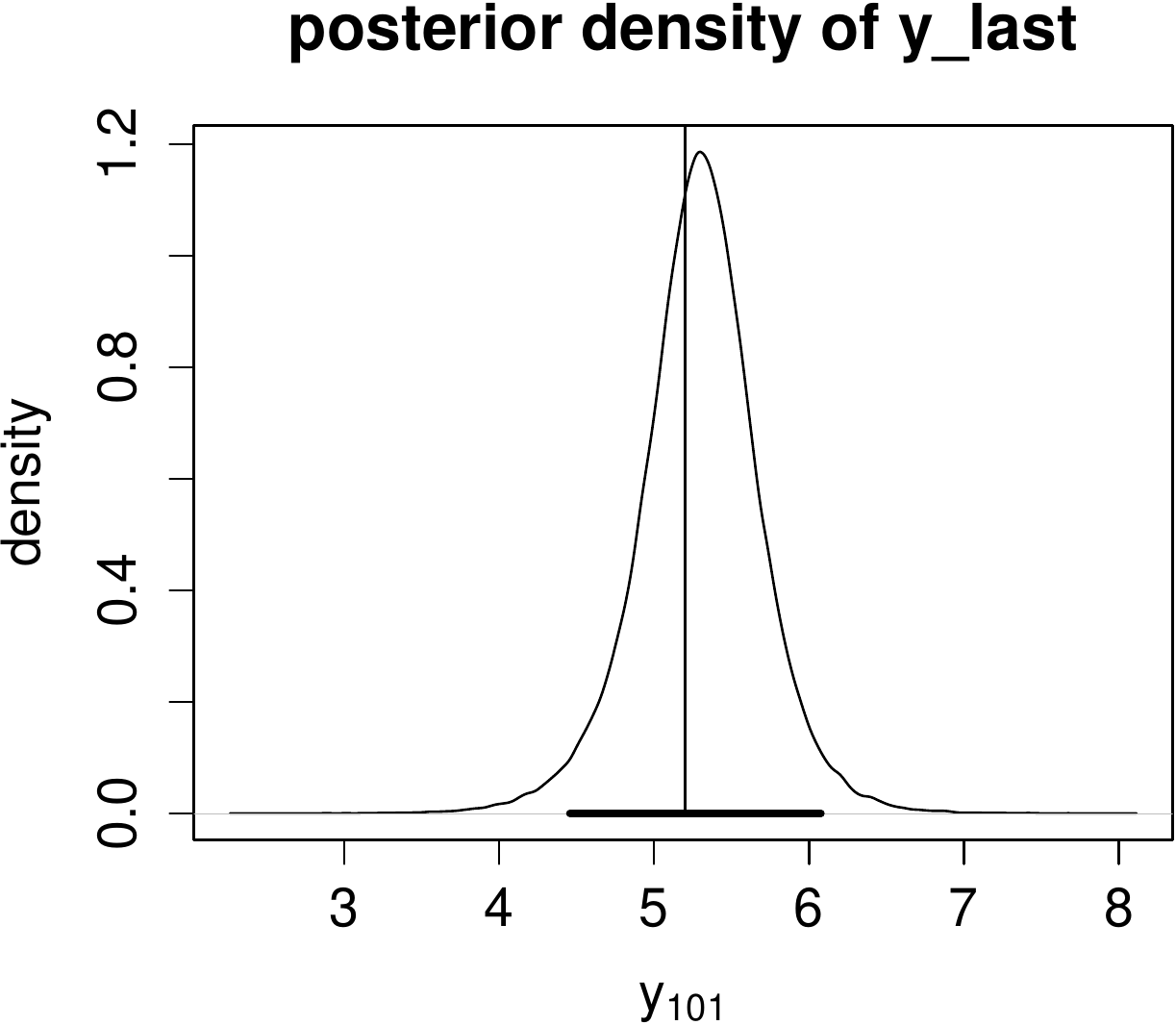}
\includegraphics[height=3in,width=3in]{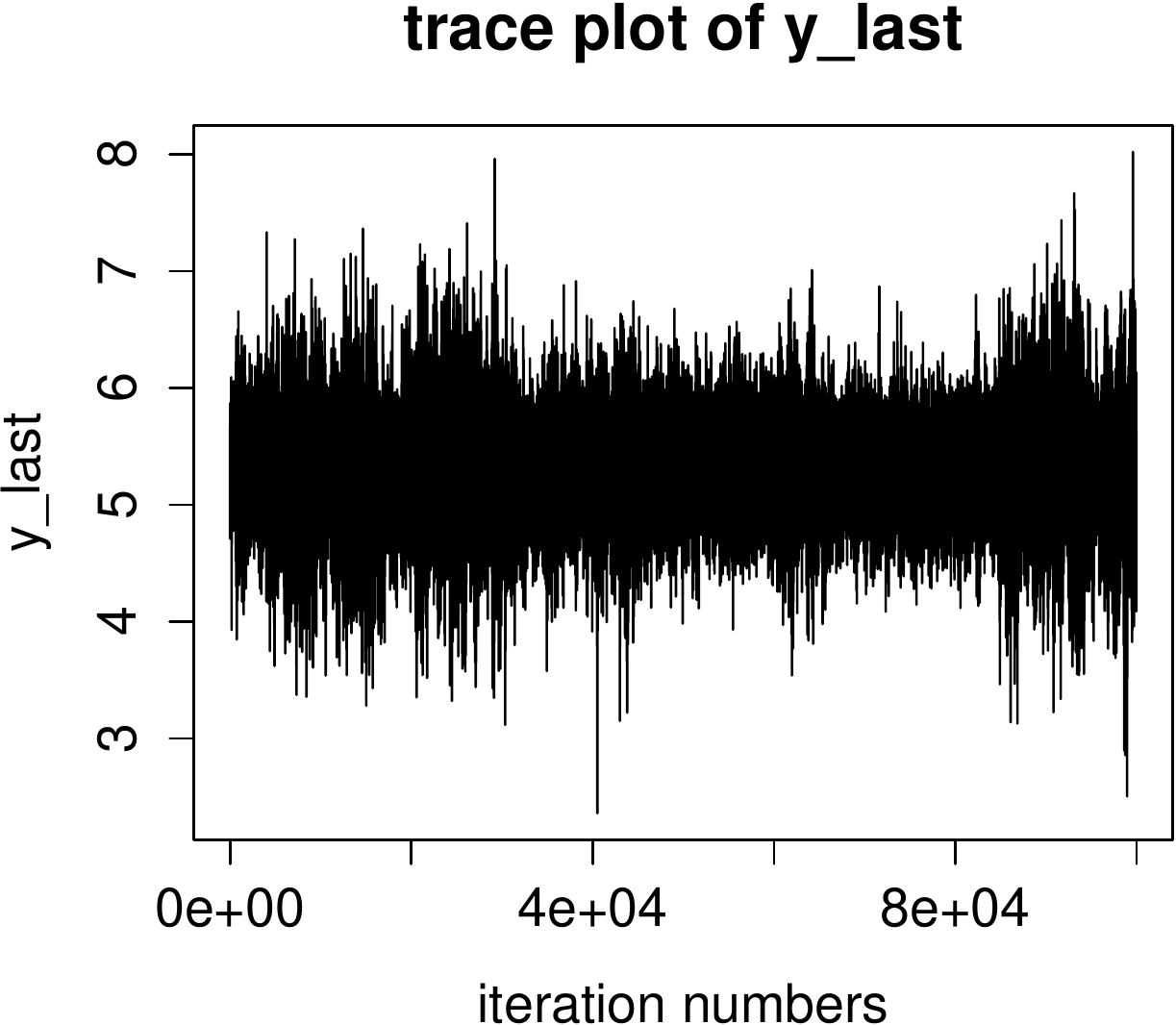}
\caption{From left\/: The first panel displays the posterior predictive density of the $101$-th observation for 
the wind speed data. The thick horizontal line denotes the 95\% highest posterior density credible interval 
and the vertical line denotes the true value. The second panel depicts the trace plot of $y_{101}$ for the 
last 1,00,000 iterations.}
\label{Fig:Post predictive of y_last of wind_speed}
\end{figure}

Figures \ref{Fig:Post of beta_f for wind speed data} and \ref{Fig:Post of beta_g for wind speed data} 
provide the posterior densities of four components of the vector
$\bi{\beta}_f$ and $\bi{\beta}_g$, respectively. The posterior densities of $\sigma_e$ and $\sigma_f$ 
are shown in Figure \ref{Fig:Post of sigma_f and sigma_e for wind speed data}. The posterior 
predictive density of $y_{101}$ is provided in Figure \ref{Fig:Post predictive of y_last of wind_speed}.
It is seen that the true value falls well within the 95\% highest posterior density credible interval, 
which shows how well our model and the 
prior distributions of the parameters succeed in describing the uncertainty present in the data. 
(see, for instance, \ctn{Box73} and \ctn{Bickel07}). 
A trace plot for $y_{101}$ is also
displayed in Figure \ref{Fig:Post predictive of y_last of wind_speed} as a sample demonstration
of MCMC convergence. 
We depict the marginal posterior densities of the latent variables in Figure
\ref{Fig:latent_x for wind speed}, where progressively higher intensities of the color denote regions of progressively
higher posterior densities. It can be seen that the true values of the latent
variable, that is, the true values of wind direction (in radians) fall mostly in the corresponding 
high probability regions. 
Indeed, it is really encouraging to observe that our model and methods successfully capture the highly non-linear 
trend, even with a sharp discontinuity at around $t=10$, denoting a change point, present in the 
original wind direction data. This has been possible because of our nonparametric ideas and also because our 
model allows the unknown
observational and the evolutionary
function based on Gaussian processes to change with time.
To sum up, it can be inferred that our model and methodologies not only capture the true 
wind directions in the respective high posterior probability regions, but ensure that the posterior probabilities concentrate
on relatively small regions, which, in turn, allows us to identify the trend present in the actual process with much precision.

\subsection{Ozone level data}
\label{subsec:Ozone level data}
\subsubsection{A brief description of the data set}
\label{subsec:data_description}
\begin{figure}[htp]
\centering
\includegraphics[height=3in,width=3in]{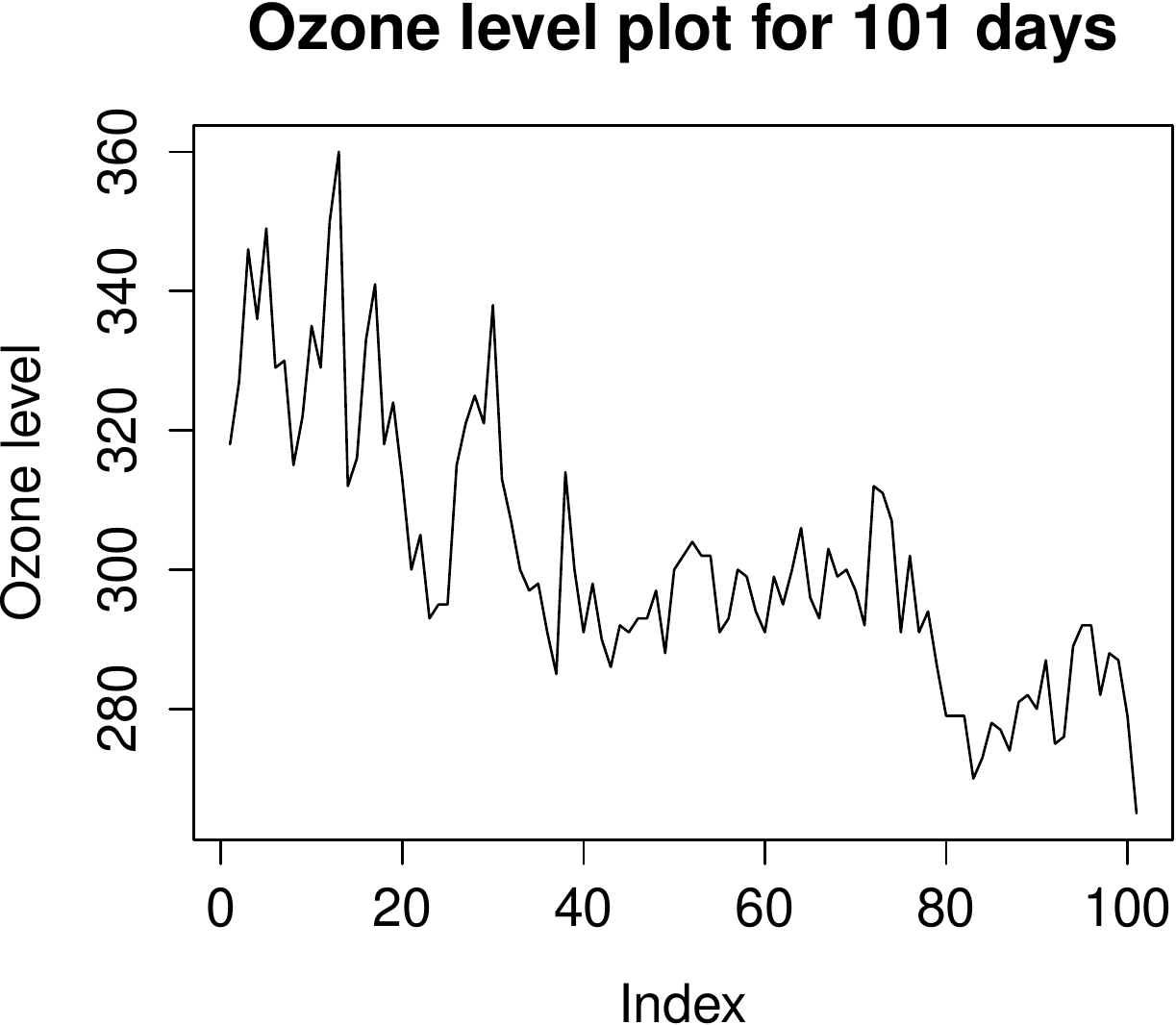}
\includegraphics[height=3in,width=3in]{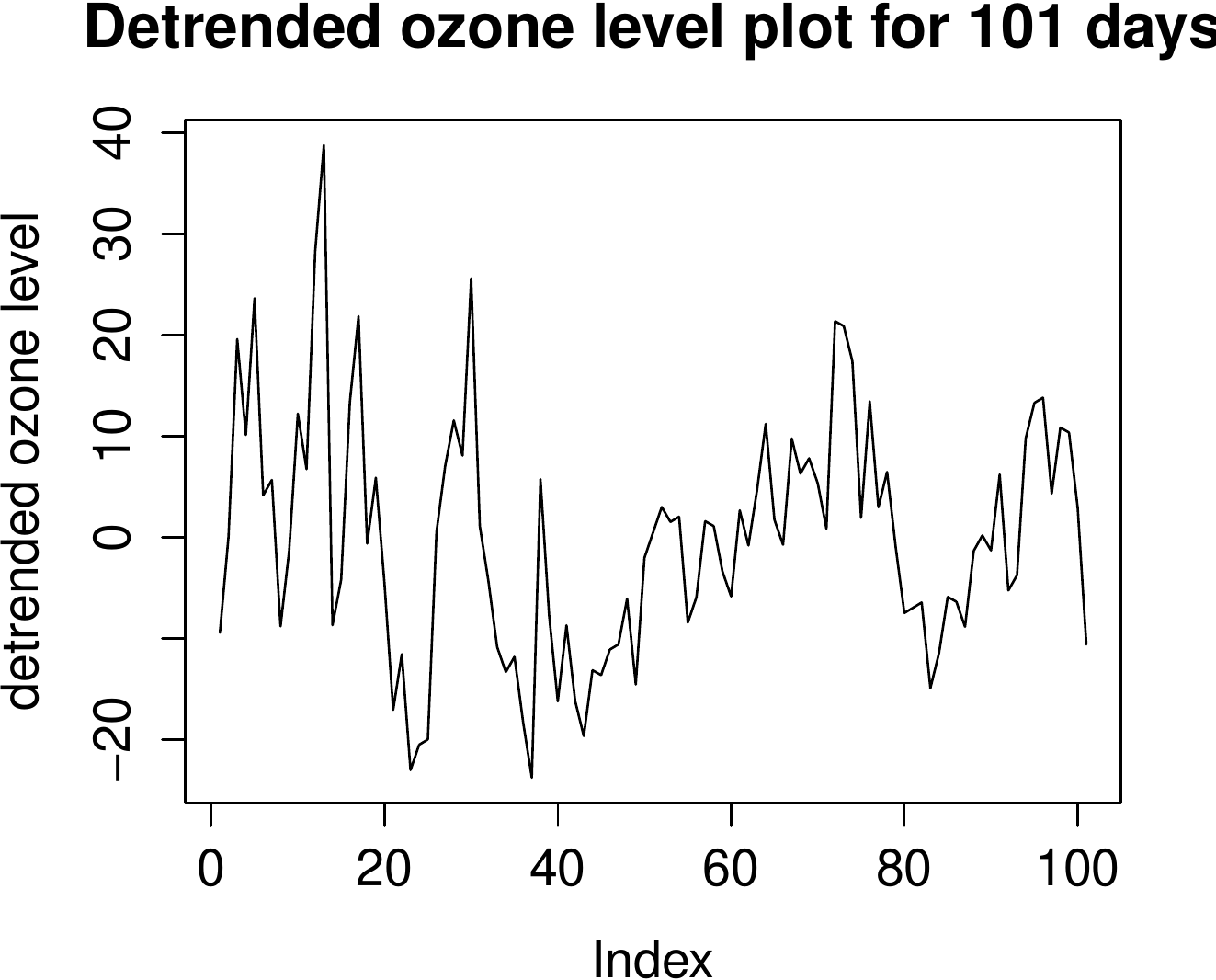}
\caption{Plot of ozone level for 101 days in Boulder, Colorado.}
\label{Fig:Ozone data plot}
\end{figure}

We now apply our model and methodologies to a real data set obtained from the website \\
http://www.esrl.noaa.gov/gmd/grad/neubrew/OmiDataTimeSeries.jsp. 
The data concerns the ozone level present in the atmosphere at a particular location and at a particular year. 
For our analysis we select a location with latitude 40.125 and longitude 105.238, which corresponds to Boulder, Colorado. 
We collected 101 observations starting from May 15, 2013. 
The plot of the ozone level data is provided in Figure \ref{Fig:Ozone data plot}.
Although it is expected that the ozone level present 
in the atmosphere depends upon the direction of wind flow (see \ctn{Jamma06}), the data on the direction of wind flow 
is not available at that particular location and time. Therefore, we expect that our general, nonparametric model
and the associated methods will be quite useful in this situation. We retain 100 observations for our analysis and keep aside 
the last observation for the purpose of prediction. Before applying our model and methods, we first de-trend the
data-set. Plot of detrended ozone data is displayed in Figure \ref{Fig:Ozone data plot} along with the plot of observed ozone data for 101 days.

\subsubsection{Prior choices}

We keep the same choices of the prior parameters as done in case of ozone level data. 
The MLEs of $\sigma_{\eta}$ and $\sigma_g$, obtained by the simulated annealing method discussed
in Section \ref{subsec:impropriety}, are 0.0493 and 0.2269, respectively. 

\subsubsection{MCMC implementation}
\label{subsec:mcmc_real_data}

With these choices of prior parameters we implement our MCMC algorithm detailed in Section S-2
of the supplement with the random walk scales chosen 
on the basis of informal trial and error method associated with many pilot runs of our MCMC algorithm. 
As mentioned in \ref{subsec:mcmc_details}, here also we use the same mixture of von-Mises to update $x_t$, $t=1,\ldots,100$.
Our final MCMC run is based on $2,50,000$ iterations of MCMC, of which we discarded the first $2,00,000$ iterations 
as the burn-in period. The time taken for $2,50,000$ iterations of MCMC on our desktop computer with $i7$ processors, 
is about 21 hours.

\subsubsection{Results of ozone level data}
\label{subsec:real_data_results}
\begin{figure}[htp]
\centering
\includegraphics[height=1.5in,width=1.5in]{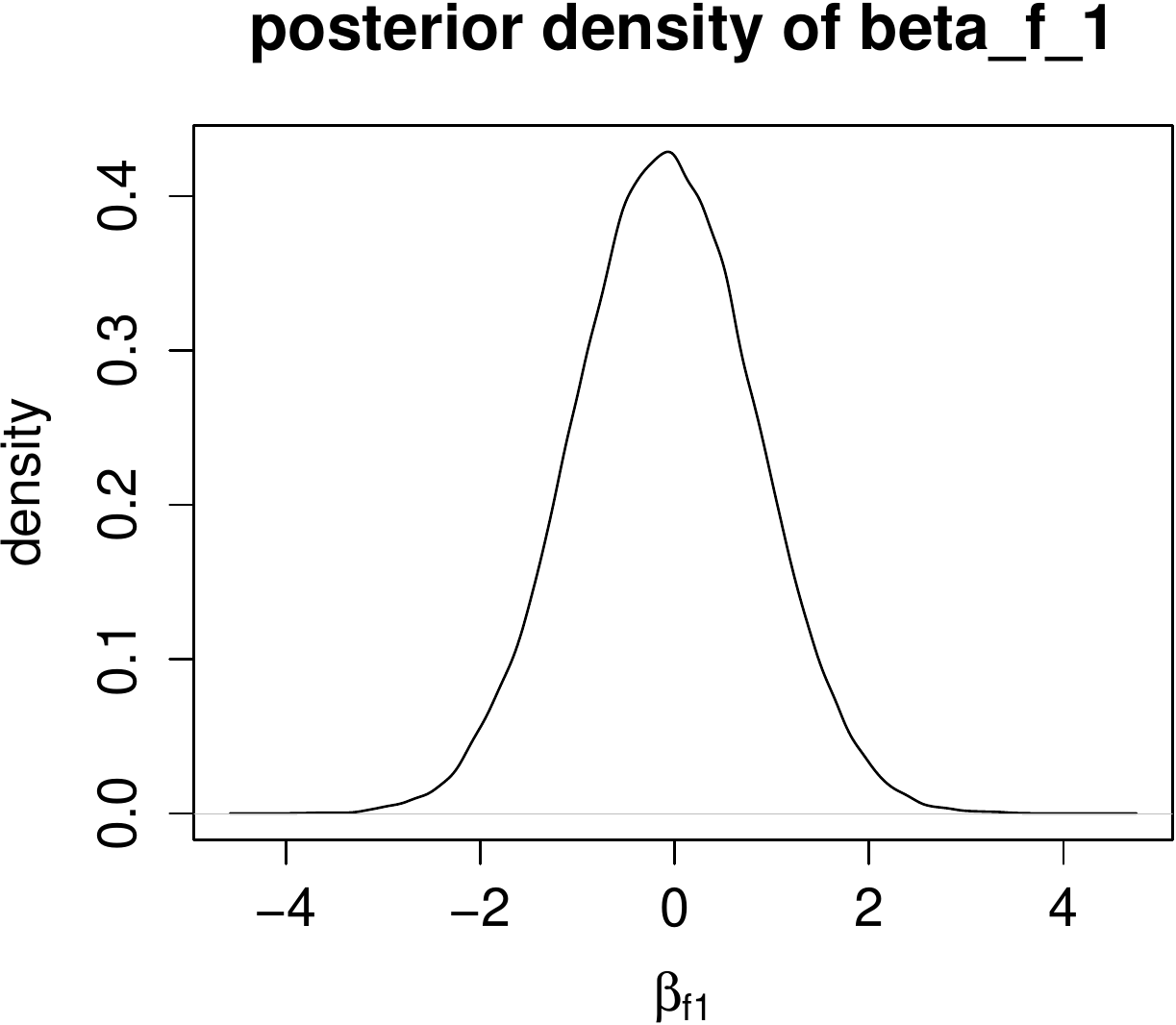}
\includegraphics[height=1.5in,width=1.5in]{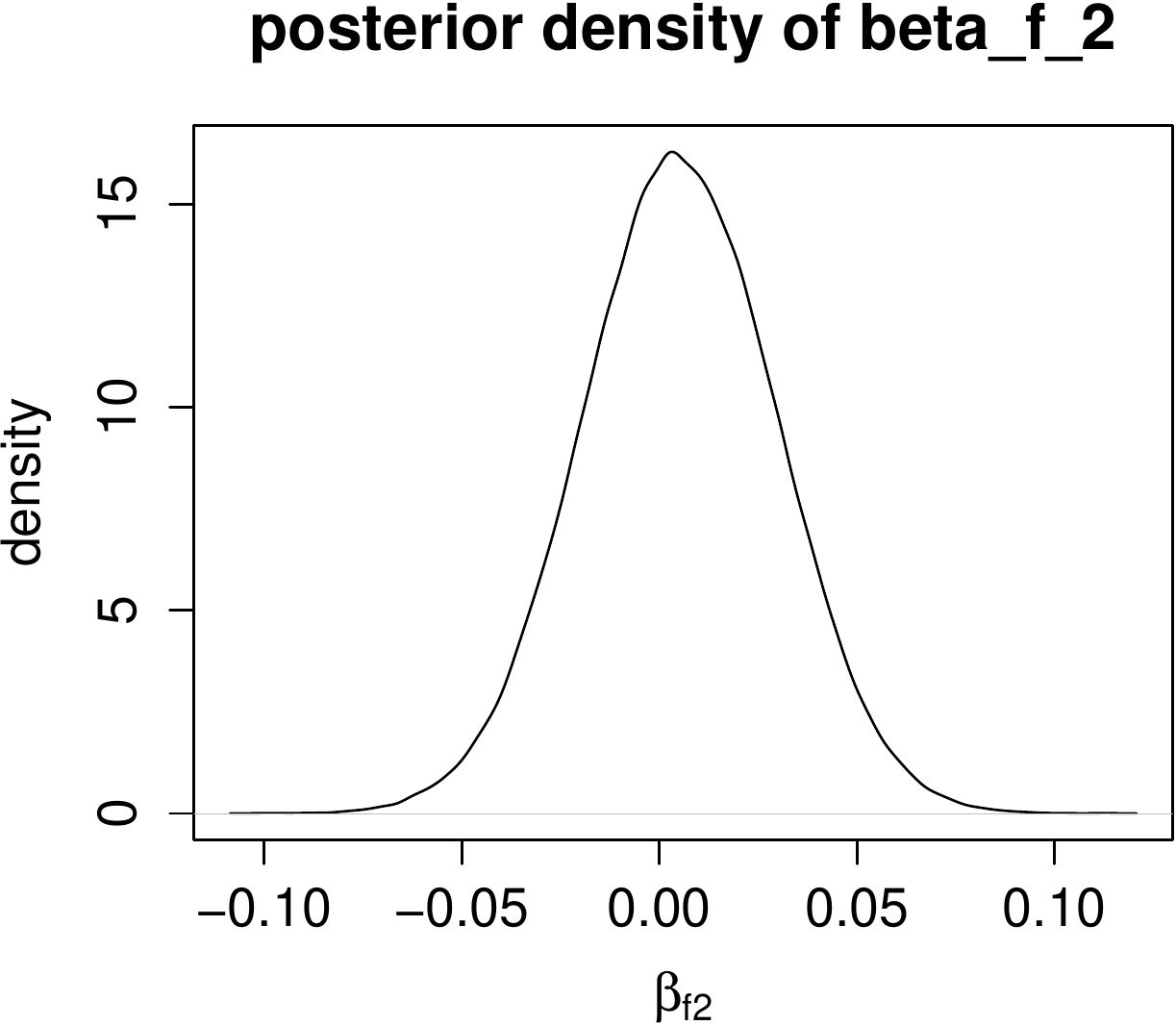}
\includegraphics[height=1.5in,width=1.5in]{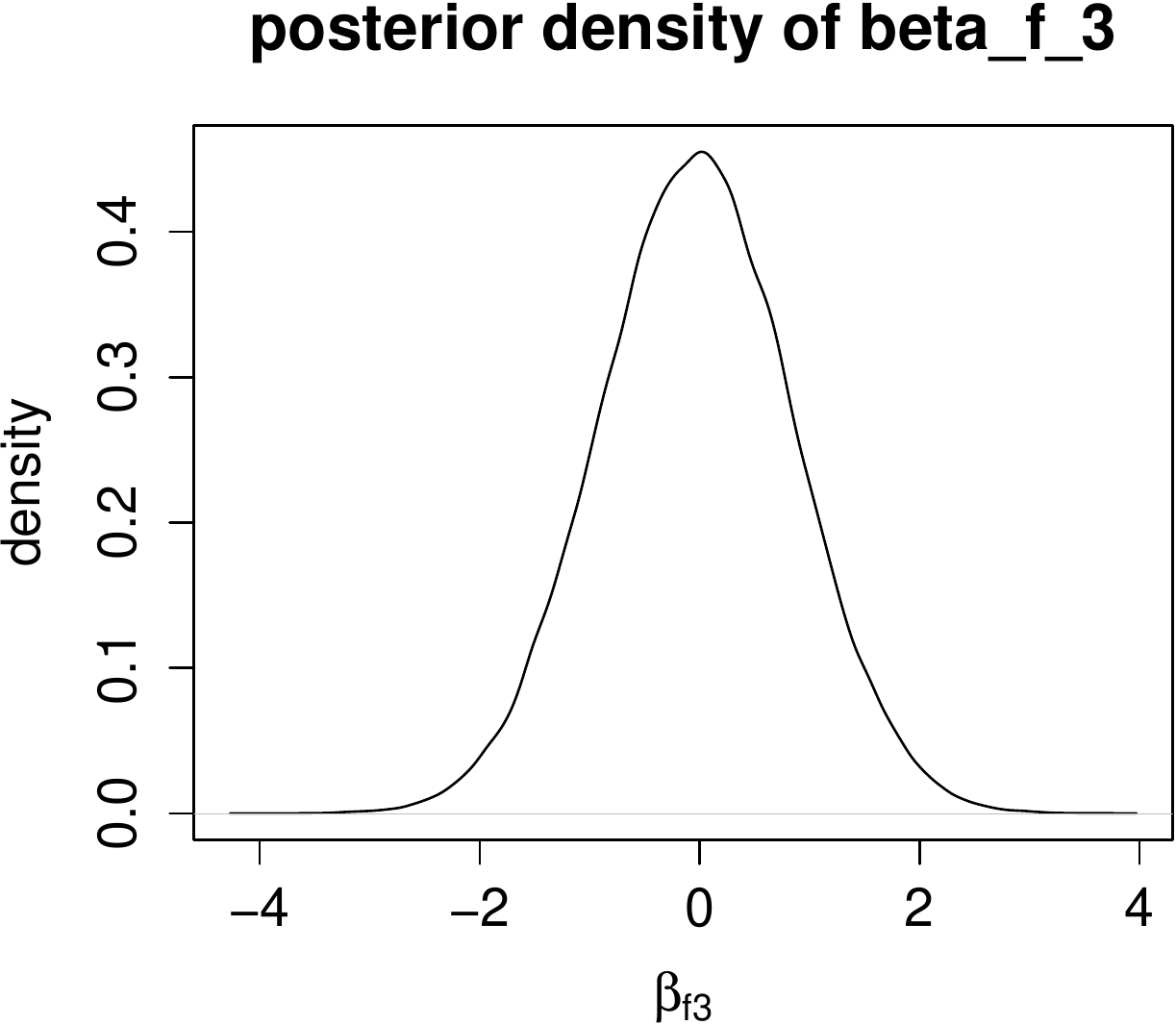}
\includegraphics[height=1.5in,width=1.5in]{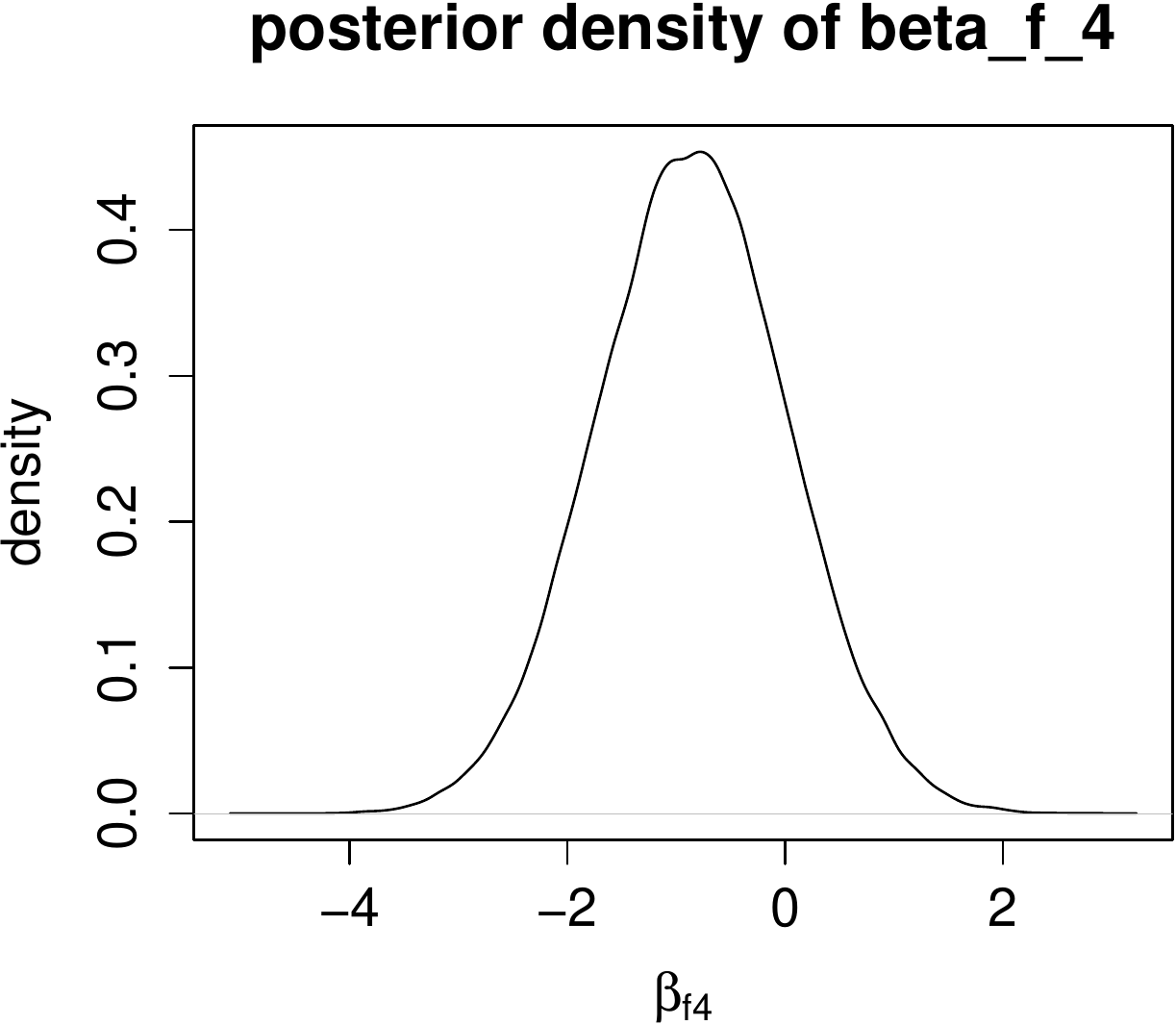}
\caption{Posterior densities of the three components of $\bi{\beta}_{f}$ for the ozone data.}
\label{Fig:Post of beta_f for ozone data}
\end{figure}

\begin{figure}[htp]
\centering
\includegraphics[height=2in,width=2in]{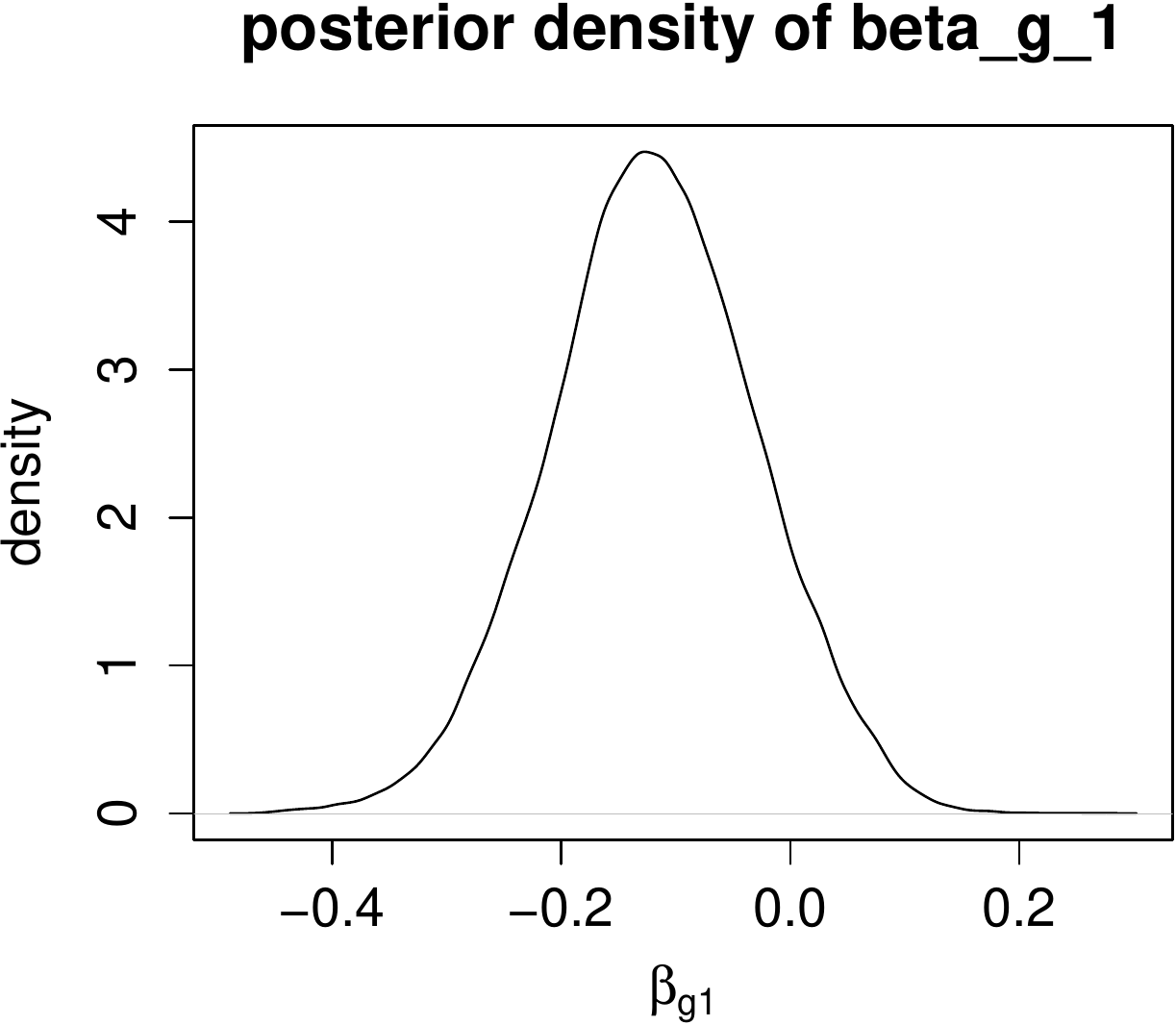}
\includegraphics[height=2in,width=2in]{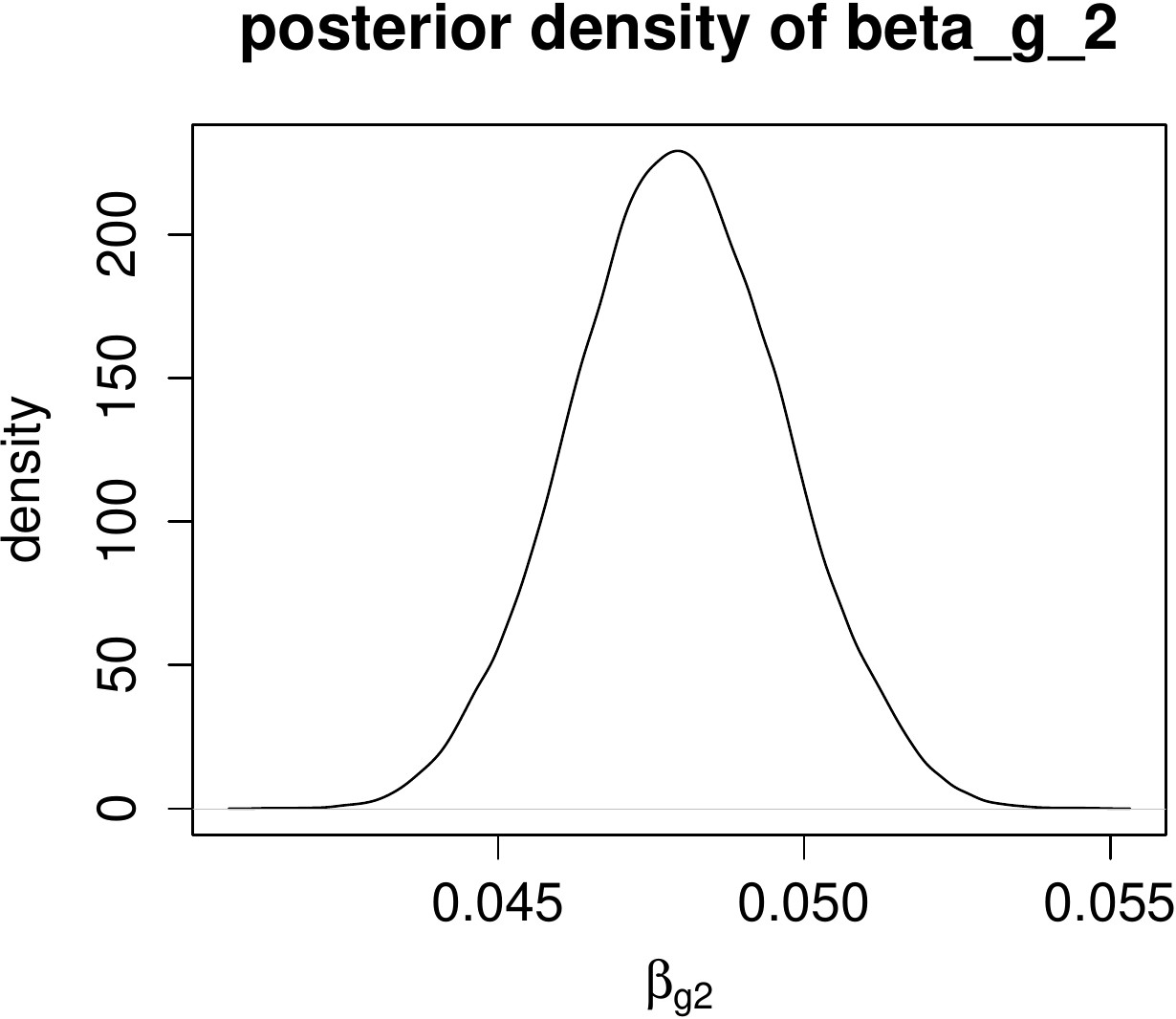}
\caption{Posterior densities of the first two components of $\bi{\beta}_{g}$ for the ozone data.}
\label{Fig:Post of beta_g for ozone data}
\end{figure}

\begin{figure}[htp]
\centering
\includegraphics[height=2in,width=2in]{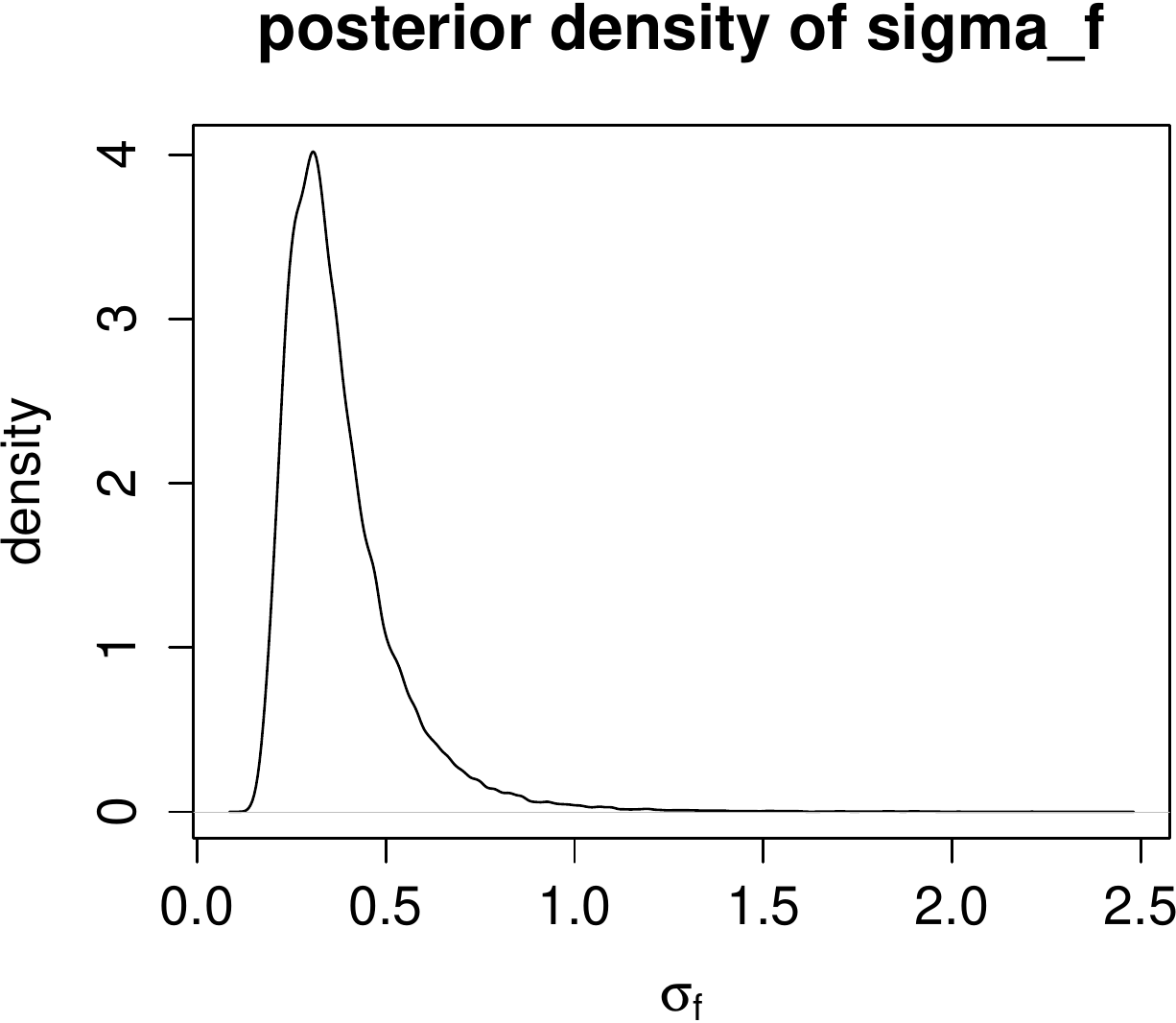}
\includegraphics[height=2in,width=2in]{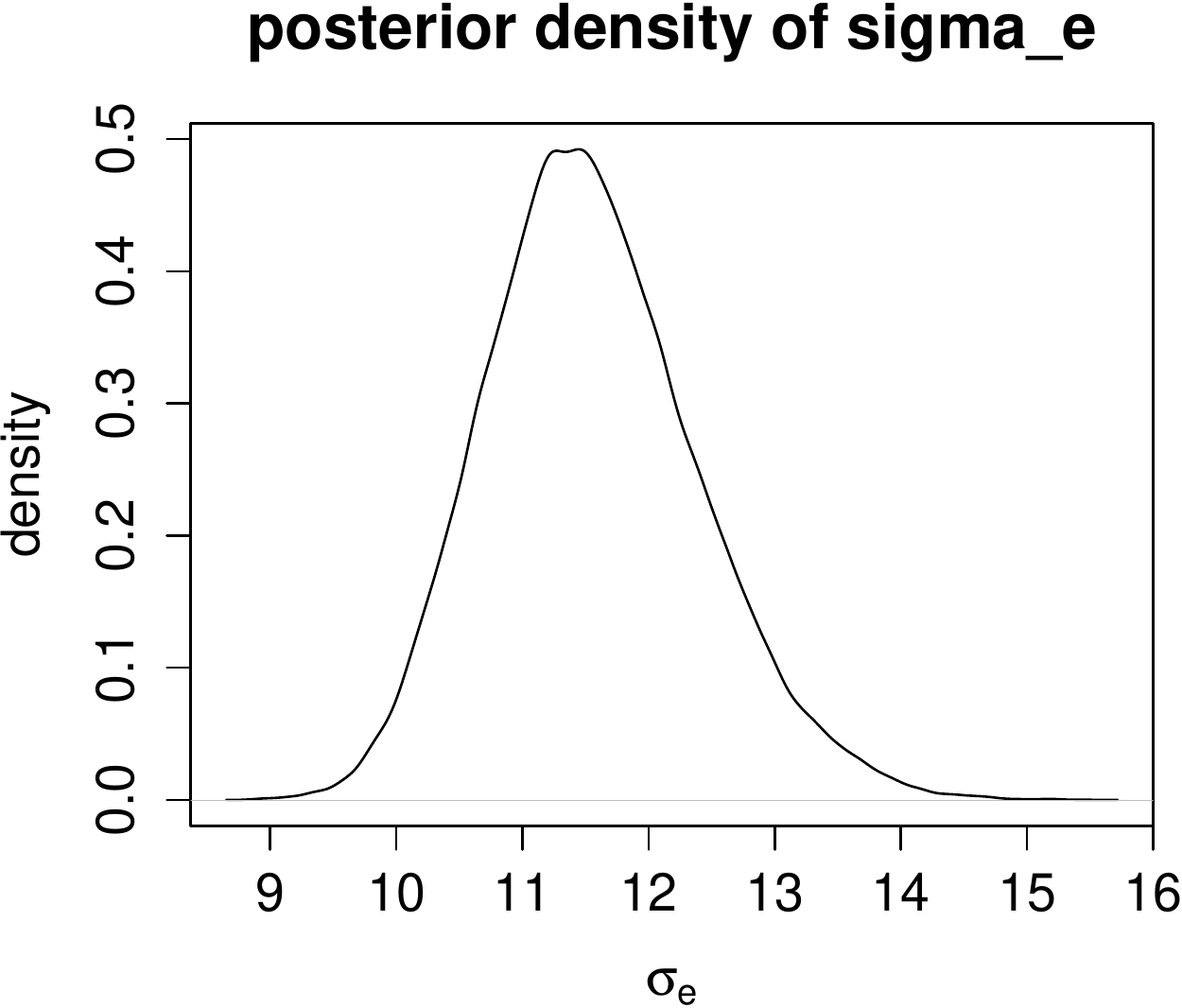}
\caption{Posterior densities of $\sigma_{f}$ and $\sigma_e$ for the ozone data.}
\label{Fig:Post of sigma_f and sigma_e for ozone data}
\end{figure}

\begin{figure}[htp]
\centering
\includegraphics[height=3in,width=5in]{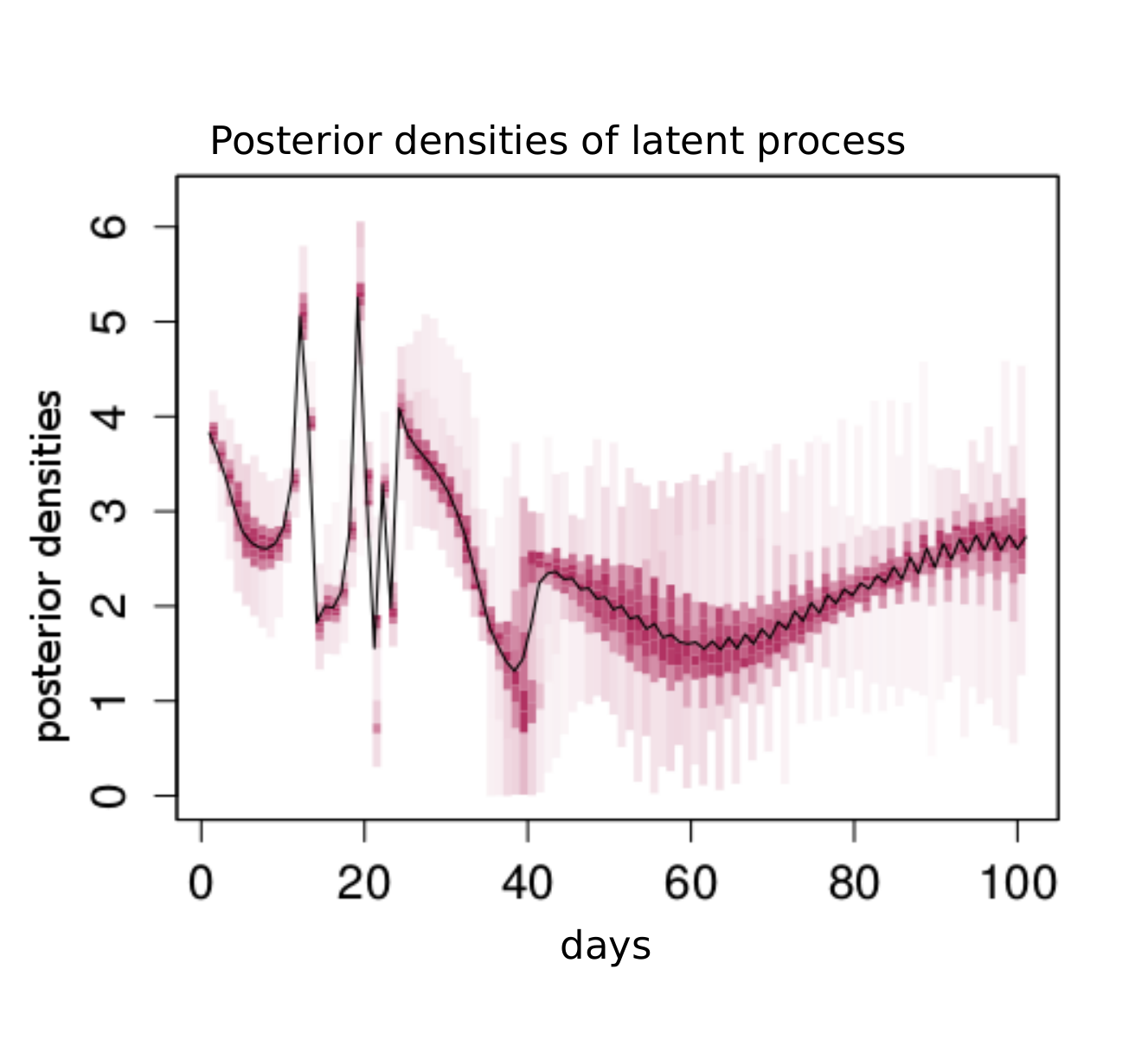}
\caption{Depiction of the marginal posterior distributions of the latent variables 
using progressively intense colors for progressively higher densities, 
and the median for the latent process of the ozone data.}
\label{Fig:latent_x for real data}
\end{figure}

\begin{figure}[htp]
\centering
\includegraphics[height=3in,width=3in]{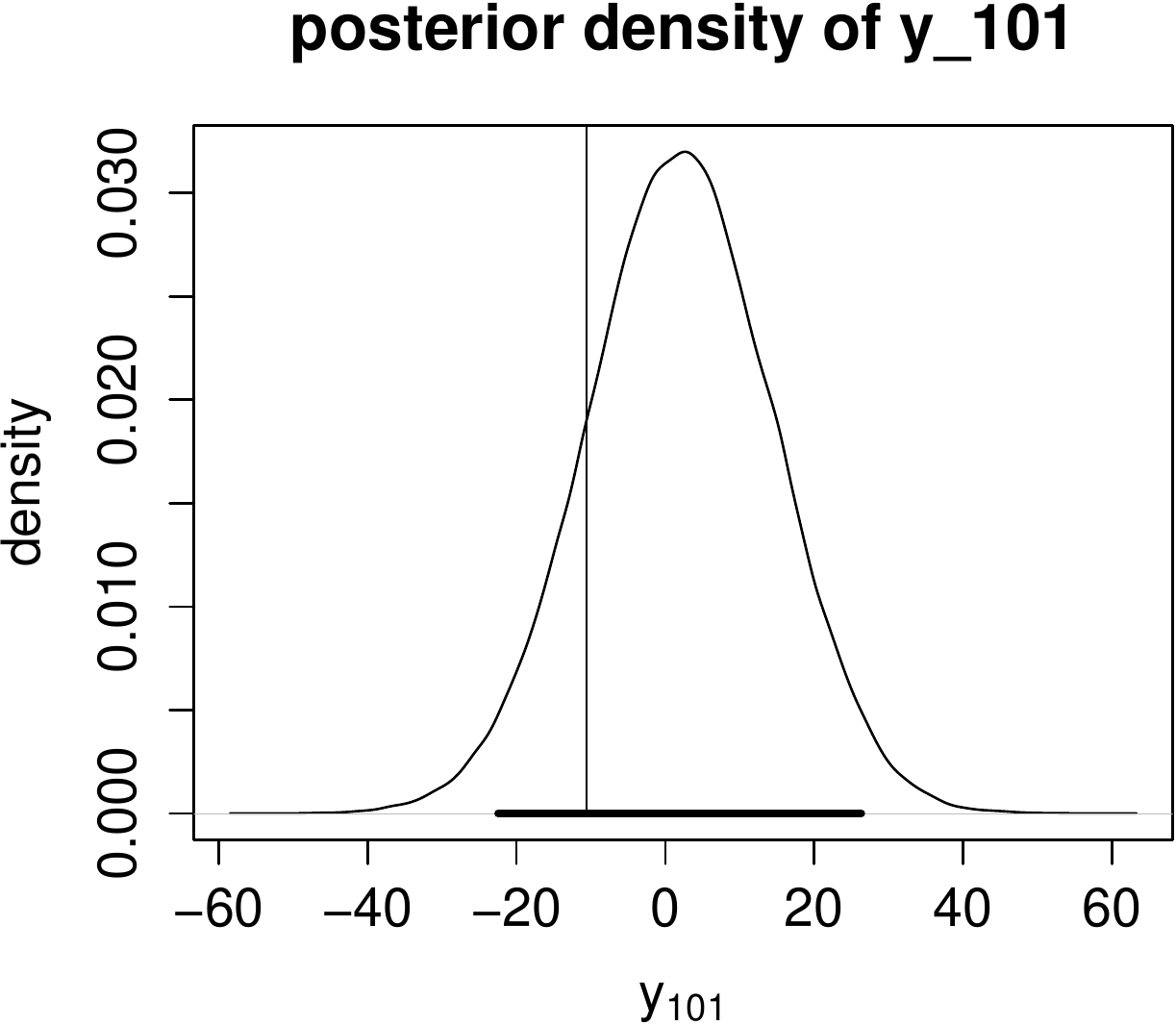}
\includegraphics[height=3in,width=3in]{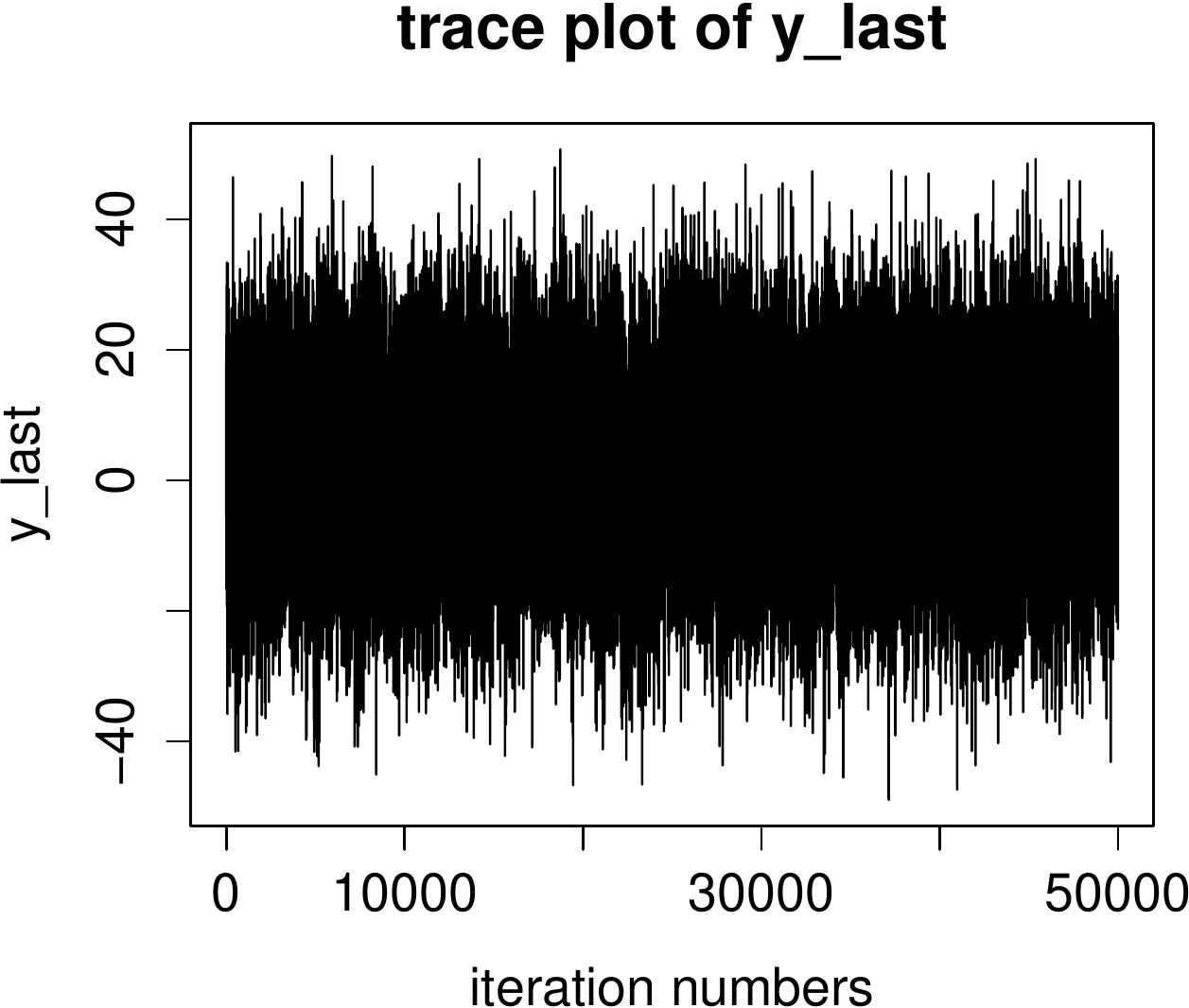}
\caption{Posterior predictive density of the $101$-th observation for the ozone data. 
The thick horizontal line denotes the 95\% highest posterior density credible interval and 
the vertical line denotes the true value.}
\label{Fig:Post predictive of y_last}
\end{figure}

The posterior densities of the four components of $\bi{\beta}_f$ and the two components of $\bi{\beta}_g$
are provided in Figures \ref{Fig:Post of beta_f for ozone data} and \ref{Fig:Post of beta_g for ozone data}, respectively. 
The posterior densities of $\sigma_e$ and $\sigma_f$ are shown in Figure \ref{Fig:Post of sigma_f and sigma_e for ozone data}. 
Figure \ref{Fig:latent_x for real data} shows the marginal posterior distributions  
associated with the latent circular process depicted by progressively intense colors, along
with the posterior median indicated by black line.
Finally, the posterior predictive density corresponding to $y_{101}$ is provided in Figure \ref{Fig:Post predictive of y_last}. 
Here the thin vertical line denotes the true value of the $101$-th observation and the thick line represents the 
95\% highest density region of the posterior predictive density. As in our previous experiments, here also the true value 
falls well within the 95\% highest posterior density credible interval. 
Also, as in our previous experiments,  
trace plot of $y_{101}$ for the last 50,000 observations, illustrate the convergence of our MCMC iterations.

\section{Discussion and conclusion}
\label{conclusion}
In this paper we have proposed a novel nonparametric dynamic state space model where the 
latent process is in the circular manifold. 
We assumed that both the observational and the evolutionary functions are time-varying, but have unknown functional forms,
which we model nonparametrically via appropriate Gaussian processes. For this purpose
we derived a suitable Gaussian processes with both linear and circular arguments using kernel convolution.

Previously, some research has been carried out on Gaussian process with circular argument; 
see, for example, \ctn{Dufour76}, \ctn{Gneiting98}. However, most of the previous works considered 
the circular variable as the only argument. The main issue with the procedure of \ctn{Dufour76} 
is that the covariance function turns out to be an infinite sum, and therefore, one has to approximate 
the infinite sum with proper truncation while applying to data. Hence, the question of error of approximation 
lurks in their procedure. \ctn{Gneiting98} provided sufficient conditions under which 
any correlation function on the real line can be treated as a correlation function on circles. 
For that purpose \ctn{Gneiting98} had to bound the argument of the correlation function on a finite interval, 
and therefore, the correlation can not tend to zero for the underlying Gaussian process.   

The kernel convolution method has been used in \ctn{Shafie03}, although they derived the 
Gaussian process on two linear arguments, 
one in ${\mathbb R}$ (real line) and the other in ${\mathbb R}^{+}$ (positive part of the real line). 
\ctn{Adler81} and \ctn{Adler07} dealt with Gaussian processes on manifolds in great details. 
However, they focused on Gaussian processes with arguments only on single manifold. Here we mention that 
although we also use the kernel convolution approach to forming appropriate Gaussian processes, 
our case is substantially different in that our Gaussian process
construction is based on both linear and circular arguments. Moreover, we have chosen our kernel 
appropriately such that the Gaussian process satisfies all desirable smoothness properties. 
The most elegant property of our Gaussian process is that the covariance function becomes $0$ whenever 
$|\theta_1-\theta_2|$ = $\pi/2$.  This implies that whenever two angular observations have orthogonal directions,
their correlation turns out to be $0$ irrespective of the difference in time. Obviously, we also have shown 
that as $|t_1-t_2|$ $\rightarrow$ $\infty$ then the covariance function tends to $0$, that is, 
as the difference in time goes to $\infty$, the correlation goes to $0$.

The main aim of our research is to predict single or multiple future observations given the dynamic data at hand. 
That is, considering the Bayesian paradigm, our main objective is to obtain posterior predictive distributions. 
To achieve the posterior predictive distributions, appropriate MCMC simulation techniques needed to be devised. 
The main MCMC challenge for this model is to simulate the complete latent process; aided by the
look up table concept of \ctn{Bhattacharya07} (see also \ctn{Ghosh14}), appropriately adapted to suit
the circular context, we could create an MCMC algorithm that has demonstrated very reasonable performances
in both simulated and real data situations.

Our model and methods are applied to a simulated data where the data is generated 
from a highly nonlinear model, which is completely different from our own model. This simulation is done 
purposefully to demonstrate that our method is applicable to any nonlinear dynamic model where the 
latent process is in the circular manifold. It is also successfully shown that the future observation is well 
within the 95\% credible region of the posterior predictive density. Quite importantly, almost 
the complete set of true latent variables fell well within their respective high marginal posterior probability 
density regions. The encouraging results are expected 
to provide any practitioner with some degree of latitude in applying our model in any practical context.

To demonstrate the effectiveness of our model 
and methodologies in capturing the underlying latent circular process in real data scenarios, 
we implemented our model on 
wind speed and direction data of a particular location for a particular period of time. 
In this experiment we took 100 observations on wind speed data for implementing our method. 
We pretended that the data on wind direction were unknown. Quite importantly it is noticed that the 
high probability region of the posterior densities associated with the latent process covered most 
of the observed wind direction values, and the underlying highly non-linear and 
discontinuous trend associated with the wind directions has been quite precisely captured. 

Finally, we applied our model to the level of ozone present in the atmosphere 
for a particular location 
over a period of time consisting of hundred observations, where wind direction data, expected
to be associated with the ozone data, are not recorded. 
Even in this real example, our ideas yielded quite encouraging results. In particular, our posterior predictive
density for the set-aside ``future" observation successfully captured the true, set-aside value within 
the 95\% highest posterior density credible interval. 

These two real data analyses ensure 
that our model and methodologies are equally good in predicting the future observations of the observed data 
and in capturing the underlying latent circular process generating the linear observed data.

In fine, we remark that in this paper we assumed the observations $y_t$, $t=1,\ldots, T$, to be in $\mathbb R$. 
However, it is straightforward to extend our theory to $\mathbb R^p$ by suitably adjusting the kernel convolution technique. 
The technique can be extended even to cases where the latent $x_t$, $t=1,\ldots,T$, are also multidimensional. 
To keep the size of the paper reasonable we skip the multivariate part for this paper. 
 

\section*{Acknowledgment}
The authors are thankful to Moumita Das for very useful discussions. The authors are also thankful to the 
anonymous referees and an Associate Editor for their valuable comments and suggestions which helped 
improve our paper significantly.


\appendix
\section{Appendix}
\label{appendix I}
\subsection{Gaussian process on linear and angular component and its properties}
\label{Gaussian-univar}
To define a Gaussian process on linear and angular component we use the well known kernel convolution method. 
Let $k$ be any $d$-dimensional kernel such that

\[
 \int k^2(\mathbf{t})\,d\mathbf{t} < \infty. 
\]

Here we choose two kernels as follows (in case $d=1$)

\[
 k_1(t)=\frac{1}{\psi}\pi^{-1/4} e^{-\frac{1}{2\psi^2}{t}^2},
\]
where $\psi$ $>$ $0$, and
\[
 k_2(t)=\pi^{-1/2}\cos(t)\, I(0\leq t \leq \pi),
\]
a trigonometric kernel.
Based on above two choices of the kernel we propose a new Gaussian process for time and angle as arguments as follows. 
\allowdisplaybreaks
{
\begin{align*}
 X(t,\theta)&=\mu(t,\theta)+\left(\int_{-\infty}^{\infty}\psi^{-1}\pi^{-1/4} e^{-\frac{1}{2\psi^2}(y-t)^2}\,dW(y)\right)\left(\int_{0}^{\pi}\pi^{-1/2}\cos(u-\theta)\,dW(u)\right)
\\[1ex]
&=\mu(t,\theta)+\psi^{-1}\pi^{-3/4}\int_{-\infty}^{\infty}\int_{0}^{\pi}e^{-\frac{1}{2\psi^2}(y-t)^2}\cos(u-\theta)\,dW(u)\,dW(y),
\end{align*}
}
where $\mu(t,\theta)$ is the mean of the process which may depend on time $t$ and angle $\theta$ 
(as in our case mean is assumed to be of the form $\bi{h}(\cdot,\cdot)'\bi{\beta}$, with $\bi{h}(t,\theta)'$ = $(1,t,\cos(\theta),\sin(\theta))$);
$W(\cdot)$ is the one dimensional standard Wiener process. 
Next, we determine the structure of the covariance of our Gaussian process thus constructed.

\subsection{Covariance structure of our Gaussian process}
\label{subsec:covariance}
With these separable kernels we calculate the covariance function of $X(t_1,\theta_1)$ and $X(t_2,\theta_2)$ for  fixed $(t_1,\theta_1)$ and $(t_2,\theta_2)$ as
\allowdisplaybreaks{
\begin{align*}
 \mbox{cov}(X(t_1,\theta_1),X(t_2,\theta_2))&=\psi^{-2}\pi^{-6/4}E\left\{\left(\int_{-\infty}^{\infty}\int_{0}^{\pi}e^{-\frac{1}{2\psi^2}(y-t_1)^2} \cos(u-\theta_1)\,dW(u)\,dW(y)\right)\right.
\\[1ex]
&
\qquad \left.
\left(\int_{-\infty}^{\infty}\int_{0}^{\pi}e^{-\frac{1}{2\psi^2}(y-t_2)^2}\cos(u-\theta_2)\,dW(u)\,dW(y)\right)\right\}
\\[1ex]
&=\psi^{-2}\pi^{-6/4}\int_{-\infty}^{\infty} e^{-\frac{1}{2\psi^2}\left\{(y-t_1)^2+(y-t_2)^2\right\}}\,dy\,\int_{0}^{\pi} \cos (u-\theta_1) \cos (u-\theta_2)\,du
\\[1ex]
&=\frac{1}{2}\psi^{-2}\pi^{-6/4} e^{-\frac{1}{2\psi^2}(t_1^2+t_2^2)}\int_{-\infty}^{\infty} e^{-\frac{1}{2\psi^2}2(y^2-y(t_1+t_2))}\,dy
\\[1ex]
&\qquad \int_{0}^{\pi} \left[\cos (- (\theta_1-\theta_2)) +\cos (2u-(\theta_1+\theta_2))\right]\,du
\\[1ex]
&=\frac{1}{2}\psi^{-2}\pi^{-6/4} e^{-\frac{1}{2\psi^2}(t_1^2+t_2^2)+\frac{1}{4\psi^2}(t_1+t_2)^2}\left\{\int_{-\infty}^{\infty} e^{-\frac{1}{\psi^2}(y-\frac{t_1+t_2}{2})^2}\,dy\right\}
\\[1ex]
&
\qquad
\left\{\pi \cos (|\theta_1-\theta_2|)
+\int_{0}^{\pi}\cos(2u-(\theta_1+\theta_2))\,du\right\}
\\[1ex]
&=
\frac{1}{2}\psi^{-2}\pi^{-6/4} \psi \sqrt{\pi} e^{-\frac{1}{4\psi^2}(t_1-t_2)^2}\pi\cos(|\theta_1-\theta_2|)
\\[1ex]
&=\frac{1}{2}\psi^{-1} e^{-\frac{1}{4\psi^2} |t_1-t_2|^2}\cos(|\theta_1-\theta_2|)
\\[1ex]
& = \sigma^2 \exp \{-\sigma^4 |t_1-t_2|^2\}\cos(|\theta_1-\theta_2|),
\end{align*}
}
where $\sigma^2$ = $\frac{\psi^{-1}}{2}$.

Here it is important to remind the reader that in this paper our motive of introducing the kernels $k_1$ and $k_2$
is entirely different from the other existing works involving circular and spherical data, 
where the goals are density estimation, nonparametric regression
and smoothing (see, for example, \ctn{Hall87}, \ctn{Marzio09a}, \ctn{Marzio09b}, \ctn{Marzio11},
\ctn{Marzio12a}, \ctn{Marzio12b}, \ctn{Marzio14}). Hence, our kernels need not satisfy the optimality
properties required for the aforementioned works.

Indeed, here our goal is to construct an appropriate Gaussian process model for random functions
having both time and angle as arguments. 
The Gaussian process is required to possess desired properties,
such as stationarity in time and angle, zero correlation when the directions are orthogonal
and/or when the time difference tends to infinity, along with desired continuity and smoothness properties. 
Moreover, quite importantly, a closed form
of the covariance of the Gaussian process is also required, which, as we discuss in Section \ref{conclusion} of our paper, 
is difficult to obtain in general. With our kernels $k_1$ and $k_2$, all these properties have been achieved, 
and in this sense, they are optimal.

\input{supp_arxiv}

\newpage
\bibliographystyle{natbib}
\bibliography{irmcmc}
 
\end{document}

%% file: supp_arxiv.tex
\renewcommand\thefigure{S-\arabic{figure}}
\renewcommand\thetable{S-\arabic{table}}
\renewcommand\thesection{S-\arabic{section}}

\setcounter{section}{0}
\setcounter{figure}{0}
\setcounter{table}{0}

\begin{center}
{\bf \Large Supplementary Material}
\end{center}

Throughout, we refer to our main paper \ctn{Mazumder14a} as MB. 

\section{Smoothness properties of our Gaussian process with linear-circular arguments}
\label{sec:smoothness}
Here we assume that $\mu(t,\theta)$ is twice differentiable with respect to $t$ and $\theta$, and that 
the derivatives are bounded. Formally, we assume that
$\frac{\partial ^2 \mu(t,\theta)}{\partial t^2}$, $\frac{\partial ^2 \mu(t,\theta)}{\partial \theta^2}$, 
$\frac{\partial ^2 \mu(t,\theta)}{\partial t \partial \theta}$ (= $\frac{\partial ^2 \mu(t,\theta)}{\partial \theta \partial t }$) 
exist and are bounded.
We denote the covariance function 
$\sigma^2 \exp \{-\sigma^4 |t_1-t_2|^2\}\cos(|\theta_1-\theta_2|)$
(where $\sigma^2$ = $\frac{\psi^{-1}}{2}$)
by $K(|t_1-t_2|,|\theta_1-\theta_2|)$.

\subsection{Mean square continuity:}
\begin{enumerate}
\item
\textbf{With respect to time $t$}
\begin{align*}
&E[X(t+h,\theta)-X(t,\theta)]^2
\\[1ex]
=&E[X(t+h,\theta)]^2+E[X(t,\theta)]^2-2E[X(t+h,\theta)X(t,\theta)]
\\[1ex]
=&K(0,0)+K(0,0)-2K(h,0)
\\[1ex]
=&2(K(0,0)-K(h,0))
\end{align*}
Now as $h$ $\rightarrow$ 0, $E[X(t+h,\theta)-X(t,\theta)]^2$ $\rightarrow$ 0 because of the fact that $K(h,0)$ is continuous in $h$.
\item
\textbf{With respect to angle $\theta$:}
\allowdisplaybreaks{
\begin{align*}
&E[X(t,\theta+\alpha)-X(t,\theta)]^2
\\[1ex]
=&E[X(t,\theta+\alpha)]^2+E[X(t,\theta)]^2-2E[X(t,\theta+\alpha)X(t,\theta)]
\\[1ex]
=&K(0,0)+K(0,0)-2K(0,\alpha)
\\[1ex]
=&2(K(0,0)-K(0,\alpha))
\end{align*}
}
Now as $\alpha$ $\rightarrow$ 0, $E[X(t,\theta+\alpha)-X(t,\theta)]^2$ $\rightarrow$ 0 because of the fact that $K(0,\alpha)$ is continuous in $\alpha$.
\item
\textbf{With respect to time $t$ and angle $\theta$:}
\begin{align*}
&E[X(t+h,\theta+\alpha)-X(t,\theta)]^2
\\[1ex]
=&E[X(t+h,\theta+\alpha)]^2+E[X(t,\theta)]^2-2E[X(t+h,\theta+\alpha)X(t,\theta)]
\\[1ex]
=&K(0,0)+K(0,0)-2K(h,\alpha)
\\[1ex]
=&2(K(0,0)-K(h,\alpha))
\end{align*}
Now as ($h$, $\alpha$) $\rightarrow$ (0,0) then $E[X(t+h,\theta+\alpha)-X(t,\theta)]^2$ $\rightarrow$ 0 because of the fact that $K(h,\alpha)$ is continuous in $h$ and $\alpha$.
\end{enumerate}

\subsection{Mean square differentiability} 
A process $X(\mathbf{u})$, $\mathbf{u}\in \mathbf{R}^{d}$, is said to be 
{\it Mean Square Differentiable} at $\mathbf{u}_{0}$ if for any direction $\mathbf{p}$ there exists 
a process $L_{\mbox{\scriptsize{$\mathbf{u}_{0}$}}}(\mathbf{p})$, linear in $\mathbf{p}$, such that
\[
 X(\mathbf{u}_{0}+\mathbf{p}) = X(\mathbf{u}_{0}) + L_{\mbox{\scriptsize{$\mathbf{u}_{0}$}}}(\mathbf{p}) + R(\mathbf{u}_{0},\mathbf{p}),
\]
where $\mathbf{p}\in \mathbf{R}^{d}$, and $R(\mathbf{u}_{0},\mathbf{p})$ satisfies the following
\[
\frac{R(\mathbf{u}_{0},\mathbf{p})}{||\mathbf{p}||} \rightarrow 0, \mbox{ in } L^2,
\]
with $||\cdot||$ being the usual Euclidean norm (for details see \ctn{Banerjee03}).
\par
However, we have $t\in \mathbb R^{+}$ and $\theta\in [0,2\pi]$, so we can not directly apply 
the definition of mean square differentiability that is appropriate for 
$\mathbb R^{d}$. For our purpose we define a new metric on time and angular space as

\[
 d(t_1,t_2,\theta_1,\theta_2) = |t_1-t_2| + |\theta_1-\theta_2|,
\]
(recall that we have used the angular distance as a metric on the angular space to represent the covariance as a function 
of distance in time and angle). 
Note that $d(\cdot,\cdot,\cdot,\cdot)$ satisfies all the three criteria for being a metric, that is,
\begin{align*}
 &1. \, d(t_1,t_2,\theta_1,\theta_2)\geq 0\\[1ex]
 &2. \, d(t_1,t_2,\theta_1,\theta_2) = 0 \mbox{ iff } t_1=t_2, \theta_1=\theta_2\\[1ex]
 &3. \, d(t_1,t_3,\theta_1,\theta_3)\leq [|t_1-t_2|+|\theta_1-\theta_2|] + [|t_2-t_3|+|\theta_1-\theta_2|]
\\[1ex]
&~~~~~~~~~~~~~~~~~~~~~= d(t_1,t_2,\theta_1,\theta_2)+ d(t_2,t_3,\theta_2,\theta_3)
\end{align*}
With the help of this new metric in time and angular space we define {\it Mean Square Differentiability} in time and circular domain as

\begin{definition}
A process $X(t,\theta)$ is said to be \textbf{Mean Square Differentiable} in $L^2$ sense at $(t_0,\theta_0)$ if 
for any direction $(h,\alpha)$ there exists a process $L_{\mbox{\scriptsize{$t_{0},\theta_0$}}}(h,\alpha)$, linear 
in $h,\alpha$, such that
\[
 X(t_{0}+h,\theta_0+\alpha) = X(t_0,\theta_0) + L_{\mbox{\scriptsize{$t_{0},\theta_0$}}}(h,\alpha) + R(t_0,\theta_0,h,\alpha),
\]
where $R(t_0,\theta_0,h,\alpha)$ satisfies the following condition
\[
\frac{R(t_0,\theta_0,h,\alpha)}{d(h,0,\alpha,0)} \rightarrow 0, \mbox{ in $L^2$ as $d(h,0,\alpha,0)\rightarrow \, 0$}.
\]
\end{definition}
\par
In our case, since our covariance function 
$K(|t_1-t_2|,|\theta_1-\theta_2|)$ 
has partial derivatives of all orders, the partial derivative processes of all orders
exist with covariance structures given by partial derivatives of our covariance function; see Section 2.2 of \ctn{Adler81}
for details. 
In fact, the partial derivative processes are all Gaussian processes, and hence, they are bounded in $L^2$.

Hence, we can 
apply Taylor series expansion to obtain a linear function $L_{\mbox{\scriptsize{$\mathbf{u}_{0}$}}}(\mathbf{p})$. The following 
calculation will make the things clear. Following the multivariate Taylor series expansion (using our new metric) we have
\[
 X(t_{0}+h,\theta_{0}+\alpha)= X(t_{0},\theta_{0}) + h\,\frac{\partial}{\partial t} X(t,\theta)\bigg|_{t=t_0,\theta=\theta_0} 
 + \alpha \,\frac{\partial}{\partial \theta} X(t,\theta)\bigg|_{t=t_0,\theta=\theta_0} 
 + R(t_{0},\theta_{0},h,\alpha),
\]
where $|R(t_{0},\theta_{0},h,\alpha)|$ $\leq \, M^{*}d^2(h,0,\alpha,0)$, with 
$M^{*}$ = max$\bigg\{\left|\frac{\partial^2 X(t,\theta)}{\partial t^2}\right|,\left|\frac{\partial^2 X(t,\theta)}
{\partial t \partial \theta}\right|,\left|\frac{\partial^2 X(t,\theta)}{\partial \theta \partial t}\right|,
\\[1ex]
\left|\frac{\partial^2 X(t,\theta)}{\partial \theta^2}\right|\bigg\}$
(using the analogy with multivariate Taylor series expansion in $\mathbf{R}^{d}$, recall that in the case of 
$\mathbf{R}^d$, $R(\mathbf{u}_{0},\mathbf{p})$ $\leq\, M^{*}||\mathbf{p}||^2$).

Since each of the partial derivative processes is bounded in $L^2$, it is obvious that $M^*$ is also bounded in $L^2$.
Mean square differentiability of our kernel convolved Gaussian process thus follows.

\section{MCMC-based inference}
\label{MCMC inference}
In our MCMC-based inference we include the problem of forecasting
$y_{T+1}$, given the observed data set $\bi{D}_T$. 
The posterior predictive distribution of $y_{T+1}$ given $\bi{D}_T$ is given by
\begin{align}
 \label{eq13:post y_t+1|D_t}
[y_{T+1}|\bi{D}_{T}] &= \int [y_{T+1}|\bi{D}_T,x_0,\ldots,x_{T+1},\bi{\beta}_f,\bi{\beta}_g,
\sigma^2_{\mbox{\scriptsize $\epsilon$}},\sigma^2_{\mbox{\scriptsize $\eta$}},\sigma^2_f,\sigma^2_g]\notag
\\[1ex]
&\qquad \times [x_0,\ldots,x_{T+1},\bi{\beta}_f,\bi{\beta}_g,\sigma^2_{\mbox{\scriptsize $\epsilon$}},
\sigma^2_{\mbox{\scriptsize $\eta$}},\sigma^2_g,\sigma^2_f|\bi{D}_T]\notag
\\[1ex]
&\qquad d\bi{\beta}_f d\bi{\beta}_g d\sigma^2_{\mbox{\scriptsize $\epsilon$}} d\sigma^2_{\mbox{\scriptsize $\eta$}} 
d\sigma^2_g d\sigma^2_f dx_0\ldots dx_{T+1}.
\end{align}
Thus, once we have a sample realization from the joint posterior
 
\noindent $[x_0,\ldots,x_{T+1},\bi{\beta}_f, \bi{\beta}_g, \sigma^2_{\mbox{\scriptsize $\epsilon$}},
\sigma^2_{\mbox{\scriptsize $\eta$}},\sigma^2_g,\sigma^2_f|\bi{D}_T]$, we can generate a realization from
$[y_{T+1}|\bi{D}_{T}]$ by simply simulating from 
$[y_{T+1}|\bi{D}_T,x_0,\ldots,x_{T+1},\bi{\beta}_f,\bi{\beta}_g,
\sigma^2_{\mbox{\scriptsize $\epsilon$}},\sigma^2_{\mbox{\scriptsize $\eta$}},\sigma^2_f,\sigma^2_g]$, conditional on the
realization obtained from the former joint posterior.
Observe that the conditional distribution $[y_{T+1} = f(T+1,x_{T+1}) + \epsilon_{T+1}| 
\bi{D}_T,x_0,\ldots,x_{T+1},\bi{\beta}_f,\sigma^2_{\mbox{\scriptsize $\epsilon$}}, \sigma^2_f]$ is normal with mean
\begin{equation}
 \label{eq14:cond mean Y_T+1|all}
\mu_{y_{T+1}} = \bi{h}(T+1,x_{T+1})'\bi{\beta}_f + \bi{s}_{f,D_{T}}(T+1,x_{T+1})' \bi{A}_{f,D_{T}}^{-1} (\bi{D}_T-
\bi{H}_{D_{T}}\bi{\beta}_f)
\end{equation}
and variance
\begin{equation}
 \label{eq15: cond var Y_T+1|all}
\sigma^2_{y_{T+1}} = \sigma^2_{\mbox{\scriptsize{$\epsilon$}}} + {\sigma_f^2}\left(1-(\bi{s}_{f,D_T}(T+1,x_{T+1}))' 
\bi{A}_{f,D_{T}}^{-1}\bi{s}_{f,D_T}(T+1,x_{T+1})\right).
\end{equation}

Using the auxiliary variables $K_1,\ldots,K_{T+1}$, the posterior distribution of the latent circular variables
and the other parameters can be represented as
\allowdisplaybreaks
{
\begin{align}
  \label{eq16:post of x_0 to x_T+1,parameters|D_T}
&[x_0,x_1,\ldots, x_{T+1},\bi{\beta}_f,\bi{\beta}_g,\sigma^2_{\mbox{\scriptsize $\epsilon$}},
\sigma^2_{\mbox{\scriptsize $\eta$}},\sigma^2_g,\sigma^2_f|\bi{D}_T]\notag
\\[1ex]
&= \sum_{K_1,\ldots,K_{T+1}}\int [x_0,x_1,\ldots, x_T, x_{T+1},\bi{\beta}_f,\bi{\beta}_g,\sigma^2_{\mbox{\scriptsize $\epsilon$}},
\sigma^2_{\mbox{\scriptsize $\eta$}},\sigma^2_g, \sigma^2_f, g^*(1,x_0),\bi{D}_z,K_1,\ldots, K_T,K_{T+1}|\bi{D}_T]\notag
\\[1ex]
&\qquad \times dg^*(1,x_0)d\bi{D}_z \notag
\\[1ex]
&\propto \sum_{K_1,\ldots,K_{T+1}}\int [x_0,x_1,\ldots, x_{T+1},\bi{\beta}_f,\bi{\beta}_g,\sigma^2_{\mbox{\scriptsize $\epsilon$}},
\sigma^2_{\mbox{\scriptsize $\eta$}},\sigma^2_g,\sigma^2_f, g^*(1,x_0),\bi{D}_z,K_1,\ldots, K_T, K_{T+1}, \bi{D}_T]
\notag
\\[1ex]
&\qquad \times dg^*(1,x_0)d\bi{D}_z\notag 
\\[1ex]
&=\sum_{K_1,\ldots,K_{T+1}}\int [\bi{\beta}_f][\bi{\beta}_g][\sigma^2_{\mbox{\scriptsize $\epsilon$}}]
[\sigma^2_{\mbox{\scriptsize $\eta$}}][\sigma^2_g][\sigma^2_f][x_0][g^*(1,x_0)|x_0,\bi{\beta}_g,\sigma^2_g]
[\bi{D}_z|g^*(1,x_0),x_0,\bi{\beta}_g,\sigma^2_g]
\notag
\\[1ex]
&\qquad [x_1|g^*(1,x_0),\sigma^2_{\mbox{\scriptsize $\eta$}}, K_1][K_1|g^*(1,x_0),\sigma^2_{\mbox{\scriptsize $\eta$}}]
[\bi{D}_{T}|x_1,\ldots,x_{T},\bi{\beta}_f,\sigma^2_{\mbox{\scriptsize $\epsilon$}},\sigma^2_f]
\notag
\\[1ex]
&\qquad \prod_{t=2}^{T+1} [x_t|\bi{\beta}_g,\sigma^2_{\mbox{\scriptsize $\eta$}},\sigma^2_g, \bi{D}_z,x_{t-1},K_t] 
\prod_{t=2}^{T+1} [K_t|\bi{\beta}_g,\sigma^2_{\mbox{\scriptsize $\eta$}},\sigma^2_g, \bi{D}_z,x_{t-1}]\, \, 
dg^*(1,x_0)\,d\bi{D}_z. 
\end{align}
}

In order to obtain MCMC samples from 
$[x_0,x_1,\ldots, x_{T+1},\bi{\beta}_f,\bi{\beta}_g,\sigma^2_{\mbox{\scriptsize $\epsilon$}},
\sigma^2_{\mbox{\scriptsize $\eta$}},\sigma^2_g,\sigma^2_f|\bi{D}_T]$, we first carry out MCMC simulations
from the joint posterior which is proportional to integrand (\ref{eq16:post of x_0 to x_T+1,parameters|D_T}). 
Ignoring $g^*(1,x_0)$, $D_z$ and $K_1,\ldots, K_{T+1}$ in these MCMC simulations and storing the realizations
associated with the remaining parameters yield the desired samples.


\subsection{Full conditional distributions}
\label{full conditional univar}
Here we provide the full conditional distributions of the unknowns. 
In what follows, we shall express $[g^*(1,x_0)|x_0,\bi{\beta}_g,\sigma^2_g][\bi{D}_z|g^*(1,x_0),x_0,\bi{\beta}_g,\sigma^2_g]$ 
as $[\bi{D}_z,g^*(1,x_0)|x_0,\bi{\beta}_g,\sigma^2_g]$. 
\allowdisplaybreaks
{ 
\begin{align}
\label{full conditional for beta_f}
[\bi{\beta}_f|\cdots] &\propto [\bi{\beta}_f][\bi{D}_{T}|x_1,\ldots,x_{T},\bi{\beta}_f,\sigma^2_{\mbox{\scriptsize $\epsilon$}}]
\\[1ex]
\label{full conditional for beta_g}
[\bi{\beta}_g|\cdots] &\propto [\bi{\beta}_g] [\bi{D}_z,g^*(1,x_0)|x_0,\bi{\beta}_g,\sigma^2_g] 
\prod_{t=2}^{T+1} [x_t|\bi{\beta}_g,\sigma^2_{\mbox{\scriptsize $\eta$}},\sigma^2_g, \bi{D}_z,x_{t-1},K_t] \prod_{t=2}^{T+1} \left [ K_t|\bi{\beta}_g, \sigma^2_{\mbox{\scriptsize $\eta$}} \right.
\notag
\\
&\quad 
\left. \sigma^2_g,\bi{D}_z,x_{t-1}\right ]
\\
\label{full condtional for sigma^2_epsilon}
[\sigma^2_{\mbox{\scriptsize $\epsilon$}}|\cdots] &\propto [\sigma^2_{\mbox{\scriptsize $\epsilon$}}] 
[\bi{D}_{T}|x_1,\ldots,x_{T},\bi{\beta}_f,\sigma^2_{\mbox{\scriptsize $\epsilon$}}]
\\
\label{full conditional for sigma^2_f}
[\sigma^2_{f}|\cdots] &\propto [\sigma^2_{f}] [\bi{D}_{T}|x_1,\ldots,x_{T},\bi{\beta}_f,\sigma^2_{f}]
\\
\label{full conditional for sigma^2_eta}
[\sigma^2_{\mbox{\scriptsize $\eta$}}|\cdots] &\propto 
[\sigma^2_{\mbox{\scriptsize $\eta$}}] [x_1|g^*(1,x_0),\sigma^2_{\mbox{\scriptsize $\eta$}},K_1] 
[K_1|g^*(1,x_0),\sigma^2_{\mbox{\scriptsize $\eta$}}] \prod_{t=2}^{T+1} 
[x_t|\bi{\beta}_g,\sigma^2_{\mbox{\scriptsize $\eta$}},\sigma^2_g, \bi{D}_z,x_{t-1},K_t]\notag
\\
&\quad ~ \prod_{t=2}^{T+1}  [K_t|\bi{\beta}_g,\sigma^2_g, \sigma^2_{\mbox{\scriptsize $\eta$}},\bi{D}_z,x_{t-1}]
\\
\label{full conditional for sigma^2_g}
[\sigma^2_{g}|\cdots] &\propto [\sigma^2_{g}] [\bi{D}_z,g^*(1,x_0)|x_0,\bi{\beta}_g,\sigma^2_g] 
\prod_{t=2}^{T+1} [x_t|\bi{\beta}_g,\sigma^2_{\mbox{\scriptsize $\eta$}},\sigma^2_{g},\bi{D}_z,x_{t-1},K_t] \prod_{t=2}^{T+1} \left[ K_t|\bi{\beta}_g,\sigma^2_g, \right.
\notag
\\
&\quad ~ 
\left. \sigma^2_{\mbox{\scriptsize $\eta$}},\bi{D}_z,x_{t-1}\right]
\\
\label{full conditional for x_0}
[x_0|\cdots] &\propto [x_0][\bi{D}_z,g^*(1,x_0)|x_0,\bi{\beta}_g,\sigma^2_g]
\\
\label{full conditional for g*(1,x_0)}
[g^*(1,x_0)|\cdots] &\propto [g^*(1,x_0)|x_0,\bi{\beta}_g,\sigma^2_g]
[\bi{D}_z|g^*(1,x_0),x_0,\bi{\beta}_g,\sigma^2_g] [x_1|g^*(1,x_0),x_0,\sigma^2_{\mbox{\scriptsize $\eta$}},K_1] \notag
\\
&\qquad  [K_1|g^*(1,x_0),\sigma^2_{\mbox{\scriptsize $\eta$}}]
\\
\label{full conditional for D_z}
[\bi{D}_z|\cdots] &\propto [\bi{D}_z|g^*(1,x_0),x_0,\bi{\beta}_g,\sigma^2_g] 
\prod_{t=2}^{T+1} [x_t|\bi{\beta}_g,\sigma^2_g, \sigma^2_{\mbox{\scriptsize $\eta$}},\bi{D}_z,x_{t-1},K_t]
\prod_{t=2}^{T+1} \left[ K_t|\bi{\beta}_g,\sigma^2_g, \sigma^2_{\mbox{\scriptsize $\eta$}}, \right.
\notag
\\
&\quad ~ 
\left. \bi{D}_z,x_{t-1}\right]
\\
\label{full conditional for x_1}
[x_1|\cdots] &\propto [x_1|g^*(1,x_0),\sigma^2_{\mbox{\scriptsize $\eta$}}] 
[\bi{D}_{T}|x_1,\ldots,x_{T},\bi{\beta}_f,\sigma^2_{\mbox{\scriptsize $\epsilon$}}]
\notag
\\
&\qquad [x_2|\bi{\beta}_g,\sigma^2_g, \sigma^2_{\mbox{\scriptsize $\eta$}},\bi{D}_z,x_1,K_2] 
[K_2|\bi{\beta}_g,\sigma^2_g, \sigma^2_{\mbox{\scriptsize $\eta$}},\bi{D}_z,x_{1}]
\\
\label{full conditional for x_T+1}
[x_{T+1}|\cdots] &\propto  [x_{T+1}|\bi{\beta}_g,\sigma^2_g, \sigma^2_{\mbox{\scriptsize $\eta$}},\bi{D}_z,x_T,K_{T+1}]
\\
\label{full conditional for x_{t+1}}
[x_{t+1}|\cdots] &\propto [x_{t+1}|\bi{\beta}_g,\sigma^2_g, \sigma^2_{\mbox{\scriptsize $\eta$}},\bi{D}_z,x_{t}] 
[x_{t+2}|\bi{\beta}_g,\sigma^2_g, \sigma^2_{\mbox{\scriptsize $\eta$}},\bi{D}_z,x_{t+1},K_{t+2}] 
\left[K_{t+2}|\bi{\beta}_g,\sigma^2_g, \sigma^2_{\mbox{\scriptsize $\eta$}},
\right.
\notag
\\
&\qquad
\left. \bi{D}_z, x_{t+1}\right] [\bi{D}_{T}|x_1,\ldots,x_{T},\bi{\beta}_f,\sigma^2_{\mbox{\scriptsize $\epsilon$}}],~~ t=1,\ldots, T-1
\end{align}
}
Finally, we write down the full conditional distribution of $K_t$, for $t=1,\ldots, T+1$, as
 \begin{align}
 \label{full conditional of K_1}
 [K_1|\cdots]  &\propto [K_1|g^*(1,x_0),\sigma^2_{\mbox{\scriptsize $\eta$}}] 
 [x_1|g^*(1,x_0),\bi{\beta}_g,\sigma^2_{\mbox{\scriptsize $\eta$}},K_1]
\\[1ex]
\label{full conditional of K_ts}
 [K_t|\cdots] & \propto [x_t|\bi{\beta}_g,\sigma^2_{\mbox{\scriptsize $\eta$}},\bi{D}_z,x_{t-1},K_t] 
 [K_t|\bi{\beta}_g,\sigma^2_{\mbox{\scriptsize $\eta$}},\bi{D}_z,x_{t-1}], ~ t=2,\ldots,T+1.
 \end{align}
\subsubsection{Updating $\bi{\beta}_f$ by Gibbs steps}
The full conditional of $\bi{\beta}_f$ is a multivariate normal distribution with mean
\begin{align}
\label{eq21: mean of full condtional of beta_f}
E[\bi{\beta}_f|\cdots] &=  \{\bi{H}'_{D_{T}}(\sigma^2_f \bi{A}_{f,D_{T}}
+\sigma^2_{\mbox{\scriptsize $\epsilon$}}I)^{-1}\bi{H}_{D_{T}} \,+\, 
\bi{\Sigma}_{\beta_{f,0}}\}^{-1} \notag
\\[1ex]
&\times \{\bi{H}'_{D_{T}}(\sigma^2_f \bi{A}_{f,D_{T}}+\sigma^2_{\mbox{\scriptsize $\epsilon$}}I)^{-1}\bi{D}_{T}\,+\, 
\bi{\Sigma}_{\beta_{f,0}}^{-1}\bi{\beta}_{f,0}\} 
\end{align}
and variance
\begin{align}
\label{eq22: covariance of full condional of beta_f}
V[\bi{\beta}_f|\cdots] =  \{\bi{H}'_{D_{T}}(\sigma^2_f \bi{A}_{f,D_{T}}+\sigma^2_{\mbox{\scriptsize $\epsilon$}}I)^{-1}\bi{H}_{D_{T}} \,
+\, \bi{\Sigma}_{\beta_{f,0}}\}^{-1}.
\end{align}
\subsubsection{Updating $\bi{\beta}_g$}
We first explicitly write down the right hand side of (\ref{full conditional for beta_g}). 
\begin{align}
\label{eq25:rhs of full conditional for beta_g}
&[\bi{\beta}_g] [\bi{D}_z,g^*(1,x_0)|x_0,\bi{\beta}_g]\prod_{t=2}^{T+1} 
[x_t|\bi{\beta}_g,\sigma^2_{\mbox{\scriptsize $\eta$}},\bi{D}_z,x_{t-1},K_t] 
\prod_{t=2}^{T+1} [K_t|\bi{\beta}_g,\sigma^2_{\mbox{\scriptsize $\eta$}},\bi{D}_z,x_{t-1}]
\notag
\\[1ex]
&\propto \exp{\left(-\frac{1}{2}(\bi{\beta}_g-\bi{\beta}_{g,0})'\bi{\Sigma}_{\beta_{g},0}^{-1}(\bi{\beta}_g-\bi{\beta}_{g,0})\right)}
\notag
\\[1ex]
\quad
&\exp{\left(-\frac{1}{2}[(\bi{D}_z,g^*)'-(\bi{H}_{D_{z}}\bi{\beta}_g,\bi{h}'(1,x_0))']'\bi{A}^{-1}_{D_z,g^*(1,x_0)}
[(\bi{D}_z,g^*)'-(\bi{H}_{D_{z}}\bi{\beta}_g,\bi{h}'(1,x_0))']\right)}
\notag
\\[1ex]
&\quad \exp{\left\{-\sum_{i=2}^{T+1}\frac{1}{2\sigma^2_{x_{t}}}(x_t+2\pi K_t-\mu_{x_{t}})^2\right\}} \prod_{t=2}^{T+1} I_{[0,2\pi]}(x_t)
\end{align}
Observe that the denominator of $[x_t|\bi{\beta}_g,\sigma^2_{\mbox{\scriptsize $\eta$}},\bi{D}_z,x_{t-1},K_t]$ 
cancels with the density of 

\noindent
$[K_t|\bi{\beta}_g,\sigma^2_{\mbox{\scriptsize $\eta$}},\bi{D}_z,x_{t-1}]$ 
for each $t=2,\ldots, T+1$. Also we note that the indicator function does not involve $\bi{\beta}_g$ for all $t=2,\ldots ,T+1$. 
Therefore, after simplifying the exponent terms and ignoring the indicator function we can write
\begin{equation}
\label{eq26:Gibbs full conditional of beta_g}
[\bi{\beta}_g|\cdots] \propto \exp{\left\{-\frac{1}{2}(\bi{\beta}_g-\mu_{\beta_g})'
\bi{\Sigma}_{\beta_g}^{-1}(\bi{\beta}_g-\mu_{\beta_g})\right\}},
\end{equation}
where

\begin{align}
\label{eq27: mean of full conditional of beta_g}
& \mu_{\beta_g}= E[\bi{\beta}_g|\cdots] = \left\{ \bi{\Sigma}_{\beta_{g},0}^{-1} 
+ \frac{1}{\sigma^2_g}[\bi{H}_{D_{z}}',\bi{h}(1,x_0)]\bi{A}^{-1}_{D_z,g^*(1,x_0)}[\bi{H}_{D_{z}}',\bi{h}(1,x_0)]' \right.
\notag
\\[1ex]
& \left. 
+ \sum_{t=1}^{T} \frac{\left(\bi{H}_{D_{z}}'\bi{A}^{-1}_{g,D_z} \bi{s}_{g,D_z}(t+1,x_{t})-\bi{h}(t+1,x_{t})\right)
\left(\bi{H}_{D_{z}}'\bi{A}^{-1}_{g,D_z} \bi{s}_{g,D_z}(t+1,x_{t})-\bi{h}(t+1,x_{t})\right)'}{\sigma^2_{x_{t}}}\right\}^{-1}
\notag
\\[1ex]
&\left\{\bi{\Sigma}_{\beta_{g},0}^{-1} \bi{\beta}_{g,0} + \frac{1}{\sigma^2_g} [\bi{H}_{D_{z}}',
\bi{h}(1,x_0)]\bi{A}^{-1}_{D_z,g^*(1,x_0)} [\bi{D}_z,g^*(1,x_0)] \right.
\notag
\\[1ex]
&\left. 
+ \sum_{t=1}^{T} \frac{\left(x_{t+1}+2\pi K_{t+1}-\bi{s}_{g,D_z}(t+1,x_{t})'\bi{A}^{-1}_{g,D_z} \bi{D}_{z}\right)
\left(\bi{h}(t+1,x_{t})-\bi{H}_{D_{z}}'\bi{A}^{-1}_{g,D_z} \bi{s}_{g,D_z}(t+1,x_{t})\right)}{\sigma^2_{x_{t}}}\right\} 
\end{align}
and
\begin{align}
\label{eq28: covariance of full conditional of beta_g} 
&\bi{\Sigma}_{\beta_g} = V[\bi{\beta}_g|\cdots] = \left\{ \bi{\Sigma}_{\beta_{g},0}^{-1} +\frac{1}{\sigma^2_g} [
\bi{H}_{D_{z}}',\bi{h}(1,x_0)]\bi{A}^{-1}_{D_z,g^*(1,x_0)}[\bi{H}_{D_{z}}',\bi{h}(1,x_0)]'\right.
\notag
\\[1ex]
& \left. 
+ \sum_{t=1}^{T} \frac{\left(\bi{H}_{D_{z}}'\bi{A}^{-1}_{g,D_z} \bi{s}_{g,D_z}(t+1,x_{t})-\bi{h}(t+1,x_{t})\right)
\left(\bi{H}_{D_{z}}'\bi{A}^{-1}_{g,D_z} \bi{s}_{g,D_z}(t+1,x_{t})-\bi{h}(t+1,x_{t})\right)'}{\sigma^2_{x_t}}\right\}^{-1}.
\end{align}
Hence $[\bi{\beta}_g|\cdots]$ follows a tri-variate normal distribution with mean and variance 
$\mu_{\beta_g}$ and $\bi{\Sigma}_{\beta_g}$, respectively, and therefore, we update $\beta_g$ using Gibbs sampling.
\subsubsection{Updating $\sigma^2_{f}$ and $\sigma^2_{g}$}
The mathematical form of the full conditional distributions of $\sigma^2_{f}$ and $\sigma^2_{g}$ are not tractable, so 
we update $\sigma^2_{f}$ and $\sigma^2_{g}$ by random walk Metropolis-Hastings steps.

\subsubsection{Updating $\sigma^2_{\mbox{\scriptsize $\epsilon$}}$}
The mathematical form of the full conditional distribution of $\sigma^2_{\mbox{\scriptsize $\epsilon$}}$ is not tractable, 
so we update $\sigma^2_{\mbox{\scriptsize $\epsilon$}}$ by a random walk Metropolis-Hastings step.

\subsubsection{Updating $\sigma^2_{\mbox{\scriptsize $\eta$}}$}
For full conditional distribution of $\sigma^2_{\mbox{\scriptsize $\eta$}}$ right hand side of 
(\ref{full conditional for sigma^2_eta}) simplifies a bit in the sense that the denominator of 
$[x_t|\bi{\beta}_g,\sigma^2_{\mbox{\scriptsize $\eta$}},\bi{D}_z,x_{t-1},K_t]$ cancels with the 
density of $[K_t|\bi{\beta}_g,\sigma^2_{\mbox{\scriptsize $\eta$}},\bi{D}_z,x_{t-1}]$ for $t=2,\ldots,T+1$, 
and the denominator of $[x_1|g^*(1,x_0),\bi{\beta}_g,\sigma^2_{\mbox{\scriptsize $\eta$}},K_1]$ cancels 
with the density of $[K_1|g^*(1,x_0),\bi{\beta}_g,\sigma^2_{\mbox{\scriptsize $\eta$}}]$, which, in turn, 
gives the following form:

\begin{equation}
\label{eq29: full conditional of sigma_eta}
[\sigma^2_{\mbox{\scriptsize $\eta$}}|\cdots] \propto [\sigma^2_{\mbox{\scriptsize $\eta$}}] 
\exp{\left\{-\sum_{i=2}^{T+1}\frac{1}{2\sigma^2_{x_{t}}}(x_t+2\pi K_t-\mu_{x_{t}})^2\right\}} 
\exp{\left\{ -\frac{1}{2\sigma^2_{\mbox{\scriptsize{$\eta$}}}}(x_1+2\pi K_1 - g^*)^2\right\}}.
\end{equation}
However, the above equation does not have a closed form; hence, for updating $\sigma^2_{\mbox{\scriptsize $\eta$}}$ 
as well, we use random walk Metropolis-Hastings.
\subsubsection{Updating $x_0$}
The full conditional distribution of $x_0$ is not tractable and hence again here we use random walk Metropolis-Hastings 
for updating $x_0$. Now note that $x_0$ is a circular random variable, so to update 
$x^{(old)}_0$ to 
$x^{(new)}_0$ 
we use the vonMises distribution with location parameter 
$x^{(old)}_0$.
\subsubsection{Updating $g^*(1,x_0)$}
Equation (\ref{full conditional for g*(1,x_0)}), after cancelling the denominator of 
$[x_1|g^*(1,x_0),x_0,\bi{\beta}_g,\sigma^2_{\mbox{\scriptsize $\eta$}},K_1]$ with the density of 
$[K_1|g^*(1,x_0),x_0,\bi{\beta}_g,\sigma^2_{\mbox{\scriptsize $\eta$}}]$, and ignoring the 
indicator function on $x_0$, reduces to
\begin{equation}
\label{eq30:full conditional for g*|all}
[g^*(1,x_0)|\cdots] \propto [g^*(1,x_0)|x_0,\bi{\beta}_g][\bi{D}_z|g^*(1,x_0),x_0,\bi{\beta}_g] 
\exp{\left\{-\frac{1}{2\sigma^2_{\mbox{\scriptsize{$\eta$}}}} (x_1+2\pi K_1- g^*)^2\right\}}. \notag
\end{equation}
After further simplification the full conditional distribution of $g^*(1,x_0)$ reduces to
\begin{equation}
\label{eq31:Gibbs full conditional of g*}
[g^*(1,x_0)|\cdots] \propto \exp{\left\{-\frac{1}{2\gamma_{g}^2}(g^*-\nu_{g})^2\right\}},
\end{equation}
where 
\begin{align}
\label{eq32: mean of full conditional of g*}
\nu_{g}= E[g^*(1,x_0)|\cdots] &= \left\{ \frac{1}{\sigma^2_{\mbox{\scriptsize $\eta$}}} 
+ \frac{1}{\sigma^2_g} (1+\bi{s}_{g,D_z}(1,x_0)'\bi{\Sigma}^{-1}_{g,D_z}\bi{s}_{g,D_z}(1,x_0))\right\}^{-1}\notag
\\[1ex]
&\qquad \left\{\frac{x_1+2\pi K_1}{\sigma_{\mbox{\scriptsize $\eta$}}^2}
+\frac{1}{\sigma^2_g}(\bi{h}(1,x_0)'\bi{\beta}_g+\bi{s}'_{g,D_z}\bi{\Sigma}^{-1}_{g,D_z}\bi{D}_z^* )\right\}
\end{align}
and
\begin{align}
\label{eq33: variance of full conditional of g*}
\gamma_{g}^2 = V[g^*(1,x_0)|\cdots] =  \left\{ \frac{1}{\sigma^2_{\mbox{\scriptsize $\eta$}}} 
+ \frac{1}{\sigma^2_g}(1+\bi{s}_{g,D_z}(1,x_0)'\bi{\Sigma}^{-1}_{g,D_z}\bi{s}_{g,D_z}(1,x_0))\right\},
\end{align}
with
\begin{equation}
\label{eq34: defn of D_z^*}
\bi{D}_z^*= \bi{D}_z-\bi{H}_{D_{z}}\bi{\beta}_g + \bi{h}(1,x_0)'\bi{\beta}_g\bi{s}_{g,D_{z}}, 
\end{equation}
and
\begin{equation}
\label{eq35: defn of Sigma_g,Dz}
\bi{\Sigma}_{g,D_z} = \bi{A}_{g,D_{z}} - \bi{s}_{g,D_z}(1,x_0) \bi{s}_{g,D_z}(1,x_0)'.
\end{equation}
Hence $[g^*|\cdots]$ follows a normal distribution with mean $\nu_g$ and variance $\gamma_g$. Therefore, 
we update $g^*$ using Gibbs sampling.
\subsubsection{Updating $\bi{D}_z$}
Here also we observe that in the full conditional distribution of $\bi{D}_z$, the denominator of
\newline $[x_t|\bi{\beta}_g,\sigma^2_{\mbox{\scriptsize $\eta$}},\bi{D}_z,x_{t-1},K_t]$ 
cancels with the density of $[K_t|\bi{\beta}_g,\sigma^2_{\mbox{\scriptsize $\eta$}},D_z,x_{t-1}]$ for each $t=2,\ldots,T+1$. 
After simplification it turns out that the full conditional distribution of $\bi{D}_z$ is an $n$-variate normal with mean
\begin{align}
\label{eq35:mean of full conditional of Dz}
E(\bi{D}_z|\cdots) &= \left\{\frac{\bi{\Sigma}_{g,D_z}^{-1}}{\sigma^2_g} +
\bi{A}_{g,D_z}^{-1} \left(\sum_{t=1}^{T} \frac{s_{g,D_z}(t+1,x_{t})s'_{g,D_z}(t+1,x_{t})}{\sigma_{x_{t}}^2}\right) 
\bi{A}_{g,D_z}^{-1}\right\}^{-1}\notag
\\[1ex]
\times 
&\,\left\{ \frac{\bi{\Sigma}_{g,D_z}^{-1}\bi{\mu}_{g,D_z}}{\sigma^2_g} + \bi{A}_{g,D_z}^{-1} \right.
\notag
\\[1ex]
& \left. \quad \sum_{t=1}^{T} \frac{s_{g,D_z}(t+1,x_{t}) \{x_{t+1}+2\pi K_{t+1} - \bi{\beta}'_{g} 
(\bi{h}(1,t+1,x_t)-\bi{H}'_{D_z}\bi{A}_{g,D_z}^{-1}s_{g,D_z}(t+1,x_{t}))\}}{\sigma_{x_{t}^2}}\right\}
\end{align}
and covariance matrix
\begin{align}
\label{eq36:var of full conditional of Dz}
V(\bi{D}_z|\cdots) =  \left\{\frac{\bi{\Sigma}_{g,D_z}^{-1}}{\sigma^2_g} + 
\bi{A}_{g,D_z}^{-1} \left(\sum_{t=1}^{T} \frac{s_{g,D_z}(t+1,x_{t})s'_{g,D_z}(t+1,x_{t})}{\sigma_{x_{t}}^2}\right) 
\bi{A}_{g,D_z}^{-1}\right\}^{-1}.
\end{align}
Therefore, we update $\bi{D}_z$ using Gibbs sampling.
\subsubsection{Updating $x_1$}
For the full conditional distribution of $x_1$ we write down the complete expression of  
(\ref {full conditional for x_1}) as follows:

\begin{align}
\label{eq37:expression for x1|all}
[x_1|\cdots]&\propto 
\frac{\frac{1}{\sqrt{2\pi}\sigma_{\mbox{\scriptsize $\eta$}}}\exp\left(-\frac{1}{2\sigma^2_{\mbox{\scriptsize $\eta$}}}
(x_1+2\pi K_1-g^*)^2\right)I_{[0,2\pi]}(x_1)}{\Phi\left(\frac{2\pi (K_1+1)-g^*}{\sigma_{\mbox{\scriptsize $\eta$}}}\right)
-\Phi\left(\frac{2\pi K_1-g^*}{\sigma_{\mbox{\scriptsize $\eta$}}}\right)}\notag
\\[1ex]
&\quad \exp{\{-\frac{1}{2}(\bi{D}_T-\bi{\mu}_{y_{t}})'\bi{\Sigma}_{y_{t}}^{-1}(\bi{D}_T-\bi{\mu}_{y_{t}})\}}\notag
\\[1ex]
&\quad \frac{1}{\sqrt{2\pi}\sigma_{x_{2}}}\exp\left(-\frac{1}{2\sigma^2_{x_{2}}}(x_2+2\pi K_2-\mu_{x_{2}})^2 \right),
\end{align}
where $\bi{\mu}_{y_{t}}$ and $\bi{\Sigma}_{y_{t}}$ are given by 
(10) and (11) of MB.
Here we note that the denominator of $[x_2|\bi{\beta}_g,\sigma^2_{\mbox{\scriptsize $\eta$}},\bi{D}_z,x_{1},K_2]$ 
cancels with $[K_2|\bi{\beta}_g,\sigma^2_{\mbox{\scriptsize $\eta$}},\bi{D}_z,x_{1}]$. Also we ignore the indicator 
term associated with $x_2$. We note that the term $\Phi\left(\frac{2\pi (K_1+1)-g^*}{\sigma_{\mbox{\scriptsize $\eta$}}}\right)
-\Phi\left(\frac{2\pi K_1-g^*}{\sigma_{\mbox{\scriptsize $\eta$}}}\right)$ does not involve $x_1$. Hence ignoring 
$\Phi\left(\frac{2\pi (K_1+1)-g^*}{\sigma_{\mbox{\scriptsize $\eta$}}}\right)
-\Phi\left(\frac{2\pi K_1-g^*}{\sigma_{\mbox{\scriptsize $\eta$}}}\right)$ we get 

\begin{align}
\label{eq38: final expression for x1|all}
[x_1|\cdots]&\propto 
\frac{1}{\sqrt{2\pi}\sigma_{\mbox{\scriptsize $\eta$}}}
\exp\left(-\frac{1}{2\sigma^2_{\mbox{\scriptsize $\eta$}}}(x_1+2\pi K_1-g^*)^2\right)I_{[0,2\pi]}(x_1)\notag
\\[1ex]
&\quad \exp{\{-\frac{1}{2}(\bi{D}_T-\bi{\mu}_{y_{t}})'\bi{\Sigma}_{y_{t}}^{-1}(\bi{D}_T-\bi{\mu}_{y_{t}})\}}\notag
\\[1ex]
&\quad \frac{1}{\sqrt{2\pi}\sigma_{x_{2}}}\exp\left(-\frac{1}{2\sigma^2_{x_{2}}}(x_2+2\pi K_2-\mu_{x_{2}})^2 \right),
\end{align}

However, it is not possible to get a closed form expression of $[x_1|\cdots]$, so we update it by random walk Metropolis-Hastings. 
\subsubsection{Updating $x_{t+1}, ~ t=1, \ldots, T-1$}
For $x_{t+1}$ we have the same structure as for $x_{1}$, except for some changes in the parameters. 
To be precise, the full conditional distribution can be explicitly written as
\begin{align}
\label{eq39:expression for xt+1|all}
[x_{t+1}|\cdots]&\propto \frac{\frac{1}{\sqrt{2\pi}\sigma_{x_{t+1}}}
\exp\left(-\frac{1}{2\sigma^2_{x_{t+1}}}(x_{t+1}+2\pi K_{t+1}-\mu_{x_{t+1}})^2 \right)I_{[0,2\pi]}(x_{t+1})}
{\Phi\left(\frac{2\pi (K_{t+1}+1)-\mu_{x_{t+1}}}{\sigma_{x_{t+1}}}\right)
-\Phi\left(\frac{2\pi K_{t+1}-\mu_{x_{t+1}}}{\sigma_{x_{t+1}}}\right)}\notag
\\[1ex]
&\quad \frac{1}{\sqrt{2\pi}\sigma_{x_{t+2}}}\exp\left(-\frac{1}{2\sigma^2_{x_{t+2}}}(x_{t+2}+2\pi K_{t+2}-\mu_{x_{t+2}})^2 \right)\notag
\\[1ex]
&\quad \exp{\{-\frac{1}{2}(\bi{D}_T-\bi{\mu}_{y_{t}})'\bi{\Sigma}_{y_{t}}^{-1}(\bi{D}_T-\bi{\mu}_{y_{t}})\}}.
\end{align}
We note here that $\Phi\left(\frac{2\pi (K_{t+1}+1)-\mu_{x_{t+1}}}{\sigma_{x_{t+1}}}\right)
-\Phi\left(\frac{2\pi K_{t+1}-\mu_{x_{t+1}}}{\sigma_{x_{t+1}}}\right)$ does not involve $x_{t+1}$ 
because $\mu_{x_{t+1}}$ and $\sigma_{x_{t+1}}$ depend on $x_{t}$, not on $x_{t+1}$, and hence we can ignore 
the term $\Phi\left(\frac{2\pi (K_{t+1}+1)-\mu_{x_{t+1}}}{\sigma_{x_{t+1}}}\right)
-\Phi\left(\frac{2\pi K_{t+1}-\mu_{x_{t+1}}}{\sigma_{x_{t+1}}}\right)$ and rewrite (\ref{eq39:expression for xt+1|all}) as 

\begin{align}
\label{eq39:expression2 for xt+1|all}
[x_{t+1}|\cdots]&\propto \frac{1}{\sqrt{2\pi}\sigma_{x_{t+1}}}\exp\left(-\frac{1}{2\sigma^2_{x_{t+1}}}
(x_{t+1}+2\pi K_{t+1}-\mu_{x_{t+1}})^2 \right)I_{[0,2\pi]}(x_{t+1})\notag
\\[1ex]
&\quad \frac{1}{\sqrt{2\pi}\sigma_{x_{t+2}}}\exp\left(-\frac{1}{2\sigma^2_{x_{t+2}}}(x_{t+2}+2\pi K_{t+2}-\mu_{x_{t+2}})^2 \right)\notag
\\[1ex]
&\quad \exp{\{-\frac{1}{2}(\bi{D}_T-\bi{\mu}_{y_{t}})'\bi{\Sigma}_{y_{t}}^{-1}(\bi{D}_T-\bi{\mu}_{y_{t}})\}}.
\end{align}

Here also the expression of the full conditional distribution of $x_{t+1}$ is not tractable. So, we adopt random walk 
Metropolis-Hastings to update $x_{t+1}$, for $t=1,\ldots, T$.
\subsubsection{Updating $x_{T+1}$}
The full conditional distribution of $x_{T+1}$ has probability density function 
of the form (29) of MB with parameters 
\begin{equation}
\label{eq23:mean parameter for conditional x_T+1}
\mu_{x_{T+1}} = \bi{h}(1,x_T)'\bi{\beta}_g + \bi{s}_{g,D_z}(T+1,x_{T})' \bi{A}_{g,D_{z}}^{-1} (\bi{D}_z-\bi{H}_{D_{z}}\bi{\beta}_{g})
\end{equation}                                                                                                                                           
and 
\begin{equation}
\label{eq24:variance parameter for conditional x_T+1} 
\sigma^2_{x_{T+1}} = \sigma^2_{\mbox{\scriptsize{$\eta$}}} + \sigma^2_g \{1-\bi{s}_{g,D_z}(T+1,x_T)' 
\bi{A}_{g,D_{z}}^{-1}\bi{s}_{g,D_z}(T+1,x_{T})\}.
\end{equation}
We note here that given all unknowns except $x_{T+1}$, $x_{T+1}+2\pi K_{T+1}$ follows a truncated normal 
distribution with left side truncation at $2\pi K_{T+1}$ and right side truncation at $2\pi (K_{T+1}+1)$ 
($K_{T+1}$ is constant in this case). Hence we update $x_{T+1}+2\pi K_{T+1}$ using Gibbs sampling and 
then subtract  $2\pi K_{T+1}$ from it to update $x_{T+1}$.
\subsubsection{Updating $K_t,~t=1,\ldots,T+1$}
The full conditional distribution of $K_1$ reduces to the following form
\begin{equation}
\label{eq39:final expression for K1|all}
[K_1|\cdots]\propto \frac{1}{\sqrt{2\pi}\sigma_{\mbox{\scriptsize $\eta$}}}
\exp\left(-\frac{1}{2\sigma^2_{\mbox{\scriptsize $\eta$}}}(x_1+2\pi K_1-g^*)^2\right) I_{\{\ldots,-1,0,1,\ldots\}}(K_1),
\end{equation}
and similarly the full conditional distribution of $K_t$ becomes
\begin{equation}
\label{eq40:final expression for K_t|all}
[K_t|\cdots]\propto \frac{1}{\sqrt{2\pi}\sigma_{x_{t}}}
\exp\left(-\frac{1}{2\sigma^2_{x_{t}}}(x_t+2\pi K_t-\mu_{x_{t}})^2 \right)I_{\{\ldots,-1,0,1,\ldots\}}(K_t),
\end{equation}
for $t=2,\ldots,T+1$. We update $K_t$, for $t=1,\ldots,K+1$, by random walk Metropolis-Hastings.

%% file: dycirc.bbl
\begin{thebibliography}{}

\bibitem[Adler(1981)Adler]{Adler81}
Adler, R.~J. (1981).
\newblock {\em {T}he {G}eometry of {R}andom {F}ields\/}.
\newblock Wiley, London.

\bibitem[Adler and Taylor(2007)Adler and Taylor]{Adler07}
Adler, R.~J. and Taylor, J.~E. (2007).
\newblock {\em {R}andom {F}ields and {G}eometry\/}.
\newblock Springer, Boston.

\bibitem[Banerjee and Gelfand(2003)Banerjee and Gelfand]{Banerjee03}
Banerjee, S. and Gelfand, A.~E. (2003).
\newblock {O}n {S}moothness {P}roperties of {S}patial {P}rocesses.
\newblock {\em Journal of Multivariate Analysis\/}, {\bf 84}, 85--100.

\bibitem[Bhattacharya(2007)Bhattacharya]{Bhattacharya07}
Bhattacharya, S. (2007).
\newblock {A} {S}imulation {A}pproach to {B}ayesian {E}mulation of {C}omplex
  {D}ynamic {C}omputer {M}odels.
\newblock {\em Bayesian Analysis\/}, {\bf 2}, 783--816.

\bibitem[Bickel and Doksum(2007)Bickel and Doksum]{Bickel07}
Bickel, J.~P. and Doksum, A.~K. (2007).
\newblock {\em {M}athematical {S}tatistics\/}.
\newblock Number 2nd Edition in Voume I. Pearson Prentice Hall.

\bibitem[Box and Tiao(1973)Box and Tiao]{Box73}
Box, E. P.~G. and Tiao, C.~G. (1973).
\newblock {\em {B}ayesian {I}nference in {S}tatistical {A}nalysis\/}.
\newblock Addison Wesley Publishing Co.

\bibitem[Dufour and Roy(1976)Dufour and Roy]{Dufour76}
Dufour, M.~J. and Roy, R. (1976).
\newblock {O}n {S}pectral {E}stimation for a {H}omogeneous {R}andom {P}rocess
  on the {C}ircle.
\newblock {\em Stochastic Processes and their Applications\/}, {\bf 4},
  107--120.

\bibitem[Durbin and Koopman(2001)Durbin and Koopman]{Durbin01}
Durbin, J. and Koopman, S.~J. (2001).
\newblock {\em {T}ime {S}eries {A}nalysis by {S}tate {S}pace {M}ethods\/}.
\newblock Oxford University Press, Oxford.

\bibitem[Epp {\em et~al.}(1971)Epp, Tukey, and Watson]{Epp71}
Epp, R.~J., Tukey, J.~W., and Watson, G.~S. (1971).
\newblock {T}esting {U}nit {V}ectors for {C}orrelation.
\newblock {\em Journal of Geophysical Research\/}, {\bf 76}, 8480--8483.

\bibitem[Ghosh {\em et~al.}(2014)Ghosh, Mukhopadhyay, Roy, and
  Bhattacharya]{Ghosh14}
Ghosh, A., Mukhopadhyay, S., Roy, S., and Bhattacharya, S. (2014).
\newblock {B}ayesian {I}nference in {N}onpaametric {D}ynamic {S}tate {S}pace
  {M}odels.
\newblock {\em Statistical Methodology\/}, {\bf 21}, 35--48.

\bibitem[Gneiting(1998)Gneiting]{Gneiting98}
Gneiting, T. (1998).
\newblock {S}imple {T}ests for the {V}alidity of {C}orrelation {F}unction
  {M}odels on the {C}ircle.
\newblock {\em Statistics and Probability Letters\/}, {\bf 39}, 119--122.

\bibitem[Hall {\em et~al.}(1987)Hall, Watson, and Cabrera]{Hall87}
Hall, P., Watson, G.~S., and Cabrera, J. (1987).
\newblock {K}ernel {D}ensity {E}stimation with {S}pherical {D}ata.
\newblock {\em Biometrika\/}, {\bf 74}, 751--762.

\bibitem[Hassanzadeh {\em et~al.}(2008)Hassanzadeh, Hosseinibalam, and
  Omidvari]{HHO08}
Hassanzadeh, S., Hosseinibalam, F., and Omidvari, M. (2008).
\newblock {S}tatistical methods and regression analysis of stratospheric ozone
  and meteorological variables in {I}sfahan.
\newblock {\em Physica A\/}, {\bf 387}, 2317--2327.

\bibitem[Holzmann {\em et~al.}(2006)Holzmann, Munk, Suster, and
  Zucchini]{Holzmann06}
Holzmann, H., Munk, A., Suster, M., and Zucchini, W. (2006).
\newblock {H}idden {M}arkov {M}odels for {C}ircular and {L}inear-{C}ircular
  {T}ime {S}eries.
\newblock {\em Environmental and Ecological Statistics\/}, {\bf 13}, 325--347.

\bibitem[Jammalamadaka and Lund(2006)Jammalamadaka and Lund]{Jamma06}
Jammalamadaka, R.~S. and Lund, J.~U. (2006).
\newblock {T}he {E}ffect of {W}ind {D}irection on {O}zone {L}evels: {A} {C}ase
  {S}tudy.
\newblock {\em Environmental and Ecological Statistics\/}, {\bf 13}, 287--298.

\bibitem[Liu(2001)Liu]{Liu01}
Liu, J. (2001).
\newblock {\em {M}onte {C}arlo {S}trategies in {S}cientific {C}omputing\/}.
\newblock Springer-Verlag, New York.

\bibitem[Marzio and Taylor(2009)Marzio and Taylor]{Marzio09b}
Marzio, M.~D. and Taylor, C.~C. (2009).
\newblock {U}sing {S}mall {B}ias {N}onparametric {D}ensity {E}stimators for
  {C}onfidence {I}nterval {E}stimation.
\newblock {\em Journal of Nonparametric Statistics\/}, {\bf 21}, 229--240.

\bibitem[Marzio {\em et~al.}(2009)Marzio, Panzera, and Taylor]{Marzio09a}
Marzio, M.~D., Panzera, A., and Taylor, C.~C. (2009).
\newblock {L}ocal {P}olynomial {R}egression for {C}ircular {P}redictors.
\newblock {\em Statistics and Probability Letters\/}, {\bf 79}, 2066--2075.

\bibitem[Marzio {\em et~al.}(2011)Marzio, Panzera, and Taylor]{Marzio11}
Marzio, M.~D., Panzera, A., and Taylor, C.~C. (2011).
\newblock {D}ensity {E}stimation on the {T}orus.
\newblock {\em Journal of Statistical Planning and Inference\/}, {\bf 141},
  2156--2173.

\bibitem[Marzio {\em et~al.}(2012a)Marzio, Panzera, and Taylor]{Marzio12b}
Marzio, M.~D., Panzera, A., and Taylor, C.~C. (2012a).
\newblock {N}onparametric {R}egression for {C}ircular {R}esponses.
\newblock {\em Scandinavian Journal of Statistics\/}, {\bf 40}, 238--255.

\bibitem[Marzio {\em et~al.}(2012b)Marzio, Panzera, and Taylor]{Marzio12a}
Marzio, M.~D., Panzera, A., and Taylor, C.~C. (2012b).
\newblock {N}onparametric {S}moothing and {P}rediction for {N}on-{L}inear
  {C}ircular {T}ime {S}eries.
\newblock {\em Journal of Time Series Analysis\/}, {\bf 33}, 620--630.

\bibitem[Marzio {\em et~al.}(2014)Marzio, Panzera, and Taylor]{Marzio14}
Marzio, M.~D., Panzera, A., and Taylor, C.~C. (2014).
\newblock {N}onparametric {R}egression for {S}pherical {D}ata.
\newblock {\em Journal of the American Statistical Association\/}, {\bf 109},
  748--763.

\bibitem[Mazumder and Bhattacharya(2014a)Mazumder and
  Bhattacharya]{Mazumder14a}
Mazumder, S. and Bhattacharya, S. (2014a).
\newblock {B}ayesian {N}onparametric {D}ynamic {S}tate-{S}pace {M}odeling with
  {C}ircular {L}atent {S}tates.
\newblock Submitted.

\bibitem[Mazumder and Bhattacharya(2014b)Mazumder and
  Bhattacharya]{Mazumder14b}
Mazumder, S. and Bhattacharya, S. (2014b).
\newblock {S}upplement to ``{B}ayesian {N}onparametric {D}ynamic
  {S}tate-{S}pace {M}odeling with {C}ircular {L}atent {S}tates''.
\newblock Submitted.

\bibitem[Ravindran and Ghosh(2011)Ravindran and Ghosh]{Ravindran11}
Ravindran, P. and Ghosh, S. (2011).
\newblock {B}ayesian {A}nalysis of {C}ircular {D}ata {U}sing {W}rapped
  {D}istributions.
\newblock {\em Journal of Statistical Theory and Practice\/}, {\bf 4}, 1--20.

\bibitem[Reinsel and Tiao(1987)Reinsel and Tiao]{RT87}
Reinsel, C.~G. and Tiao, C.~G. (1987).
\newblock {I}mpact of {C}hlorofluoromethanes on {S}tratospheric {O}zone: {A}
  {S}tatistical {A}nalysis of {O}zone {D}ata for {T}rends.
\newblock {\em Journal of the American Statistical Association\/}, {\bf 82},
  20--30.

\bibitem[Robert and Casella(2004)Robert and Casella]{Robert04}
Robert, C.~P. and Casella, G. (2004).
\newblock {\em {M}onte {C}arlo {S}tatistical {M}ethods\/}.
\newblock Springer-Verlag, New York.

\bibitem[Shafie {\em et~al.}(2003)Shafie, Siegmund, Sigal, and
  Worsley]{Shafie03}
Shafie, K., Siegmund, D., Sigal, B., and Worsley, K.~J. (2003).
\newblock {R}otation {S}pace {R}andom {F}ields with an {A}pplication to f{MRI}
  {D}ata.
\newblock {\em Annals of Statistics\/}, {\bf 31}, 1732--1771.

\bibitem[Shumway and Stoffer(2011)Shumway and Stoffer]{Shumway11}
Shumway, R.~H. and Stoffer, D.~S. (2011).
\newblock {\em {T}ime {S}eries {A}nalysis and {I}ts {A}pplications\/}.
\newblock Springer-Verlag, New York.

\bibitem[Smith(1989)Smith]{S89}
Smith, R.~L. (1989).
\newblock {E}xtreme {V}alue {A}nalysis of {E}nvironmental {T}ime {S}eries: {A}n
  {A}pplication to {T}rend {D}etection in {G}round-{L}evel {O}zone.
\newblock {\em Statistical Science\/}, {\bf 4}, 367--377.

\end{thebibliography}
